%% file: dsd2018.tex
\documentclass[conference]{IEEEtran}
\usepackage{graphicx,tikz}
\usepackage{url}
\usepackage{multicol}
\usepackage{amsmath}
\usepackage{algorithm2e}
\usepackage[noend]{algpseudocode}
\usepackage[english]{babel}
\usepackage{graphicx}
\usepackage{subcaption}
\newcommand\bred[1]{\textcolor{red}{\textbf{#1}}}

\begin{document}

\title{Optimization of Circuits for IBM's five-qubit Quantum Computers}

\author{\IEEEauthorblockN{Gerhard W.\ Dueck$^{\dagger}$, Anirban Pathak$^{*}$,  Md Mazder Rahman$^{\dagger}$, Abhishek Shukla$^{*}$, and Anindita Banerjee$^{*}$  }
\IEEEauthorblockA{  $^{\dagger}$Faculty of Computer Science,  University of New Brunswick, Canada\\
$^{*}$Department of Physics and Material Science Engineering, Jaypee Institute of Information Technology, Noida, INDIA}}

\maketitle

\begin{abstract} \label{section:abstract}
IBM has made several quantum computers available to researchers around the world via cloud services. 
Two architectures with five qubits, one with 16, and one with 20 qubits are available to run experiments.
The IBM architectures implement gates from the Clifford+T gate library.
However, each architecture only implements a subset of the possible CNOT gates.
In this paper, we show how Clifford+T circuits can efficiently be mapped into the two IBM quantum computers with 5 qubits. 
We further present an algorithm and a set of circuit identities that may be used to optimize the Clifford+T circuits in terms of gate count and number of levels. 
It is further shown that the optimized circuits can considerably reduce the gate count and number of levels and thus produce results with better fidelity.
\end{abstract}


\section{Introduction} \label{section:introduction}

Interest in quantum computing has received a boost with the availability of IBM Quantum Computers (www.research.ibm.com/ibm-q) that enables users to run quantum experiments.
Recently, various computational tasks (e.g., Bell state discrimination \cite{SSP2017experimental}, teleportation using optimal quantum resources \cite{SSTP2017design}, quantum permutation algorithm \cite{YZ2017optimization}, quantum cheque \cite{BBP2017experimental}, testing Mermin inequalities \cite{AL2016experimental}, etc.) have been realized using IBM Quantum Computers.
The IBM Quantum Computers work for a limited number of qubits (5, 16, or 20) and one of the problem with the current implementation is fidelity (i.e., noise in the output states)~\cite{DBLP:journals/corr/LinkeMRDFLWM17}. 
There is a direct (although not linear) correlation between the number of gates and the fidelity of the experimentally obtained output state and the desired output state which would have been obtained in the ideal scenario. %(i.e., in the absence of noise). 
In fact,  in Ref. \cite{SSP2017experimental} and other recent works using IBM quantum computers it is clearly observed that the state fidelity reduces considerably with the increase in the gate count. Further, because of this fact, each IBM quantum computer imposes a bound on the maximum number of gates that can be used in a single experiment.
It is therefore imperative to keep the number of gates in circuits a minimum. 
The objective of the present work is to design mapping algorithms for IBM quantum computers and to obtain optimized IBM circuits for some computational tasks of particular interest. We obtain several equivalent circuits for the same tasks and compared their gate count and number of levels.

The rest of the paper is structured as follows: Section~\ref{section:BG} briefly describes the fundamentals of quantum computing and basic steps of synthesizing quantum circuits from Boolean functions.
Section~\ref{section:Mapping} illustrates mapping CNOT gates into IBM's QX2 and QX4 Architectures. Section~\ref{section:Optimization} presents an optimization method with examples. 
The significance of the proposed method is shown with experiments in Section~\ref{section:Results}. 
Some additional ideas on how to further reduce the number of gates, are presented in Section~\ref{section:Ideas}. 
The paper concludes with some observations and directions for future research in Section~\ref{section:Conclusion}.

\section{Background} \label{section:BG}
Quantum systems perform logic operations according to the principles of quantum mechanics that are entirely different from their classical counterpart~\cite{MC:2000}. 
%\sout{Every quantum system can be described by a state vector in a 2-dimensional complex vector space.} 
The fundamental unit of information in quantum computation is the qubit. The state of a qubit is represented by a vector in a two-dimensional complex vector space~\cite{MC:2000} %. An arbitrary state of a qubit is described
and is expressed as follows:
\begin{equation}\label{eq}|\psi\rangle=\alpha|0\rangle+\beta|1\rangle=\left(\begin{array}{c}
\alpha\\
\beta\end{array}\right).\end{equation}
The coefficients $\alpha$ and $\beta$ are complex numbers called probability amplitudes that satisfy the constraint $\vert \alpha|^{2}+\vert \beta|^{2}=1$. The states $\vert 0\rangle$ and $\vert 1\rangle$ are known as computational basis states that are analogous to the states $0$ and $1$ respectively of a classical bit. The states of a qubit can be  the superposition of the states $\vert 0\rangle$ and $\vert 1\rangle$. With the superposition of states, an infinite state space can be found in the quantum computation~\cite{MC:2000}. Time evolution of a quantum system (transition between states) is described by a unitary transformation. 
Quantum gates are unitary matrices and operations are inherently reversible---except for measurements and the effect of noise. 
IBM architectures implement quantum circuits with the Clifford+T gate library. 
However, it is very difficult to directly implement logic functions in quantum circuits with this library. Hence, synthesis of logic functions into IBM architectures can be generally viewed as a series of the following steps:   
\begin{enumerate}
	\item Transform the given Boolean function into a reversible one (Multiple controlled Toffoli gates are most frequently used here~\cite{ar:ft});
	\item Decompose the Toffoli circuit into gates from the Clifford+T gate library~\cite{DBLP:conf/rc/MillerSD14,DBLP:conf/ismvl/MillerWS11}.
	\item Map the gates to one of the IBM architectures.
\end{enumerate}
In this paper, we address the last point. It should also be noted that some quantum algorithms will not start with a Boolean specification. However, they will also need the last step in the above flow.

\section{Mapping CNOT gates to IBM's QX2 and QX4 Architectures} \label{section:Mapping}
IBM's 5-qubit quantum computers (QX2 and QX4) \cite{IBMQ} support gates from the Clifford+T gate library~\cite{Sel2015-newsynth}. 
Not all CNOT gates can be implemented directly, due to their architectures (shown in Figure~\ref{IBM_QX2_Achichtecture} and~\ref{IBM_QX4_Achichtecture}). 
The arrows shown in these figures indicate which CNOT gates are implemented. Specifically, a qubit shown at the tail of the arrow can only work as control qubit and the qubit shown at the head of the arrow can only work as target qubit. 
Thus, in QX2, a CNOT can be implemented between Q3 and Q4, with Q3 as control qubit and Q4 as the target qubit, but the opposite cannot be done directly. Further, no CNOT gate can be implemented directly between Q1 and Q3 as they are not connected.
However, all CNOT gates can be implemented with the aid of some additional gates.
The two architectures are considered in turn.

\subsection{QX2 Architecture}
\begin{figure}
\begin{subfigure}{1.7in}
  \centering
  \includegraphics[width=1.2in]{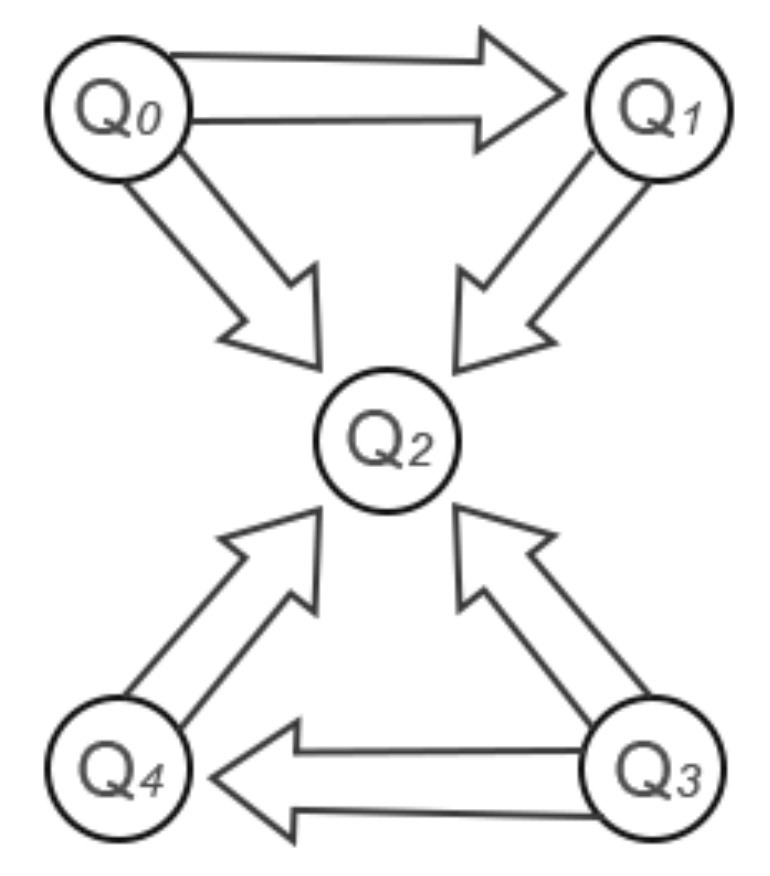}
  \caption{QX2~\cite{IBMQ2_info}.}
  \label{IBM_QX2_Achichtecture}
\end{subfigure}%
\begin{subfigure}{1.7in}
  \centering
  \includegraphics[width=1.3in]{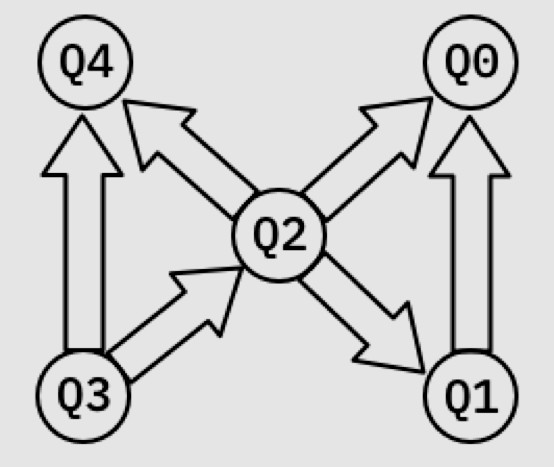}
  \caption{QX4~\cite{IBMQ4_info}.}
  \label{IBM_QX4_Achichtecture}
\end{subfigure}
\caption{Architectures of the IBM's five-qubit computers.}
\label{IBM_Achichtectures}
\end{figure}

In the {\bf QX2} architecture, shown in Figure~\ref{IBM_QX2_Achichtecture},
only six CNOTs (indicated by 6 arrows) out of ${{5}\choose{2}} = 20$  possible CNOT gates can be realized directly.
The remaining 14 gates can be realized in two different ways.

Six can be realized by interchanging the target and control.
This can be accomplished by adding four Hadamard gates (an example is shown below).

\begin{center}
	\input{invert_cnot.tikz}
\end{center}

%%%%%%%%%%%%%%%%%%

For the remaining eight CNOTs the target can be interchanged with qubit 2 and then changed back.
This is illustrated with an example as shown below. 
A similar transformation is always possible in the other cases, too, since qubit 2 can be a target for any control.

\vspace{3mm}
\noindent

	\resizebox{3.4in}{!}{\input{complex12.tikz}}
	
\vspace{3mm}
	\resizebox{2in}{!}{\input{complex3.tikz}}

This adds six Hadamard and four CNOT gates. 
The number of levels increases by at most eight.

However, there is an alternative transformation---first proposed in~\cite{DBLP:journals/corr/RahmanD15}.
As shown below, this will also add six Hadamard gates, but only three CNOT gates. 
Also, the number of levels increase by at most seven---one less than the previous method.
An additional benefit may come from the fact that the two Hadamard gates at the end of the transformation may result in further reductions.

\vspace{3mm}
	\resizebox{3.2in}{!}{\input{template.tikz}}

\subsection{QX4 Architecture}
Now consider the {\bf QX4} architecture as shown in Figure~\ref{IBM_QX4_Achichtecture}.
Again, six CNOTs  can be realized directly and
 six can be realized by interchanging the target and control (as explained above.)
 For the following CNOTs (denoted as a pair of qubits, where the first qubit is the control and the second is the target) $\{ (0,4), (1, 4), (3, 0), (3, 1), (4, 0), (4, 1)\}$ 
we may
 exchange the control with qubit 2 at a cost of 10 additional gates using the swap gate approach, 
 or with 9 additional gates using the template approach. 
 Both methods are illustrated with an example below.
 
 \vspace{3mm}
\noindent
	\resizebox{3.5in}{!}{\input{QX4swap_template.tikz}}	
 
For the CNOT pairs $ (0, 3)$ and $(1, 3)$  the target can be exchanged with qubit 2 and  the direction of the CNOT is reversed.
With some simplification (as shown below), this also adds 10 gates to the circuit.

\vspace{3mm}
\noindent
	\resizebox{3.0in}{!}{\input{QX4complex.tikz}}	
	
	\resizebox{3.0in}{!}{\input{QX4complex2.tikz}}

A similar template based transformation as with QX2 can be applied here.
Again, this  leads to one fewer gate and one fewer level.

\vspace{3mm}
\noindent
	\resizebox{3.0in}{!}{\input{QX4template.tikz}}	
	
A careful analysis reveals that an even better transformation is possible (as shown below). 
In this case the number of Hadamard gates is further reduced by two.

 \vspace{3mm}
\noindent
	\resizebox{2.7in}{!}{\input{QX4template2.tikz}}

Circuit identities described in this section  facilitate the conversion of a given circuit into a circuit that can be realized using a particular architecture of IBM's 5 qubit quantum computers (QX2 or QX4). 
However, a circuit obtained in such a way would not necessarily be optimal. Thus, to obtain better fidelity in a real experiment, we would require to develop methods for optimization of the obtained Clifford+T circuit remaining within the constrains imposed by the particular architecture. In the following section, we aim to propose such a method.
	
\section{Optimization} \label{section:Optimization}
The problem addressed in this section can be stated as follows: given a quantum circuit with gates from the Clifford+T gate library, find the optimal implementation for an IBM five-qubit quantum computer.
The only gates in such a circuit that are not supported directly, are some CNOT gates.
It has been shown in the previous section, that there are mappings for each of those gates such that they can be realized with some added cost.
The cost is measured by counting the number of gates and the number of levels.
It is easy to see, that some permutation of qubits may result in differing final cost~\cite{ZPW2018mapping}.
It is well known, that some gates in a quantum circuit may be interchanged. 
Specifically, if the matrices of two adjacent gates commute, then the gates can be interchanged.
This property has been used extensively in the optimization of reversible circuits~\cite{MDM:2005}. 
In what follows, we will use this property for the optimization of the  quantum circuits designed for the implementation in IBM's 5-qubit quantum computers.

Two adjacent gates can be interchanged if one of the following conditions applies:
\begin{itemize}
	\item They do not involve the same qubits;
	\item $T$, $S$, $S^{\dagger}$,$T^{\dagger}$, and $Z$ will commute with each other even if they act on the same qubit;
	\item $CNOT(x,y)$ commutes with $CNOT(z,w)$ if $x \ne w$ and $y \ne z$.
\end{itemize}

The following reduction rules of adjacent gates can be applied:
\begin{itemize}
	\item Any consecutive self inverse gates can be eliminated (such as $H$, $X$, $Z$, and $CNOT$);
	\item $TT=S$
	\item $SS=Z$
	\item $T^{\dagger}T^{\dagger}=S^{\dagger}$
	\item $S^{\dagger}S^{\dagger}=Z$
	\item $T^{\dagger}T = I$, $S^{\dagger}S = I$.
\end{itemize}

The reduction rules are applied by moving each gate as far as possible to the left in the circuit. 
At each position, apply a reduction rule if possible. 
This is only a heuristic, and some possible reductions may be missed.
The main algorithm is outlined below. 
The function mapIBM$(C_{in})$ will map the necessary CNOT gates to the given IBM architecture.
This function will differ according to the target architecture.
Since the target architectures only have five qubits, it is feasible to consider all $5! = 120$ permutations.
The function reduce$()$ will attempt to minimized the number of gates according to the criteria given above.

\begin{algorithm}
\TitleOfAlgo{Optimize Circuit}
	\SetAlgoLined
	\KwData{$C_{in}$: circuit to be optimized (Cliffort+T gates)}
 	\KwResult{$C_{best}$: the best circuit found}
	$C_{best}$ = mapIBM($C_{in}$)\;
	reduce($C_{best}$)\;
	\ForEach{ line permutation  $C_p$ in $C_{in}$ }{
		$C_{tmp}$ = mapIBM($C_{in}$)\;
		reduce($C_{tmp}$)\;
		\If{ cost($C_{tmp}$) $<$ cost($C_{best}$)}{
			$C_{best}$ = $C_{tmp}$
		}
	}
\end{algorithm}

\subsection{An Illustrative Example}

Li  {\it et al.}\ analyzed theoretical proposals for the implementation of approximate quantum adders~\cite{Li2017}.
These adders were optimized using a genetic algorithm. 
They show one example of realizing such a circuit with IBM's QX2 computer.
Specifically, the following  approximate quantum adder was reported in  Figure~4 of~\cite{Li2017}.

\vspace{10pt}
\noindent
\includegraphics[width=3.5in]{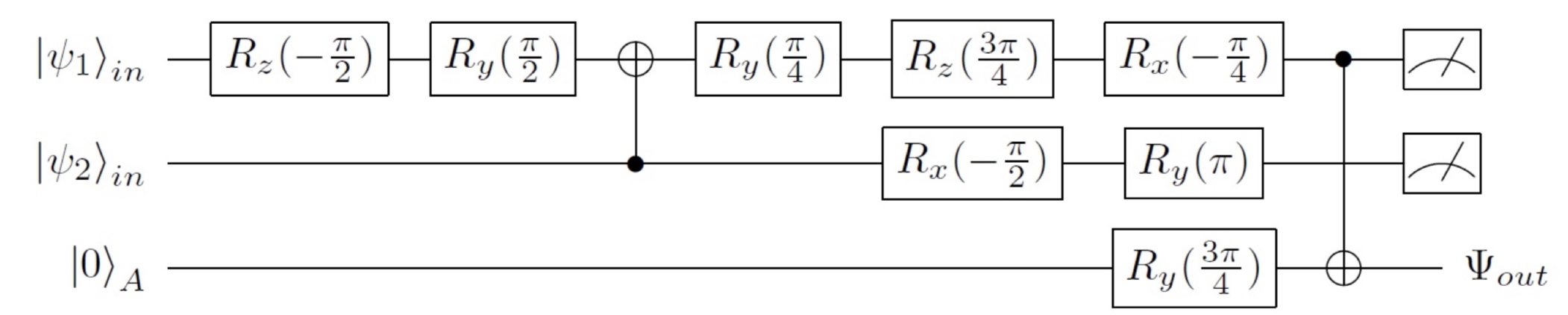}
\vspace{10pt}

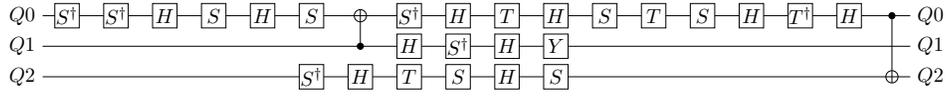
\begin{figure*}
	\centering
	\resizebox{5.0in}{!}{\input{fig4_orig.tikz}}
	\caption{The circuit with gates from the Clifford+T gate library.}
	\label{adder_org}
\end{figure*}

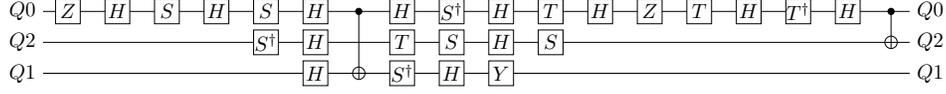
\begin{figure*}
	\centering
	\resizebox{5.0in}{!}{\input{fig4_orig_opt.tikz}}
	\caption{An adder optimized for IBM QX2. }
	\label{adder_org_opt}
\end{figure*}

This translates to a circuit with Clifford+T gates as shown in Figure~\ref{adder_org}.
In  \cite{Li2017} this is mapped to IBM QX2 quantum computer at a cost of 41 gates and 27 levels. %(see Figure~\ref{Li_fig6}).
With our proposed algorithm, the circuit shown in Figure~\ref{adder_org_opt} is obtained.
This is accomplished by interchanging the first two qubits and applying some reductions. 
Interestingly, the cost is reduced to 28 gates and 18 levels.
This is a significant reduction from the original  proposal as it requires 32\% fewer gates and 33\% fewer levels.
Such reduction has a big impact, given the poor fidelity of the QX2.
Mapping the adder to the QX4, yields a similar result with a different qubit permutation 
%, shown in Figure~\ref{adder_org_opt_QX4}.
(not shown here, due to the lack of space).

\section{Experimental Results} \label{section:Results}

The algorithms presented in this paper have been integrated into RevKit~\cite{RevKit}.
Runtimes for the experiments are negligible, and therefore not reported. 
Some experimental results are shown in Table~\ref{tab:Exp}.
For every benchmark circuit, there are two rows in the table.
The first row shows the circuit without permuting the qubits. 
There are four ways that this can be accomplished:
\begin{enumerate}
 	\item	For QX2 using the swap gate principle; 
	\item	For QX4 using the swap gate principle; 
	\item	For QX2 using the template principle; 
	\item	For QX4 using the template principle.
\end{enumerate}
The second row shows the best result from all permutations of the qubits.
For each method the number of gates as well as the number of levels are shown in the table.
The best results are shown in red.

\begin{table*}
  \begin{center}
    \caption{Experimental Results.}
    \label{tab:Exp}
  \begin{tabular}{|l|c|r|r|r|r|r|r|r|r|}
   \hline
   		& & \multicolumn{4}{c|}{Swap transformation }	&	\multicolumn{4}{c|}{Template transformation } \\ 
   Name & lines & QX2 & Dep. &  QX4 & Dep. & GX2 & Dep. &  QX4 & Dep. \\ \hline  \hline
17  \cite{Qbench} & 4 & 141 & 92 & 125 & 72 & 131 & 91 & 105 & 69 \\ \hline
 		&  & 101 & 68 & 111 & 76 & 116 & 79 & \bred{87} & \bred{56} \\ \hline
12  \cite{Qbench} & 5 & 168 & 111 & 154 & 101 & 137 & 91 & 111 & 74 \\ \hline
 		&  & 104 & 60 & 98 & 54 & 85 & 47 & \bred{83} & \bred{45} \\ \hline
a2x\_c \cite{DBLP:conf/rc/MillerSD14} & 4 & 85 & 58 & 75 & 46 & 76 & 54 & 60 & 36 \\ \hline
 		&  & 59 & 41 & 59 & 38 & 55 & 40 & \bred{40} & \bred{28} \\ \hline
a3x\_c \cite{DBLP:conf/rc/MillerSD14} & 5 & 176 & 127 & 140 & 101 & 160 & 117 & 104 & 77 \\ \hline
 		&  & 86 & 56 & 56 & 43 & 86 & 56 & \bred{56} & \bred{43} \\ \hline
7  \cite{Qbench} & 5 & 184 & 114 & 182 & 108 & 174 & 108 & 160 & 104 \\ \hline
 		&  & 122 & 73 & 130 & 73 & 125 & 76 & \bred{111} & \bred{64} \\ \hline
Full\_Adder\_c  \cite{DBLP:conf/rc/MillerSD14} & 4 & 60 & 48 & 60 & 40 & 60 & 50 & 50 & 35 \\ \hline
 		&  & 28 & 22 & 32 & 26 & 27 & \bred{22} & \bred{26} & 23 \\ \hline
Toffoli\_c  \cite{DBLP:conf/rc/MillerSD14} & 3 & 17 & 14 & 31 & 23 & 17 & 14 & 31 & 23 \\ \hline
		 &  & \bred{17} & \bred{14} & \bred{17} & \bred{14} & \bred{17} & \bred{14} & \bred{17} & \bred{14} \\ \hline
4mod5-v0\_18 \cite{ZPW2018mapping} & 5 & 209 & 134 & 211 & 130 & 192 & 118 & 184 & 111 \\ \hline
 		&  & 167 & 104 & 157 & 98 & 152 & 98 & \bred{132} & \bred{80} \\ \hline
3\_17\_e & 3 & 47 & 26 & 49 & 26 & 47 & 26 & 49 & 26 \\ \hline
 		&  & \bred{41} & \bred{23} & \bred{41} & \bred{23} & \bred{41} & \bred{23} & \bred{41} & \bred{23} \\ \hline
01  \cite{Qbench} & 5 & 149 & 89 & 157 & 92 & 141 & 83 & 135 & 84 \\ \hline
 &  & \bred{77} & \bred{38} & 77 & 40 & \bred{77} & \bred{38} & 77 & 40 \\ \hline

\hline 
\end{tabular} 
\end{center}
\end{table*}

As can be seen from Table~\ref{tab:Exp}, significant reductions can be obtained by simply permuting the qubits.
For example, the circuit {\tt 4mod5-v0\_18}, designed for QX4, was reduced from  211 (with swap transformations) to 157 gates (26\%). By applying template transformation the gate count can be further reduced to 132, for a total reduction of 37\%.
The number of levels went from 130 to 80 (a 38\% reduction).
As expected,  ``template'' transformations yield better results in most cases.

\section{Additional Optimization Ideas} \label{section:Ideas}

There are some circuit identities that can be particularly useful in optimizing circuits to be implemented in IBM's 5-qubit quantum computers. To illustrate this point, in what follows, we would develop some rewriting rules (circuit identities) that can facilitate gate reductions in Clifford+T circuits.
The first such useful circuit identity is obtained as (where both $T$ gates may be replaced with $T^\dagger$ gates):
	
	\begin{center}
		\input{cnotTcnot.tikz}
	\end{center}

We will now illustrate the effectiveness of this circuit identity with an example. Consider the full adder implementation from \cite{DBLP:conf/rc/MillerSD14} (with the last two lines interchanged):

\vspace{3mm}
\noindent
\resizebox{3.5in}{!}{
	\input{Full_Adder_c.tikz}
}
\vspace{3mm}

All gates, except the CNOT coloured in red are directly supported by the IBM QX2 architecture.
The CNOT in question can be transformed at a cost of 10 additional gates using the ``swap'' transformation or 9 additional gates using the ``template'' transformation.
However, there is a better solution as shown below.
First, interchange lines 2 and 3 right after gate 13. 
This has the effect that the output qubits are permuted---as indicated by the output labelling.
This results in the following:

\vspace{3mm}
\noindent
\resizebox{3.5in}{!}{
	\input{Full_Adder_c_a.tikz}
}
\vspace{3mm}

Now the last two CNOTs need to be flipped.
However, as explained above this can be done at no cost.
This results in the following circuit:

\vspace{3mm}
\noindent
\resizebox{3.5in}{!}{
	\input{Full_Adder_c_b.tikz}
}
\vspace{3mm}

The circuit can be further reduced by eliminating the two blue CNOT gates.
Finally, we have the following circuit with an additional 4 gates---much better than the 8 in the first approach
(which started with 10 additional gates, but a pair of CNOT gates could be removed.)

\vspace{3mm}
\noindent
\resizebox{3.5in}{!}{
	\input{Full_Adder_c_c.tikz}
}
\vspace{3mm}

\section{Conclusion} \label{section:Conclusion}

Optimization of circuits is always an important step in the design of computing devices.
With %\textcolor{blue}{the} 
integrated semiconductor circuits, optimizations may have different objectives: chip area, power consumption, and speed are the most common  parameters that require to be optimized.
With quantum computing, an important parameter that can quantitatively describe the quality of the corresponding quantum circuit is the fidelity of the obtained output state and the desired output state. %The more is the state fidelity the better is the circuit. 
In this context, it would be apt to note that the
 authors of~\cite{SSTP2017design} have shown that a circuit with a larger number of gates has lower fidelity than an equivalent one with fewer gates. 
This is due to the decoherence of quantum states and the noise in the system (channel noise and lower than unity gate fidelity).
%They have used experiments to verify this claim.
Experimental results support this claim.

In this paper, we have presented tools that can help the circuit designer to find circuits with potentially fewer gates, and thus help increase the fidelity of of the results obtained in the experiments.
Even if a Clifford+T circuit---of the function to be implemented---is available, it may not be possible to implement it directly on one of the IBM quantum computers.
This is due to the fact, that not all CNOT gates are available. 
Here, we have provided a clear prescription for obtaining Clifford+T circuit which is implementable in the target IBM quantum computer.

Finally, we show some ways in which additional optimizations are possible.
The next step will be to scale the tools presented here to quantum computers with 16 and 20 qubits.
The translation of CNOT gates and the application of reduction rules will be pretty straightforward.
However, it is not feasible
to check all 16!\ or 20! permutations that would be required for the larger quantum computers. 
Consequently, smart ways of reducing the search space must be developed. 
We may further note that  the method adapted in this work is quite general can easily be extended to improve the performance of the IBM quantum computer based experiments reported in Refs. \cite{SSP2017experimental, SSTP2017design, YZ2017optimization, BBP2017experimental, AL2016experimental} and references therein by optimizing the corresponding circuits and thus improving the fidelity. 
Finally, we conclude the paper with an expectation that this work will influence the performance (actually improve the fidelity) of many future experiments involving IBM quantum computers.

\bibliographystyle{IEEEtran}
\bibliography{lit_myrev}

\end{document}

%% file: invert_cnot.tikz
\begin{tikzpicture}[scale=1.000000,x=1pt,y=1pt]
\filldraw[color=white] (0.000000, -7.500000) rectangle (111.000000, 22.500000);
% Drawing wires
% Line 1: l0 W i0 o0
\draw[color=black] (0.000000,15.000000) -- (111.000000,15.000000);
\draw[color=black] (0.000000,15.000000) node[left] {$i0$};
% Line 2: l1 W i1 o1
\draw[color=black] (0.000000,0.000000) -- (111.000000,0.000000);
\draw[color=black] (0.000000,0.000000) node[left] {$i1$};
% Done with wires; drawing gates
% Line 3: l1 +l0
\draw (9.000000,15.000000) -- (9.000000,0.000000);
\filldraw (9.000000, 0.000000) circle(1.500000pt);
\begin{scope}
\draw[fill=white] (9.000000, 15.000000) circle(3.000000pt);
\clip (9.000000, 15.000000) circle(3.000000pt);
\draw (6.000000, 15.000000) -- (12.000000, 15.000000);
\draw (9.000000, 12.000000) -- (9.000000, 18.000000);
\end{scope}
% Line 4: =
\draw[fill=white,color=white] (24.000000, -6.000000) rectangle (39.000000, 21.000000);
\draw (31.500000, 7.500000) node {$=$};
% Line 5: l0 H
\begin{scope}
\draw[fill=white] (57.000000, 15.000000) +(-45.000000:8.485281pt and 8.485281pt) -- +(45.000000:8.485281pt and 8.485281pt) -- +(135.000000:8.485281pt and 8.485281pt) -- +(225.000000:8.485281pt and 8.485281pt) -- cycle;
\clip (57.000000, 15.000000) +(-45.000000:8.485281pt and 8.485281pt) -- +(45.000000:8.485281pt and 8.485281pt) -- +(135.000000:8.485281pt and 8.485281pt) -- +(225.000000:8.485281pt and 8.485281pt) -- cycle;
\draw (57.000000, 15.000000) node {$H$};
\end{scope}
% Line 6: l1 H
\begin{scope}
\draw[fill=white] (57.000000, -0.000000) +(-45.000000:8.485281pt and 8.485281pt) -- +(45.000000:8.485281pt and 8.485281pt) -- +(135.000000:8.485281pt and 8.485281pt) -- +(225.000000:8.485281pt and 8.485281pt) -- cycle;
\clip (57.000000, -0.000000) +(-45.000000:8.485281pt and 8.485281pt) -- +(45.000000:8.485281pt and 8.485281pt) -- +(135.000000:8.485281pt and 8.485281pt) -- +(225.000000:8.485281pt and 8.485281pt) -- cycle;
\draw (57.000000, -0.000000) node {$H$};
\end{scope}
% Line 7: l0 +l1
\draw (78.000000,15.000000) -- (78.000000,0.000000);
\filldraw (78.000000, 15.000000) circle(1.500000pt);
\begin{scope}
\draw[fill=white] (78.000000, 0.000000) circle(3.000000pt);
\clip (78.000000, 0.000000) circle(3.000000pt);
\draw (75.000000, 0.000000) -- (81.000000, 0.000000);
\draw (78.000000, -3.000000) -- (78.000000, 3.000000);
\end{scope}
% Line 8: l0 H
\begin{scope}
\draw[fill=white] (99.000000, 15.000000) +(-45.000000:8.485281pt and 8.485281pt) -- +(45.000000:8.485281pt and 8.485281pt) -- +(135.000000:8.485281pt and 8.485281pt) -- +(225.000000:8.485281pt and 8.485281pt) -- cycle;
\clip (99.000000, 15.000000) +(-45.000000:8.485281pt and 8.485281pt) -- +(45.000000:8.485281pt and 8.485281pt) -- +(135.000000:8.485281pt and 8.485281pt) -- +(225.000000:8.485281pt and 8.485281pt) -- cycle;
\draw (99.000000, 15.000000) node {$H$};
\end{scope}
% Line 9: l1 H
\begin{scope}
\draw[fill=white] (99.000000, -0.000000) +(-45.000000:8.485281pt and 8.485281pt) -- +(45.000000:8.485281pt and 8.485281pt) -- +(135.000000:8.485281pt and 8.485281pt) -- +(225.000000:8.485281pt and 8.485281pt) -- cycle;
\clip (99.000000, -0.000000) +(-45.000000:8.485281pt and 8.485281pt) -- +(45.000000:8.485281pt and 8.485281pt) -- +(135.000000:8.485281pt and 8.485281pt) -- +(225.000000:8.485281pt and 8.485281pt) -- cycle;
\draw (99.000000, -0.000000) node {$H$};
\end{scope}
% Done with gates; drawing ending labels
\draw[color=black] (111.000000,15.000000) node[right] {$o0$};
\draw[color=black] (111.000000,0.000000) node[right] {$o1$};
% Done with ending labels; drawing cut lines and comments
% Done with comments
\end{tikzpicture}

%% file: complex12.tikz
\begin{tikzpicture}[scale=1.000000,x=1pt,y=1pt]
\filldraw[color=white] (0.000000, -7.500000) rectangle (348.000000, 52.500000);
% Drawing wires
% Line 1: l0 W i0
\draw[color=black] (0.000000,45.000000) -- (348.000000,45.000000);
\draw[color=black] (0.000000,45.000000) node[left] {$i0$};
% Line 2: l1 W i1
\draw[color=black] (0.000000,30.000000) -- (348.000000,30.000000);
\draw[color=black] (0.000000,30.000000) node[left] {$i1$};
% Line 3: l2 W i2
\draw[color=black] (0.000000,15.000000) -- (348.000000,15.000000);
\draw[color=black] (0.000000,15.000000) node[left] {$i2$};
% Line 4: l3 W i3
\draw[color=black] (0.000000,0.000000) -- (348.000000,0.000000);
\draw[color=black] (0.000000,0.000000) node[left] {$i3$};
% Done with wires; drawing gates
% Line 5: l3 +l0
\draw (9.000000,45.000000) -- (9.000000,0.000000);
\filldraw (9.000000, 0.000000) circle(1.500000pt);
\begin{scope}
\draw[fill=white] (9.000000, 45.000000) circle(3.000000pt);
\clip (9.000000, 45.000000) circle(3.000000pt);
\draw (6.000000, 45.000000) -- (12.000000, 45.000000);
\draw (9.000000, 42.000000) -- (9.000000, 48.000000);
\end{scope}
% Line 6: =
\draw[fill=white,color=white] (24.000000, -6.000000) rectangle (39.000000, 51.000000);
\draw (31.500000, 22.500000) node {$=$};
% Line 7: l0 l2 SWAP
\draw (54.000000,45.000000) -- (54.000000,15.000000);
\begin{scope}
\draw (51.878680, 42.878680) -- (56.121320, 47.121320);
\draw (51.878680, 47.121320) -- (56.121320, 42.878680);
\end{scope}
\begin{scope}
\draw (51.878680, 12.878680) -- (56.121320, 17.121320);
\draw (51.878680, 17.121320) -- (56.121320, 12.878680);
\end{scope}
% Line 8: l3 +l2
\draw (72.000000,15.000000) -- (72.000000,0.000000);
\filldraw (72.000000, 0.000000) circle(1.500000pt);
\begin{scope}
\draw[fill=white] (72.000000, 15.000000) circle(3.000000pt);
\clip (72.000000, 15.000000) circle(3.000000pt);
\draw (69.000000, 15.000000) -- (75.000000, 15.000000);
\draw (72.000000, 12.000000) -- (72.000000, 18.000000);
\end{scope}
% Line 9: l0 l2 SWAP
\draw (90.000000,45.000000) -- (90.000000,15.000000);
\begin{scope}
\draw (87.878680, 42.878680) -- (92.121320, 47.121320);
\draw (87.878680, 47.121320) -- (92.121320, 42.878680);
\end{scope}
\begin{scope}
\draw (87.878680, 12.878680) -- (92.121320, 17.121320);
\draw (87.878680, 17.121320) -- (92.121320, 12.878680);
\end{scope}
% Line 10: =
\draw[fill=white,color=white] (105.000000, -6.000000) rectangle (120.000000, 51.000000);
\draw (112.500000, 22.500000) node {$=$};
% Line 11: l0 +l2
\draw (135.000000,45.000000) -- (135.000000,15.000000);
\filldraw (135.000000, 45.000000) circle(1.500000pt);
\begin{scope}
\draw[fill=white] (135.000000, 15.000000) circle(3.000000pt);
\clip (135.000000, 15.000000) circle(3.000000pt);
\draw (132.000000, 15.000000) -- (138.000000, 15.000000);
\draw (135.000000, 12.000000) -- (135.000000, 18.000000);
\end{scope}
% Line 12: l0 H
\begin{scope}
\draw[fill=white] (156.000000, 45.000000) +(-45.000000:8.485281pt and 8.485281pt) -- +(45.000000:8.485281pt and 8.485281pt) -- +(135.000000:8.485281pt and 8.485281pt) -- +(225.000000:8.485281pt and 8.485281pt) -- cycle;
\clip (156.000000, 45.000000) +(-45.000000:8.485281pt and 8.485281pt) -- +(45.000000:8.485281pt and 8.485281pt) -- +(135.000000:8.485281pt and 8.485281pt) -- +(225.000000:8.485281pt and 8.485281pt) -- cycle;
\draw (156.000000, 45.000000) node {$H$};
\end{scope}
% Line 13: l2 H
\begin{scope}
\draw[fill=white] (156.000000, 15.000000) +(-45.000000:8.485281pt and 8.485281pt) -- +(45.000000:8.485281pt and 8.485281pt) -- +(135.000000:8.485281pt and 8.485281pt) -- +(225.000000:8.485281pt and 8.485281pt) -- cycle;
\clip (156.000000, 15.000000) +(-45.000000:8.485281pt and 8.485281pt) -- +(45.000000:8.485281pt and 8.485281pt) -- +(135.000000:8.485281pt and 8.485281pt) -- +(225.000000:8.485281pt and 8.485281pt) -- cycle;
\draw (156.000000, 15.000000) node {$H$};
\end{scope}
% Line 14: l0 +l2
\draw (177.000000,45.000000) -- (177.000000,15.000000);
\filldraw (177.000000, 45.000000) circle(1.500000pt);
\begin{scope}
\draw[fill=white] (177.000000, 15.000000) circle(3.000000pt);
\clip (177.000000, 15.000000) circle(3.000000pt);
\draw (174.000000, 15.000000) -- (180.000000, 15.000000);
\draw (177.000000, 12.000000) -- (177.000000, 18.000000);
\end{scope}
% Line 15: l0 H
\begin{scope}
\draw[fill=white] (198.000000, 45.000000) +(-45.000000:8.485281pt and 8.485281pt) -- +(45.000000:8.485281pt and 8.485281pt) -- +(135.000000:8.485281pt and 8.485281pt) -- +(225.000000:8.485281pt and 8.485281pt) -- cycle;
\clip (198.000000, 45.000000) +(-45.000000:8.485281pt and 8.485281pt) -- +(45.000000:8.485281pt and 8.485281pt) -- +(135.000000:8.485281pt and 8.485281pt) -- +(225.000000:8.485281pt and 8.485281pt) -- cycle;
\draw (198.000000, 45.000000) node {$H$};
\end{scope}
% Line 16: l2 H
\begin{scope}
\draw[fill=white] (198.000000, 15.000000) +(-45.000000:8.485281pt and 8.485281pt) -- +(45.000000:8.485281pt and 8.485281pt) -- +(135.000000:8.485281pt and 8.485281pt) -- +(225.000000:8.485281pt and 8.485281pt) -- cycle;
\clip (198.000000, 15.000000) +(-45.000000:8.485281pt and 8.485281pt) -- +(45.000000:8.485281pt and 8.485281pt) -- +(135.000000:8.485281pt and 8.485281pt) -- +(225.000000:8.485281pt and 8.485281pt) -- cycle;
\draw (198.000000, 15.000000) node {$H$};
\end{scope}
% Line 17: l0 +l2
\draw (219.000000,45.000000) -- (219.000000,15.000000);
\filldraw (219.000000, 45.000000) circle(1.500000pt);
\begin{scope}
\draw[fill=white] (219.000000, 15.000000) circle(3.000000pt);
\clip (219.000000, 15.000000) circle(3.000000pt);
\draw (216.000000, 15.000000) -- (222.000000, 15.000000);
\draw (219.000000, 12.000000) -- (219.000000, 18.000000);
\end{scope}
% Line 18: l3 +l2
\draw (237.000000,15.000000) -- (237.000000,0.000000);
\filldraw (237.000000, 0.000000) circle(1.500000pt);
\begin{scope}
\draw[fill=white] (237.000000, 15.000000) circle(3.000000pt);
\clip (237.000000, 15.000000) circle(3.000000pt);
\draw (234.000000, 15.000000) -- (240.000000, 15.000000);
\draw (237.000000, 12.000000) -- (237.000000, 18.000000);
\end{scope}
% Line 19: l0 +l2
\draw (255.000000,45.000000) -- (255.000000,15.000000);
\filldraw (255.000000, 45.000000) circle(1.500000pt);
\begin{scope}
\draw[fill=white] (255.000000, 15.000000) circle(3.000000pt);
\clip (255.000000, 15.000000) circle(3.000000pt);
\draw (252.000000, 15.000000) -- (258.000000, 15.000000);
\draw (255.000000, 12.000000) -- (255.000000, 18.000000);
\end{scope}
% Line 20: l0 H
\begin{scope}
\draw[fill=white] (276.000000, 45.000000) +(-45.000000:8.485281pt and 8.485281pt) -- +(45.000000:8.485281pt and 8.485281pt) -- +(135.000000:8.485281pt and 8.485281pt) -- +(225.000000:8.485281pt and 8.485281pt) -- cycle;
\clip (276.000000, 45.000000) +(-45.000000:8.485281pt and 8.485281pt) -- +(45.000000:8.485281pt and 8.485281pt) -- +(135.000000:8.485281pt and 8.485281pt) -- +(225.000000:8.485281pt and 8.485281pt) -- cycle;
\draw (276.000000, 45.000000) node {$H$};
\end{scope}
% Line 21: l2 H
\begin{scope}
\draw[fill=white] (276.000000, 15.000000) +(-45.000000:8.485281pt and 8.485281pt) -- +(45.000000:8.485281pt and 8.485281pt) -- +(135.000000:8.485281pt and 8.485281pt) -- +(225.000000:8.485281pt and 8.485281pt) -- cycle;
\clip (276.000000, 15.000000) +(-45.000000:8.485281pt and 8.485281pt) -- +(45.000000:8.485281pt and 8.485281pt) -- +(135.000000:8.485281pt and 8.485281pt) -- +(225.000000:8.485281pt and 8.485281pt) -- cycle;
\draw (276.000000, 15.000000) node {$H$};
\end{scope}
% Line 22: l0 +l2
\draw (297.000000,45.000000) -- (297.000000,15.000000);
\filldraw (297.000000, 45.000000) circle(1.500000pt);
\begin{scope}
\draw[fill=white] (297.000000, 15.000000) circle(3.000000pt);
\clip (297.000000, 15.000000) circle(3.000000pt);
\draw (294.000000, 15.000000) -- (300.000000, 15.000000);
\draw (297.000000, 12.000000) -- (297.000000, 18.000000);
\end{scope}
% Line 23: l0 H
\begin{scope}
\draw[fill=white] (318.000000, 45.000000) +(-45.000000:8.485281pt and 8.485281pt) -- +(45.000000:8.485281pt and 8.485281pt) -- +(135.000000:8.485281pt and 8.485281pt) -- +(225.000000:8.485281pt and 8.485281pt) -- cycle;
\clip (318.000000, 45.000000) +(-45.000000:8.485281pt and 8.485281pt) -- +(45.000000:8.485281pt and 8.485281pt) -- +(135.000000:8.485281pt and 8.485281pt) -- +(225.000000:8.485281pt and 8.485281pt) -- cycle;
\draw (318.000000, 45.000000) node {$H$};
\end{scope}
% Line 24: l2 H
\begin{scope}
\draw[fill=white] (318.000000, 15.000000) +(-45.000000:8.485281pt and 8.485281pt) -- +(45.000000:8.485281pt and 8.485281pt) -- +(135.000000:8.485281pt and 8.485281pt) -- +(225.000000:8.485281pt and 8.485281pt) -- cycle;
\clip (318.000000, 15.000000) +(-45.000000:8.485281pt and 8.485281pt) -- +(45.000000:8.485281pt and 8.485281pt) -- +(135.000000:8.485281pt and 8.485281pt) -- +(225.000000:8.485281pt and 8.485281pt) -- cycle;
\draw (318.000000, 15.000000) node {$H$};
\end{scope}
% Line 25: l0 +l2
\draw (339.000000,45.000000) -- (339.000000,15.000000);
\filldraw (339.000000, 45.000000) circle(1.500000pt);
\begin{scope}
\draw[fill=white] (339.000000, 15.000000) circle(3.000000pt);
\clip (339.000000, 15.000000) circle(3.000000pt);
\draw (336.000000, 15.000000) -- (342.000000, 15.000000);
\draw (339.000000, 12.000000) -- (339.000000, 18.000000);
\end{scope}
% Done with gates; drawing ending labels
% Done with ending labels; drawing cut lines and comments
% Done with comments
\end{tikzpicture}

%% file: complex3.tikz
\begin{tikzpicture}[scale=1.000000,x=1pt,y=1pt]
\filldraw[color=white] (0.000000, -7.500000) rectangle (213.000000, 52.500000);
% Drawing wires
% Line 1: l0 W
\draw[color=black] (0.000000,45.000000) -- (213.000000,45.000000);
% Line 2: l1 W
\draw[color=black] (0.000000,30.000000) -- (213.000000,30.000000);
% Line 3: l2 W
\draw[color=black] (0.000000,15.000000) -- (213.000000,15.000000);
% Line 4: l3 W
\draw[color=black] (0.000000,0.000000) -- (213.000000,0.000000);
% Done with wires; drawing gates
% Line 5: =
\draw[fill=white,color=white] (6.000000, -6.000000) rectangle (21.000000, 51.000000);
\draw (13.500000, 22.500000) node {$=$};
% Line 6: l0 +l2
\draw (36.000000,45.000000) -- (36.000000,15.000000);
\filldraw (36.000000, 45.000000) circle(1.500000pt);
\begin{scope}
\draw[fill=white] (36.000000, 15.000000) circle(3.000000pt);
\clip (36.000000, 15.000000) circle(3.000000pt);
\draw (33.000000, 15.000000) -- (39.000000, 15.000000);
\draw (36.000000, 12.000000) -- (36.000000, 18.000000);
\end{scope}
% Line 7: l0 H
\begin{scope}
\draw[fill=white] (57.000000, 45.000000) +(-45.000000:8.485281pt and 8.485281pt) -- +(45.000000:8.485281pt and 8.485281pt) -- +(135.000000:8.485281pt and 8.485281pt) -- +(225.000000:8.485281pt and 8.485281pt) -- cycle;
\clip (57.000000, 45.000000) +(-45.000000:8.485281pt and 8.485281pt) -- +(45.000000:8.485281pt and 8.485281pt) -- +(135.000000:8.485281pt and 8.485281pt) -- +(225.000000:8.485281pt and 8.485281pt) -- cycle;
\draw (57.000000, 45.000000) node {$H$};
\end{scope}
% Line 8: l2 H
\begin{scope}
\draw[fill=white] (57.000000, 15.000000) +(-45.000000:8.485281pt and 8.485281pt) -- +(45.000000:8.485281pt and 8.485281pt) -- +(135.000000:8.485281pt and 8.485281pt) -- +(225.000000:8.485281pt and 8.485281pt) -- cycle;
\clip (57.000000, 15.000000) +(-45.000000:8.485281pt and 8.485281pt) -- +(45.000000:8.485281pt and 8.485281pt) -- +(135.000000:8.485281pt and 8.485281pt) -- +(225.000000:8.485281pt and 8.485281pt) -- cycle;
\draw (57.000000, 15.000000) node {$H$};
\end{scope}
% Line 9: l0 +l2
\draw (78.000000,45.000000) -- (78.000000,15.000000);
\filldraw (78.000000, 45.000000) circle(1.500000pt);
\begin{scope}
\draw[fill=white] (78.000000, 15.000000) circle(3.000000pt);
\clip (78.000000, 15.000000) circle(3.000000pt);
\draw (75.000000, 15.000000) -- (81.000000, 15.000000);
\draw (78.000000, 12.000000) -- (78.000000, 18.000000);
\end{scope}
% Line 10: l2 H
\begin{scope}
\draw[fill=white] (99.000000, 15.000000) +(-45.000000:8.485281pt and 8.485281pt) -- +(45.000000:8.485281pt and 8.485281pt) -- +(135.000000:8.485281pt and 8.485281pt) -- +(225.000000:8.485281pt and 8.485281pt) -- cycle;
\clip (99.000000, 15.000000) +(-45.000000:8.485281pt and 8.485281pt) -- +(45.000000:8.485281pt and 8.485281pt) -- +(135.000000:8.485281pt and 8.485281pt) -- +(225.000000:8.485281pt and 8.485281pt) -- cycle;
\draw (99.000000, 15.000000) node {$H$};
\end{scope}
% Line 11: l3 +l2
\draw (120.000000,15.000000) -- (120.000000,0.000000);
\filldraw (120.000000, 0.000000) circle(1.500000pt);
\begin{scope}
\draw[fill=white] (120.000000, 15.000000) circle(3.000000pt);
\clip (120.000000, 15.000000) circle(3.000000pt);
\draw (117.000000, 15.000000) -- (123.000000, 15.000000);
\draw (120.000000, 12.000000) -- (120.000000, 18.000000);
\end{scope}
% Line 12: l2 H
\begin{scope}
\draw[fill=white] (141.000000, 15.000000) +(-45.000000:8.485281pt and 8.485281pt) -- +(45.000000:8.485281pt and 8.485281pt) -- +(135.000000:8.485281pt and 8.485281pt) -- +(225.000000:8.485281pt and 8.485281pt) -- cycle;
\clip (141.000000, 15.000000) +(-45.000000:8.485281pt and 8.485281pt) -- +(45.000000:8.485281pt and 8.485281pt) -- +(135.000000:8.485281pt and 8.485281pt) -- +(225.000000:8.485281pt and 8.485281pt) -- cycle;
\draw (141.000000, 15.000000) node {$H$};
\end{scope}
% Line 13: l0 +l2
\draw (162.000000,45.000000) -- (162.000000,15.000000);
\filldraw (162.000000, 45.000000) circle(1.500000pt);
\begin{scope}
\draw[fill=white] (162.000000, 15.000000) circle(3.000000pt);
\clip (162.000000, 15.000000) circle(3.000000pt);
\draw (159.000000, 15.000000) -- (165.000000, 15.000000);
\draw (162.000000, 12.000000) -- (162.000000, 18.000000);
\end{scope}
% Line 14: l0 H
\begin{scope}
\draw[fill=white] (183.000000, 45.000000) +(-45.000000:8.485281pt and 8.485281pt) -- +(45.000000:8.485281pt and 8.485281pt) -- +(135.000000:8.485281pt and 8.485281pt) -- +(225.000000:8.485281pt and 8.485281pt) -- cycle;
\clip (183.000000, 45.000000) +(-45.000000:8.485281pt and 8.485281pt) -- +(45.000000:8.485281pt and 8.485281pt) -- +(135.000000:8.485281pt and 8.485281pt) -- +(225.000000:8.485281pt and 8.485281pt) -- cycle;
\draw (183.000000, 45.000000) node {$H$};
\end{scope}
% Line 15: l2 H
\begin{scope}
\draw[fill=white] (183.000000, 15.000000) +(-45.000000:8.485281pt and 8.485281pt) -- +(45.000000:8.485281pt and 8.485281pt) -- +(135.000000:8.485281pt and 8.485281pt) -- +(225.000000:8.485281pt and 8.485281pt) -- cycle;
\clip (183.000000, 15.000000) +(-45.000000:8.485281pt and 8.485281pt) -- +(45.000000:8.485281pt and 8.485281pt) -- +(135.000000:8.485281pt and 8.485281pt) -- +(225.000000:8.485281pt and 8.485281pt) -- cycle;
\draw (183.000000, 15.000000) node {$H$};
\end{scope}
% Line 16: l0 +l2
\draw (204.000000,45.000000) -- (204.000000,15.000000);
\filldraw (204.000000, 45.000000) circle(1.500000pt);
\begin{scope}
\draw[fill=white] (204.000000, 15.000000) circle(3.000000pt);
\clip (204.000000, 15.000000) circle(3.000000pt);
\draw (201.000000, 15.000000) -- (207.000000, 15.000000);
\draw (204.000000, 12.000000) -- (204.000000, 18.000000);
\end{scope}
% Done with gates; drawing ending labels
% Done with ending labels; drawing cut lines and comments
% Done with comments
\end{tikzpicture}

%% file: template.tikz
\begin{tikzpicture}[scale=1.000000,x=1pt,y=1pt]
\filldraw[color=white] (0.000000, -7.500000) rectangle (312.000000, 52.500000);
% Drawing wires
% Line 1: l0 W i0
\draw[color=black] (0.000000,45.000000) -- (312.000000,45.000000);
\draw[color=black] (0.000000,45.000000) node[left] {$i0$};
% Line 2: l1 W i1
\draw[color=black] (0.000000,30.000000) -- (312.000000,30.000000);
\draw[color=black] (0.000000,30.000000) node[left] {$i1$};
% Line 3: l2 W i2
\draw[color=black] (0.000000,15.000000) -- (312.000000,15.000000);
\draw[color=black] (0.000000,15.000000) node[left] {$i2$};
% Line 4: l3 W i3
\draw[color=black] (0.000000,0.000000) -- (312.000000,0.000000);
\draw[color=black] (0.000000,0.000000) node[left] {$i3$};
% Done with wires; drawing gates
% Line 5: l3 +l0
\draw (9.000000,45.000000) -- (9.000000,0.000000);
\filldraw (9.000000, 0.000000) circle(1.500000pt);
\begin{scope}
\draw[fill=white] (9.000000, 45.000000) circle(3.000000pt);
\clip (9.000000, 45.000000) circle(3.000000pt);
\draw (6.000000, 45.000000) -- (12.000000, 45.000000);
\draw (9.000000, 42.000000) -- (9.000000, 48.000000);
\end{scope}
% Line 6: =
\draw[fill=white,color=white] (24.000000, -6.000000) rectangle (39.000000, 51.000000);
\draw (31.500000, 22.500000) node {$=$};
% Line 7: l3 +l2
\draw (54.000000,15.000000) -- (54.000000,0.000000);
\filldraw (54.000000, 0.000000) circle(1.500000pt);
\begin{scope}
\draw[fill=white] (54.000000, 15.000000) circle(3.000000pt);
\clip (54.000000, 15.000000) circle(3.000000pt);
\draw (51.000000, 15.000000) -- (57.000000, 15.000000);
\draw (54.000000, 12.000000) -- (54.000000, 18.000000);
\end{scope}
% Line 8: l2 +l0
\draw (72.000000,45.000000) -- (72.000000,15.000000);
\filldraw (72.000000, 15.000000) circle(1.500000pt);
\begin{scope}
\draw[fill=white] (72.000000, 45.000000) circle(3.000000pt);
\clip (72.000000, 45.000000) circle(3.000000pt);
\draw (69.000000, 45.000000) -- (75.000000, 45.000000);
\draw (72.000000, 42.000000) -- (72.000000, 48.000000);
\end{scope}
% Line 9: l3 +l2
\draw (90.000000,15.000000) -- (90.000000,0.000000);
\filldraw (90.000000, 0.000000) circle(1.500000pt);
\begin{scope}
\draw[fill=white] (90.000000, 15.000000) circle(3.000000pt);
\clip (90.000000, 15.000000) circle(3.000000pt);
\draw (87.000000, 15.000000) -- (93.000000, 15.000000);
\draw (90.000000, 12.000000) -- (90.000000, 18.000000);
\end{scope}
% Line 10: l2 +l0
\draw (108.000000,45.000000) -- (108.000000,15.000000);
\filldraw (108.000000, 15.000000) circle(1.500000pt);
\begin{scope}
\draw[fill=white] (108.000000, 45.000000) circle(3.000000pt);
\clip (108.000000, 45.000000) circle(3.000000pt);
\draw (105.000000, 45.000000) -- (111.000000, 45.000000);
\draw (108.000000, 42.000000) -- (108.000000, 48.000000);
\end{scope}
% Line 11: =
\draw[fill=white,color=white] (123.000000, -6.000000) rectangle (138.000000, 51.000000);
\draw (130.500000, 22.500000) node {$=$};
% Line 12: l3 +l2
\draw (153.000000,15.000000) -- (153.000000,0.000000);
\filldraw (153.000000, 0.000000) circle(1.500000pt);
\begin{scope}
\draw[fill=white] (153.000000, 15.000000) circle(3.000000pt);
\clip (153.000000, 15.000000) circle(3.000000pt);
\draw (150.000000, 15.000000) -- (156.000000, 15.000000);
\draw (153.000000, 12.000000) -- (153.000000, 18.000000);
\end{scope}
% Line 14: l2 H
\begin{scope}
\draw[fill=white] (174.000000, 15.000000) +(-45.000000:8.485281pt and 8.485281pt) -- +(45.000000:8.485281pt and 8.485281pt) -- +(135.000000:8.485281pt and 8.485281pt) -- +(225.000000:8.485281pt and 8.485281pt) -- cycle;
\clip (174.000000, 15.000000) +(-45.000000:8.485281pt and 8.485281pt) -- +(45.000000:8.485281pt and 8.485281pt) -- +(135.000000:8.485281pt and 8.485281pt) -- +(225.000000:8.485281pt and 8.485281pt) -- cycle;
\draw (174.000000, 15.000000) node {$H$};
\end{scope}
% Line 15: l0 H
\begin{scope}
\draw[fill=white] (174.000000, 45.000000) +(-45.000000:8.485281pt and 8.485281pt) -- +(45.000000:8.485281pt and 8.485281pt) -- +(135.000000:8.485281pt and 8.485281pt) -- +(225.000000:8.485281pt and 8.485281pt) -- cycle;
\clip (174.000000, 45.000000) +(-45.000000:8.485281pt and 8.485281pt) -- +(45.000000:8.485281pt and 8.485281pt) -- +(135.000000:8.485281pt and 8.485281pt) -- +(225.000000:8.485281pt and 8.485281pt) -- cycle;
\draw (174.000000, 45.000000) node {$H$};
\end{scope}
% Line 17: l0 +l2
\draw (195.000000,45.000000) -- (195.000000,15.000000);
\filldraw (195.000000, 45.000000) circle(1.500000pt);
\begin{scope}
\draw[fill=white] (195.000000, 15.000000) circle(3.000000pt);
\clip (195.000000, 15.000000) circle(3.000000pt);
\draw (192.000000, 15.000000) -- (198.000000, 15.000000);
\draw (195.000000, 12.000000) -- (195.000000, 18.000000);
\end{scope}
% Line 18: l2 H
\begin{scope}
\draw[fill=white] (216.000000, 15.000000) +(-45.000000:8.485281pt and 8.485281pt) -- +(45.000000:8.485281pt and 8.485281pt) -- +(135.000000:8.485281pt and 8.485281pt) -- +(225.000000:8.485281pt and 8.485281pt) -- cycle;
\clip (216.000000, 15.000000) +(-45.000000:8.485281pt and 8.485281pt) -- +(45.000000:8.485281pt and 8.485281pt) -- +(135.000000:8.485281pt and 8.485281pt) -- +(225.000000:8.485281pt and 8.485281pt) -- cycle;
\draw (216.000000, 15.000000) node {$H$};
\end{scope}
% Line 19: l3 +l2
\draw (237.000000,15.000000) -- (237.000000,0.000000);
\filldraw (237.000000, 0.000000) circle(1.500000pt);
\begin{scope}
\draw[fill=white] (237.000000, 15.000000) circle(3.000000pt);
\clip (237.000000, 15.000000) circle(3.000000pt);
\draw (234.000000, 15.000000) -- (240.000000, 15.000000);
\draw (237.000000, 12.000000) -- (237.000000, 18.000000);
\end{scope}
% Line 20: l2 H
\begin{scope}
\draw[fill=white] (258.000000, 15.000000) +(-45.000000:8.485281pt and 8.485281pt) -- +(45.000000:8.485281pt and 8.485281pt) -- +(135.000000:8.485281pt and 8.485281pt) -- +(225.000000:8.485281pt and 8.485281pt) -- cycle;
\clip (258.000000, 15.000000) +(-45.000000:8.485281pt and 8.485281pt) -- +(45.000000:8.485281pt and 8.485281pt) -- +(135.000000:8.485281pt and 8.485281pt) -- +(225.000000:8.485281pt and 8.485281pt) -- cycle;
\draw (258.000000, 15.000000) node {$H$};
\end{scope}
% Line 21: l0 +l2
\draw (279.000000,45.000000) -- (279.000000,15.000000);
\filldraw (279.000000, 45.000000) circle(1.500000pt);
\begin{scope}
\draw[fill=white] (279.000000, 15.000000) circle(3.000000pt);
\clip (279.000000, 15.000000) circle(3.000000pt);
\draw (276.000000, 15.000000) -- (282.000000, 15.000000);
\draw (279.000000, 12.000000) -- (279.000000, 18.000000);
\end{scope}
% Line 23: l2 H
\begin{scope}
\draw[fill=white] (300.000000, 15.000000) +(-45.000000:8.485281pt and 8.485281pt) -- +(45.000000:8.485281pt and 8.485281pt) -- +(135.000000:8.485281pt and 8.485281pt) -- +(225.000000:8.485281pt and 8.485281pt) -- cycle;
\clip (300.000000, 15.000000) +(-45.000000:8.485281pt and 8.485281pt) -- +(45.000000:8.485281pt and 8.485281pt) -- +(135.000000:8.485281pt and 8.485281pt) -- +(225.000000:8.485281pt and 8.485281pt) -- cycle;
\draw (300.000000, 15.000000) node {$H$};
\end{scope}
% Line 24: l0 H
\begin{scope}
\draw[fill=white] (300.000000, 45.000000) +(-45.000000:8.485281pt and 8.485281pt) -- +(45.000000:8.485281pt and 8.485281pt) -- +(135.000000:8.485281pt and 8.485281pt) -- +(225.000000:8.485281pt and 8.485281pt) -- cycle;
\clip (300.000000, 45.000000) +(-45.000000:8.485281pt and 8.485281pt) -- +(45.000000:8.485281pt and 8.485281pt) -- +(135.000000:8.485281pt and 8.485281pt) -- +(225.000000:8.485281pt and 8.485281pt) -- cycle;
\draw (300.000000, 45.000000) node {$H$};
\end{scope}
% Done with gates; drawing ending labels
% Done with ending labels; drawing cut lines and comments
% Done with comments
\end{tikzpicture}

%% file: QX4swap_template.tikz
\begin{tikzpicture}[scale=1.000000,x=1pt,y=1pt]
\filldraw[color=white] (0.000000, -7.500000) rectangle (426.000000, 37.500000);
% Drawing wires
% Line 1: l0 W Q0
\draw[color=black] (0.000000,30.000000) -- (426.000000,30.000000);
\draw[color=black] (0.000000,30.000000) node[left] {$Q0$};
% Line 2: l2 W Q2
\draw[color=black] (0.000000,15.000000) -- (426.000000,15.000000);
\draw[color=black] (0.000000,15.000000) node[left] {$Q2$};
% Line 3: l4 W Q4
\draw[color=black] (0.000000,0.000000) -- (426.000000,0.000000);
\draw[color=black] (0.000000,0.000000) node[left] {$Q4$};
% Done with wires; drawing gates
% Line 5: l0 +l4
\draw (9.000000,30.000000) -- (9.000000,0.000000);
\filldraw (9.000000, 30.000000) circle(1.500000pt);
\begin{scope}
\draw[fill=white] (9.000000, 0.000000) circle(3.000000pt);
\clip (9.000000, 0.000000) circle(3.000000pt);
\draw (6.000000, 0.000000) -- (12.000000, 0.000000);
\draw (9.000000, -3.000000) -- (9.000000, 3.000000);
\end{scope}
% Line 6: =
\draw[fill=white,color=white] (24.000000, -6.000000) rectangle (39.000000, 36.000000);
\draw (31.500000, 15.000000) node {$=$};
% Line 7: l2 +l0
\draw (54.000000,30.000000) -- (54.000000,15.000000);
\filldraw (54.000000, 15.000000) circle(1.500000pt);
\begin{scope}
\draw[fill=white] (54.000000, 30.000000) circle(3.000000pt);
\clip (54.000000, 30.000000) circle(3.000000pt);
\draw (51.000000, 30.000000) -- (57.000000, 30.000000);
\draw (54.000000, 27.000000) -- (54.000000, 33.000000);
\end{scope}
% Line 8: l0 H
\begin{scope}
\draw[fill=white] (75.000000, 30.000000) +(-45.000000:8.485281pt and 8.485281pt) -- +(45.000000:8.485281pt and 8.485281pt) -- +(135.000000:8.485281pt and 8.485281pt) -- +(225.000000:8.485281pt and 8.485281pt) -- cycle;
\clip (75.000000, 30.000000) +(-45.000000:8.485281pt and 8.485281pt) -- +(45.000000:8.485281pt and 8.485281pt) -- +(135.000000:8.485281pt and 8.485281pt) -- +(225.000000:8.485281pt and 8.485281pt) -- cycle;
\draw (75.000000, 30.000000) node {$H$};
\end{scope}
% Line 9: l2 H
\begin{scope}
\draw[fill=white] (75.000000, 15.000000) +(-45.000000:8.485281pt and 8.485281pt) -- +(45.000000:8.485281pt and 8.485281pt) -- +(135.000000:8.485281pt and 8.485281pt) -- +(225.000000:8.485281pt and 8.485281pt) -- cycle;
\clip (75.000000, 15.000000) +(-45.000000:8.485281pt and 8.485281pt) -- +(45.000000:8.485281pt and 8.485281pt) -- +(135.000000:8.485281pt and 8.485281pt) -- +(225.000000:8.485281pt and 8.485281pt) -- cycle;
\draw (75.000000, 15.000000) node {$H$};
\end{scope}
% Line 10: l2 +l0
\draw (96.000000,30.000000) -- (96.000000,15.000000);
\filldraw (96.000000, 15.000000) circle(1.500000pt);
\begin{scope}
\draw[fill=white] (96.000000, 30.000000) circle(3.000000pt);
\clip (96.000000, 30.000000) circle(3.000000pt);
\draw (93.000000, 30.000000) -- (99.000000, 30.000000);
\draw (96.000000, 27.000000) -- (96.000000, 33.000000);
\end{scope}
% Line 11: l2 H
\begin{scope}
\draw[fill=white] (117.000000, 15.000000) +(-45.000000:8.485281pt and 8.485281pt) -- +(45.000000:8.485281pt and 8.485281pt) -- +(135.000000:8.485281pt and 8.485281pt) -- +(225.000000:8.485281pt and 8.485281pt) -- cycle;
\clip (117.000000, 15.000000) +(-45.000000:8.485281pt and 8.485281pt) -- +(45.000000:8.485281pt and 8.485281pt) -- +(135.000000:8.485281pt and 8.485281pt) -- +(225.000000:8.485281pt and 8.485281pt) -- cycle;
\draw (117.000000, 15.000000) node {$H$};
\end{scope}
% Line 12: l2 +l4
\draw (138.000000,15.000000) -- (138.000000,0.000000);
\filldraw (138.000000, 15.000000) circle(1.500000pt);
\begin{scope}
\draw[fill=white] (138.000000, 0.000000) circle(3.000000pt);
\clip (138.000000, 0.000000) circle(3.000000pt);
\draw (135.000000, 0.000000) -- (141.000000, 0.000000);
\draw (138.000000, -3.000000) -- (138.000000, 3.000000);
\end{scope}
% Line 13: l2 H
\begin{scope}
\draw[fill=white] (159.000000, 15.000000) +(-45.000000:8.485281pt and 8.485281pt) -- +(45.000000:8.485281pt and 8.485281pt) -- +(135.000000:8.485281pt and 8.485281pt) -- +(225.000000:8.485281pt and 8.485281pt) -- cycle;
\clip (159.000000, 15.000000) +(-45.000000:8.485281pt and 8.485281pt) -- +(45.000000:8.485281pt and 8.485281pt) -- +(135.000000:8.485281pt and 8.485281pt) -- +(225.000000:8.485281pt and 8.485281pt) -- cycle;
\draw (159.000000, 15.000000) node {$H$};
\end{scope}
% Line 14: l2 +l0
\draw (180.000000,30.000000) -- (180.000000,15.000000);
\filldraw (180.000000, 15.000000) circle(1.500000pt);
\begin{scope}
\draw[fill=white] (180.000000, 30.000000) circle(3.000000pt);
\clip (180.000000, 30.000000) circle(3.000000pt);
\draw (177.000000, 30.000000) -- (183.000000, 30.000000);
\draw (180.000000, 27.000000) -- (180.000000, 33.000000);
\end{scope}
% Line 15: l0 H
\begin{scope}
\draw[fill=white] (201.000000, 30.000000) +(-45.000000:8.485281pt and 8.485281pt) -- +(45.000000:8.485281pt and 8.485281pt) -- +(135.000000:8.485281pt and 8.485281pt) -- +(225.000000:8.485281pt and 8.485281pt) -- cycle;
\clip (201.000000, 30.000000) +(-45.000000:8.485281pt and 8.485281pt) -- +(45.000000:8.485281pt and 8.485281pt) -- +(135.000000:8.485281pt and 8.485281pt) -- +(225.000000:8.485281pt and 8.485281pt) -- cycle;
\draw (201.000000, 30.000000) node {$H$};
\end{scope}
% Line 16: l2 H
\begin{scope}
\draw[fill=white] (201.000000, 15.000000) +(-45.000000:8.485281pt and 8.485281pt) -- +(45.000000:8.485281pt and 8.485281pt) -- +(135.000000:8.485281pt and 8.485281pt) -- +(225.000000:8.485281pt and 8.485281pt) -- cycle;
\clip (201.000000, 15.000000) +(-45.000000:8.485281pt and 8.485281pt) -- +(45.000000:8.485281pt and 8.485281pt) -- +(135.000000:8.485281pt and 8.485281pt) -- +(225.000000:8.485281pt and 8.485281pt) -- cycle;
\draw (201.000000, 15.000000) node {$H$};
\end{scope}
% Line 17: l2 +l0
\draw (222.000000,30.000000) -- (222.000000,15.000000);
\filldraw (222.000000, 15.000000) circle(1.500000pt);
\begin{scope}
\draw[fill=white] (222.000000, 30.000000) circle(3.000000pt);
\clip (222.000000, 30.000000) circle(3.000000pt);
\draw (219.000000, 30.000000) -- (225.000000, 30.000000);
\draw (222.000000, 27.000000) -- (222.000000, 33.000000);
\end{scope}
% Line 18: =
\draw[fill=white,color=white] (237.000000, -6.000000) rectangle (252.000000, 36.000000);
\draw (244.500000, 15.000000) node {$=$};
% Line 19: l0 H
\begin{scope}
\draw[fill=white] (270.000000, 30.000000) +(-45.000000:8.485281pt and 8.485281pt) -- +(45.000000:8.485281pt and 8.485281pt) -- +(135.000000:8.485281pt and 8.485281pt) -- +(225.000000:8.485281pt and 8.485281pt) -- cycle;
\clip (270.000000, 30.000000) +(-45.000000:8.485281pt and 8.485281pt) -- +(45.000000:8.485281pt and 8.485281pt) -- +(135.000000:8.485281pt and 8.485281pt) -- +(225.000000:8.485281pt and 8.485281pt) -- cycle;
\draw (270.000000, 30.000000) node {$H$};
\end{scope}
% Line 20: l2 H
\begin{scope}
\draw[fill=white] (270.000000, 15.000000) +(-45.000000:8.485281pt and 8.485281pt) -- +(45.000000:8.485281pt and 8.485281pt) -- +(135.000000:8.485281pt and 8.485281pt) -- +(225.000000:8.485281pt and 8.485281pt) -- cycle;
\clip (270.000000, 15.000000) +(-45.000000:8.485281pt and 8.485281pt) -- +(45.000000:8.485281pt and 8.485281pt) -- +(135.000000:8.485281pt and 8.485281pt) -- +(225.000000:8.485281pt and 8.485281pt) -- cycle;
\draw (270.000000, 15.000000) node {$H$};
\end{scope}
% Line 21: l2 +l0
\draw (291.000000,30.000000) -- (291.000000,15.000000);
\filldraw (291.000000, 15.000000) circle(1.500000pt);
\begin{scope}
\draw[fill=white] (291.000000, 30.000000) circle(3.000000pt);
\clip (291.000000, 30.000000) circle(3.000000pt);
\draw (288.000000, 30.000000) -- (294.000000, 30.000000);
\draw (291.000000, 27.000000) -- (291.000000, 33.000000);
\end{scope}
% Line 22: l2 H
\begin{scope}
\draw[fill=white] (312.000000, 15.000000) +(-45.000000:8.485281pt and 8.485281pt) -- +(45.000000:8.485281pt and 8.485281pt) -- +(135.000000:8.485281pt and 8.485281pt) -- +(225.000000:8.485281pt and 8.485281pt) -- cycle;
\clip (312.000000, 15.000000) +(-45.000000:8.485281pt and 8.485281pt) -- +(45.000000:8.485281pt and 8.485281pt) -- +(135.000000:8.485281pt and 8.485281pt) -- +(225.000000:8.485281pt and 8.485281pt) -- cycle;
\draw (312.000000, 15.000000) node {$H$};
\end{scope}
% Line 23: l2 +l4
\draw (333.000000,15.000000) -- (333.000000,0.000000);
\filldraw (333.000000, 15.000000) circle(1.500000pt);
\begin{scope}
\draw[fill=white] (333.000000, 0.000000) circle(3.000000pt);
\clip (333.000000, 0.000000) circle(3.000000pt);
\draw (330.000000, 0.000000) -- (336.000000, 0.000000);
\draw (333.000000, -3.000000) -- (333.000000, 3.000000);
\end{scope}
% Line 24: l2 H
\begin{scope}
\draw[fill=white] (354.000000, 15.000000) +(-45.000000:8.485281pt and 8.485281pt) -- +(45.000000:8.485281pt and 8.485281pt) -- +(135.000000:8.485281pt and 8.485281pt) -- +(225.000000:8.485281pt and 8.485281pt) -- cycle;
\clip (354.000000, 15.000000) +(-45.000000:8.485281pt and 8.485281pt) -- +(45.000000:8.485281pt and 8.485281pt) -- +(135.000000:8.485281pt and 8.485281pt) -- +(225.000000:8.485281pt and 8.485281pt) -- cycle;
\draw (354.000000, 15.000000) node {$H$};
\end{scope}
% Line 25: l0 +l2
\draw (375.000000,30.000000) -- (375.000000,15.000000);
\filldraw (375.000000, 30.000000) circle(1.500000pt);
\begin{scope}
\draw[fill=white] (375.000000, 15.000000) circle(3.000000pt);
\clip (375.000000, 15.000000) circle(3.000000pt);
\draw (372.000000, 15.000000) -- (378.000000, 15.000000);
\draw (375.000000, 12.000000) -- (375.000000, 18.000000);
\end{scope}
% Line 26: l0 H
\begin{scope}
\draw[fill=white] (396.000000, 30.000000) +(-45.000000:8.485281pt and 8.485281pt) -- +(45.000000:8.485281pt and 8.485281pt) -- +(135.000000:8.485281pt and 8.485281pt) -- +(225.000000:8.485281pt and 8.485281pt) -- cycle;
\clip (396.000000, 30.000000) +(-45.000000:8.485281pt and 8.485281pt) -- +(45.000000:8.485281pt and 8.485281pt) -- +(135.000000:8.485281pt and 8.485281pt) -- +(225.000000:8.485281pt and 8.485281pt) -- cycle;
\draw (396.000000, 30.000000) node {$H$};
\end{scope}
% Line 27: l2 H
\begin{scope}
\draw[fill=white] (396.000000, 15.000000) +(-45.000000:8.485281pt and 8.485281pt) -- +(45.000000:8.485281pt and 8.485281pt) -- +(135.000000:8.485281pt and 8.485281pt) -- +(225.000000:8.485281pt and 8.485281pt) -- cycle;
\clip (396.000000, 15.000000) +(-45.000000:8.485281pt and 8.485281pt) -- +(45.000000:8.485281pt and 8.485281pt) -- +(135.000000:8.485281pt and 8.485281pt) -- +(225.000000:8.485281pt and 8.485281pt) -- cycle;
\draw (396.000000, 15.000000) node {$H$};
\end{scope}
% Line 28: l2 +l4
\draw (417.000000,15.000000) -- (417.000000,0.000000);
\filldraw (417.000000, 15.000000) circle(1.500000pt);
\begin{scope}
\draw[fill=white] (417.000000, 0.000000) circle(3.000000pt);
\clip (417.000000, 0.000000) circle(3.000000pt);
\draw (414.000000, 0.000000) -- (420.000000, 0.000000);
\draw (417.000000, -3.000000) -- (417.000000, 3.000000);
\end{scope}
% Done with gates; drawing ending labels
% Done with ending labels; drawing cut lines and comments
% Done with comments
\end{tikzpicture}

%% file: QX4complex.tikz
\begin{tikzpicture}[scale=1.000000,x=1pt,y=1pt]
\filldraw[color=white] (0.000000, -7.500000) rectangle (252.000000, 37.500000);
% Drawing wires
% Line 1: l1 W Q1
\draw[color=black] (0.000000,30.000000) -- (252.000000,30.000000);
\draw[color=black] (0.000000,30.000000) node[left] {$Q1$};
% Line 2: l2 W Q2
\draw[color=black] (0.000000,15.000000) -- (252.000000,15.000000);
\draw[color=black] (0.000000,15.000000) node[left] {$Q2$};
% Line 3: l3 W Q3
\draw[color=black] (0.000000,0.000000) -- (252.000000,0.000000);
\draw[color=black] (0.000000,0.000000) node[left] {$Q3$};
% Done with wires; drawing gates
% Line 4: l1 +l3
\draw (9.000000,30.000000) -- (9.000000,0.000000);
\filldraw (9.000000, 30.000000) circle(1.500000pt);
\begin{scope}
\draw[fill=white] (9.000000, 0.000000) circle(3.000000pt);
\clip (9.000000, 0.000000) circle(3.000000pt);
\draw (6.000000, 0.000000) -- (12.000000, 0.000000);
\draw (9.000000, -3.000000) -- (9.000000, 3.000000);
\end{scope}
% Line 5: =
\draw[fill=white,color=white] (24.000000, -6.000000) rectangle (39.000000, 36.000000);
\draw (31.500000, 15.000000) node {$=$};
% Line 6: l1 l2 SWAP
\draw (54.000000,30.000000) -- (54.000000,15.000000);
\begin{scope}
\draw (51.878680, 27.878680) -- (56.121320, 32.121320);
\draw (51.878680, 32.121320) -- (56.121320, 27.878680);
\end{scope}
\begin{scope}
\draw (51.878680, 12.878680) -- (56.121320, 17.121320);
\draw (51.878680, 17.121320) -- (56.121320, 12.878680);
\end{scope}
% Line 7: l2 +l3
\draw (72.000000,15.000000) -- (72.000000,0.000000);
\filldraw (72.000000, 15.000000) circle(1.500000pt);
\begin{scope}
\draw[fill=white] (72.000000, 0.000000) circle(3.000000pt);
\clip (72.000000, 0.000000) circle(3.000000pt);
\draw (69.000000, 0.000000) -- (75.000000, 0.000000);
\draw (72.000000, -3.000000) -- (72.000000, 3.000000);
\end{scope}
% Line 8: l1 l2 SWAP
\draw (90.000000,30.000000) -- (90.000000,15.000000);
\begin{scope}
\draw (87.878680, 27.878680) -- (92.121320, 32.121320);
\draw (87.878680, 32.121320) -- (92.121320, 27.878680);
\end{scope}
\begin{scope}
\draw (87.878680, 12.878680) -- (92.121320, 17.121320);
\draw (87.878680, 17.121320) -- (92.121320, 12.878680);
\end{scope}
% Line 9: =
\draw[fill=white,color=white] (105.000000, -6.000000) rectangle (120.000000, 36.000000);
\draw (112.500000, 15.000000) node {$=$};
% Line 10: l2 +l1
\draw (135.000000,30.000000) -- (135.000000,15.000000);
\filldraw (135.000000, 15.000000) circle(1.500000pt);
\begin{scope}
\draw[fill=white] (135.000000, 30.000000) circle(3.000000pt);
\clip (135.000000, 30.000000) circle(3.000000pt);
\draw (132.000000, 30.000000) -- (138.000000, 30.000000);
\draw (135.000000, 27.000000) -- (135.000000, 33.000000);
\end{scope}
% Line 11: l1 +l2
\draw (153.000000,30.000000) -- (153.000000,15.000000);
\filldraw (153.000000, 30.000000) circle(1.500000pt);
\begin{scope}
\draw[fill=white] (153.000000, 15.000000) circle(3.000000pt);
\clip (153.000000, 15.000000) circle(3.000000pt);
\draw (150.000000, 15.000000) -- (156.000000, 15.000000);
\draw (153.000000, 12.000000) -- (153.000000, 18.000000);
\end{scope}
% Line 12: l2 +l1
\draw (171.000000,30.000000) -- (171.000000,15.000000);
\filldraw (171.000000, 15.000000) circle(1.500000pt);
\begin{scope}
\draw[fill=white] (171.000000, 30.000000) circle(3.000000pt);
\clip (171.000000, 30.000000) circle(3.000000pt);
\draw (168.000000, 30.000000) -- (174.000000, 30.000000);
\draw (171.000000, 27.000000) -- (171.000000, 33.000000);
\end{scope}
% Line 13: l2 +l3
\draw (189.000000,15.000000) -- (189.000000,0.000000);
\filldraw (189.000000, 15.000000) circle(1.500000pt);
\begin{scope}
\draw[fill=white] (189.000000, 0.000000) circle(3.000000pt);
\clip (189.000000, 0.000000) circle(3.000000pt);
\draw (186.000000, 0.000000) -- (192.000000, 0.000000);
\draw (189.000000, -3.000000) -- (189.000000, 3.000000);
\end{scope}
% Line 14: l2 +l1
\draw (207.000000,30.000000) -- (207.000000,15.000000);
\filldraw (207.000000, 15.000000) circle(1.500000pt);
\begin{scope}
\draw[fill=white] (207.000000, 30.000000) circle(3.000000pt);
\clip (207.000000, 30.000000) circle(3.000000pt);
\draw (204.000000, 30.000000) -- (210.000000, 30.000000);
\draw (207.000000, 27.000000) -- (207.000000, 33.000000);
\end{scope}
% Line 15: l1 +l2
\draw (225.000000,30.000000) -- (225.000000,15.000000);
\filldraw (225.000000, 30.000000) circle(1.500000pt);
\begin{scope}
\draw[fill=white] (225.000000, 15.000000) circle(3.000000pt);
\clip (225.000000, 15.000000) circle(3.000000pt);
\draw (222.000000, 15.000000) -- (228.000000, 15.000000);
\draw (225.000000, 12.000000) -- (225.000000, 18.000000);
\end{scope}
% Line 16: l2 +l1
\draw (243.000000,30.000000) -- (243.000000,15.000000);
\filldraw (243.000000, 15.000000) circle(1.500000pt);
\begin{scope}
\draw[fill=white] (243.000000, 30.000000) circle(3.000000pt);
\clip (243.000000, 30.000000) circle(3.000000pt);
\draw (240.000000, 30.000000) -- (246.000000, 30.000000);
\draw (243.000000, 27.000000) -- (243.000000, 33.000000);
\end{scope}
% Done with gates; drawing ending labels
% Done with ending labels; drawing cut lines and comments
% Done with comments
\end{tikzpicture}

%% file: QX4complex2.tikz
\begin{tikzpicture}[scale=1.000000,x=1pt,y=1pt]
\filldraw[color=white] (0.000000, -7.500000) rectangle (294.000000, 37.500000);
% Drawing wires
% Line 1: l1 W
\draw[color=black] (0.000000,30.000000) -- (294.000000,30.000000);
% Line 2: l2 W
\draw[color=black] (0.000000,15.000000) -- (294.000000,15.000000);
% Line 3: l3 W
\draw[color=black] (0.000000,0.000000) -- (294.000000,0.000000);
% Done with wires; drawing gates
% Line 4: =
\draw[fill=white,color=white] (6.000000, -6.000000) rectangle (21.000000, 36.000000);
\draw (13.500000, 15.000000) node {$=$};
% Line 5: l2 +l1
\draw (36.000000,30.000000) -- (36.000000,15.000000);
\filldraw (36.000000, 15.000000) circle(1.500000pt);
\begin{scope}
\draw[fill=white] (36.000000, 30.000000) circle(3.000000pt);
\clip (36.000000, 30.000000) circle(3.000000pt);
\draw (33.000000, 30.000000) -- (39.000000, 30.000000);
\draw (36.000000, 27.000000) -- (36.000000, 33.000000);
\end{scope}
% Line 6: l1 +l2
\draw (54.000000,30.000000) -- (54.000000,15.000000);
\filldraw (54.000000, 30.000000) circle(1.500000pt);
\begin{scope}
\draw[fill=white] (54.000000, 15.000000) circle(3.000000pt);
\clip (54.000000, 15.000000) circle(3.000000pt);
\draw (51.000000, 15.000000) -- (57.000000, 15.000000);
\draw (54.000000, 12.000000) -- (54.000000, 18.000000);
\end{scope}
% Line 8: l2 +l3
\draw (72.000000,15.000000) -- (72.000000,0.000000);
\filldraw (72.000000, 15.000000) circle(1.500000pt);
\begin{scope}
\draw[fill=white] (72.000000, 0.000000) circle(3.000000pt);
\clip (72.000000, 0.000000) circle(3.000000pt);
\draw (69.000000, 0.000000) -- (75.000000, 0.000000);
\draw (72.000000, -3.000000) -- (72.000000, 3.000000);
\end{scope}
% Line 10: l1 +l2
\draw (90.000000,30.000000) -- (90.000000,15.000000);
\filldraw (90.000000, 30.000000) circle(1.500000pt);
\begin{scope}
\draw[fill=white] (90.000000, 15.000000) circle(3.000000pt);
\clip (90.000000, 15.000000) circle(3.000000pt);
\draw (87.000000, 15.000000) -- (93.000000, 15.000000);
\draw (90.000000, 12.000000) -- (90.000000, 18.000000);
\end{scope}
% Line 11: l2 +l1
\draw (108.000000,30.000000) -- (108.000000,15.000000);
\filldraw (108.000000, 15.000000) circle(1.500000pt);
\begin{scope}
\draw[fill=white] (108.000000, 30.000000) circle(3.000000pt);
\clip (108.000000, 30.000000) circle(3.000000pt);
\draw (105.000000, 30.000000) -- (111.000000, 30.000000);
\draw (108.000000, 27.000000) -- (108.000000, 33.000000);
\end{scope}
% Line 12: =
\draw[fill=white,color=white] (123.000000, -6.000000) rectangle (138.000000, 36.000000);
\draw (130.500000, 15.000000) node {$=$};
% Line 13: l2 +l1
\draw (153.000000,30.000000) -- (153.000000,15.000000);
\filldraw (153.000000, 15.000000) circle(1.500000pt);
\begin{scope}
\draw[fill=white] (153.000000, 30.000000) circle(3.000000pt);
\clip (153.000000, 30.000000) circle(3.000000pt);
\draw (150.000000, 30.000000) -- (156.000000, 30.000000);
\draw (153.000000, 27.000000) -- (153.000000, 33.000000);
\end{scope}
% Line 14: l1 H
\begin{scope}
\draw[fill=white] (174.000000, 30.000000) +(-45.000000:8.485281pt and 8.485281pt) -- +(45.000000:8.485281pt and 8.485281pt) -- +(135.000000:8.485281pt and 8.485281pt) -- +(225.000000:8.485281pt and 8.485281pt) -- cycle;
\clip (174.000000, 30.000000) +(-45.000000:8.485281pt and 8.485281pt) -- +(45.000000:8.485281pt and 8.485281pt) -- +(135.000000:8.485281pt and 8.485281pt) -- +(225.000000:8.485281pt and 8.485281pt) -- cycle;
\draw (174.000000, 30.000000) node {$H$};
\end{scope}
% Line 15: l2 H
\begin{scope}
\draw[fill=white] (174.000000, 15.000000) +(-45.000000:8.485281pt and 8.485281pt) -- +(45.000000:8.485281pt and 8.485281pt) -- +(135.000000:8.485281pt and 8.485281pt) -- +(225.000000:8.485281pt and 8.485281pt) -- cycle;
\clip (174.000000, 15.000000) +(-45.000000:8.485281pt and 8.485281pt) -- +(45.000000:8.485281pt and 8.485281pt) -- +(135.000000:8.485281pt and 8.485281pt) -- +(225.000000:8.485281pt and 8.485281pt) -- cycle;
\draw (174.000000, 15.000000) node {$H$};
\end{scope}
% Line 17: l2 +l1
\draw (198.000000,30.000000) -- (198.000000,15.000000);
\filldraw (198.000000, 15.000000) circle(1.500000pt);
\begin{scope}
\draw[fill=white] (198.000000, 30.000000) circle(3.000000pt);
\clip (198.000000, 30.000000) circle(3.000000pt);
\draw (195.000000, 30.000000) -- (201.000000, 30.000000);
\draw (198.000000, 27.000000) -- (198.000000, 33.000000);
\end{scope}
% Line 19: l3 H
\begin{scope}
\draw[fill=white] (198.000000, -0.000000) +(-45.000000:8.485281pt and 8.485281pt) -- +(45.000000:8.485281pt and 8.485281pt) -- +(135.000000:8.485281pt and 8.485281pt) -- +(225.000000:8.485281pt and 8.485281pt) -- cycle;
\clip (198.000000, -0.000000) +(-45.000000:8.485281pt and 8.485281pt) -- +(45.000000:8.485281pt and 8.485281pt) -- +(135.000000:8.485281pt and 8.485281pt) -- +(225.000000:8.485281pt and 8.485281pt) -- cycle;
\draw (198.000000, -0.000000) node {$H$};
\end{scope}
% Line 21: l3 +l2
\draw (219.000000,15.000000) -- (219.000000,0.000000);
\filldraw (219.000000, 0.000000) circle(1.500000pt);
\begin{scope}
\draw[fill=white] (219.000000, 15.000000) circle(3.000000pt);
\clip (219.000000, 15.000000) circle(3.000000pt);
\draw (216.000000, 15.000000) -- (222.000000, 15.000000);
\draw (219.000000, 12.000000) -- (219.000000, 18.000000);
\end{scope}
% Line 23: l3 H
\begin{scope}
\draw[fill=white] (240.000000, -0.000000) +(-45.000000:8.485281pt and 8.485281pt) -- +(45.000000:8.485281pt and 8.485281pt) -- +(135.000000:8.485281pt and 8.485281pt) -- +(225.000000:8.485281pt and 8.485281pt) -- cycle;
\clip (240.000000, -0.000000) +(-45.000000:8.485281pt and 8.485281pt) -- +(45.000000:8.485281pt and 8.485281pt) -- +(135.000000:8.485281pt and 8.485281pt) -- +(225.000000:8.485281pt and 8.485281pt) -- cycle;
\draw (240.000000, -0.000000) node {$H$};
\end{scope}
% Line 25: l2 +l1
\draw (240.000000,30.000000) -- (240.000000,15.000000);
\filldraw (240.000000, 15.000000) circle(1.500000pt);
\begin{scope}
\draw[fill=white] (240.000000, 30.000000) circle(3.000000pt);
\clip (240.000000, 30.000000) circle(3.000000pt);
\draw (237.000000, 30.000000) -- (243.000000, 30.000000);
\draw (240.000000, 27.000000) -- (240.000000, 33.000000);
\end{scope}
% Line 27: l1 H
\begin{scope}
\draw[fill=white] (264.000000, 30.000000) +(-45.000000:8.485281pt and 8.485281pt) -- +(45.000000:8.485281pt and 8.485281pt) -- +(135.000000:8.485281pt and 8.485281pt) -- +(225.000000:8.485281pt and 8.485281pt) -- cycle;
\clip (264.000000, 30.000000) +(-45.000000:8.485281pt and 8.485281pt) -- +(45.000000:8.485281pt and 8.485281pt) -- +(135.000000:8.485281pt and 8.485281pt) -- +(225.000000:8.485281pt and 8.485281pt) -- cycle;
\draw (264.000000, 30.000000) node {$H$};
\end{scope}
% Line 28: l2 H
\begin{scope}
\draw[fill=white] (264.000000, 15.000000) +(-45.000000:8.485281pt and 8.485281pt) -- +(45.000000:8.485281pt and 8.485281pt) -- +(135.000000:8.485281pt and 8.485281pt) -- +(225.000000:8.485281pt and 8.485281pt) -- cycle;
\clip (264.000000, 15.000000) +(-45.000000:8.485281pt and 8.485281pt) -- +(45.000000:8.485281pt and 8.485281pt) -- +(135.000000:8.485281pt and 8.485281pt) -- +(225.000000:8.485281pt and 8.485281pt) -- cycle;
\draw (264.000000, 15.000000) node {$H$};
\end{scope}
% Line 29: l2 +l1
\draw (285.000000,30.000000) -- (285.000000,15.000000);
\filldraw (285.000000, 15.000000) circle(1.500000pt);
\begin{scope}
\draw[fill=white] (285.000000, 30.000000) circle(3.000000pt);
\clip (285.000000, 30.000000) circle(3.000000pt);
\draw (282.000000, 30.000000) -- (288.000000, 30.000000);
\draw (285.000000, 27.000000) -- (285.000000, 33.000000);
\end{scope}
% Done with gates; drawing ending labels
% Done with ending labels; drawing cut lines and comments
% Done with comments
\end{tikzpicture}

%% file: QX4template.tikz
\begin{tikzpicture}[scale=1.000000,x=1pt,y=1pt]
\filldraw[color=white] (0.000000, -7.500000) rectangle (276.000000, 37.500000);
% Drawing wires
% Line 1: l1 W Q1
\draw[color=black] (0.000000,30.000000) -- (276.000000,30.000000);
\draw[color=black] (0.000000,30.000000) node[left] {$Q1$};
% Line 2: l2 W Q2
\draw[color=black] (0.000000,15.000000) -- (276.000000,15.000000);
\draw[color=black] (0.000000,15.000000) node[left] {$Q2$};
% Line 3: l3 W Q3
\draw[color=black] (0.000000,0.000000) -- (276.000000,0.000000);
\draw[color=black] (0.000000,0.000000) node[left] {$Q3$};
% Done with wires; drawing gates
% Line 4: l1 +l3
\draw (9.000000,30.000000) -- (9.000000,0.000000);
\filldraw (9.000000, 30.000000) circle(1.500000pt);
\begin{scope}
\draw[fill=white] (9.000000, 0.000000) circle(3.000000pt);
\clip (9.000000, 0.000000) circle(3.000000pt);
\draw (6.000000, 0.000000) -- (12.000000, 0.000000);
\draw (9.000000, -3.000000) -- (9.000000, 3.000000);
\end{scope}
% Line 5: =
\draw[fill=white,color=white] (24.000000, -6.000000) rectangle (39.000000, 36.000000);
\draw (31.500000, 15.000000) node {$=$};
% Line 6: l1 +l2
\draw (54.000000,30.000000) -- (54.000000,15.000000);
\filldraw (54.000000, 30.000000) circle(1.500000pt);
\begin{scope}
\draw[fill=white] (54.000000, 15.000000) circle(3.000000pt);
\clip (54.000000, 15.000000) circle(3.000000pt);
\draw (51.000000, 15.000000) -- (57.000000, 15.000000);
\draw (54.000000, 12.000000) -- (54.000000, 18.000000);
\end{scope}
% Line 7: l2 +l3
\draw (72.000000,15.000000) -- (72.000000,0.000000);
\filldraw (72.000000, 15.000000) circle(1.500000pt);
\begin{scope}
\draw[fill=white] (72.000000, 0.000000) circle(3.000000pt);
\clip (72.000000, 0.000000) circle(3.000000pt);
\draw (69.000000, 0.000000) -- (75.000000, 0.000000);
\draw (72.000000, -3.000000) -- (72.000000, 3.000000);
\end{scope}
% Line 8: l1 +l2
\draw (90.000000,30.000000) -- (90.000000,15.000000);
\filldraw (90.000000, 30.000000) circle(1.500000pt);
\begin{scope}
\draw[fill=white] (90.000000, 15.000000) circle(3.000000pt);
\clip (90.000000, 15.000000) circle(3.000000pt);
\draw (87.000000, 15.000000) -- (93.000000, 15.000000);
\draw (90.000000, 12.000000) -- (90.000000, 18.000000);
\end{scope}
% Line 9: l2 +l3
\draw (108.000000,15.000000) -- (108.000000,0.000000);
\filldraw (108.000000, 15.000000) circle(1.500000pt);
\begin{scope}
\draw[fill=white] (108.000000, 0.000000) circle(3.000000pt);
\clip (108.000000, 0.000000) circle(3.000000pt);
\draw (105.000000, 0.000000) -- (111.000000, 0.000000);
\draw (108.000000, -3.000000) -- (108.000000, 3.000000);
\end{scope}
% Line 10: =
\draw[fill=white,color=white] (123.000000, -6.000000) rectangle (138.000000, 36.000000);
\draw (130.500000, 15.000000) node {$=$};
% Line 11: l1 H
\begin{scope}
\draw[fill=white] (156.000000, 30.000000) +(-45.000000:8.485281pt and 8.485281pt) -- +(45.000000:8.485281pt and 8.485281pt) -- +(135.000000:8.485281pt and 8.485281pt) -- +(225.000000:8.485281pt and 8.485281pt) -- cycle;
\clip (156.000000, 30.000000) +(-45.000000:8.485281pt and 8.485281pt) -- +(45.000000:8.485281pt and 8.485281pt) -- +(135.000000:8.485281pt and 8.485281pt) -- +(225.000000:8.485281pt and 8.485281pt) -- cycle;
\draw (156.000000, 30.000000) node {$H$};
\end{scope}
% Line 12: l2 H
\begin{scope}
\draw[fill=white] (156.000000, 15.000000) +(-45.000000:8.485281pt and 8.485281pt) -- +(45.000000:8.485281pt and 8.485281pt) -- +(135.000000:8.485281pt and 8.485281pt) -- +(225.000000:8.485281pt and 8.485281pt) -- cycle;
\clip (156.000000, 15.000000) +(-45.000000:8.485281pt and 8.485281pt) -- +(45.000000:8.485281pt and 8.485281pt) -- +(135.000000:8.485281pt and 8.485281pt) -- +(225.000000:8.485281pt and 8.485281pt) -- cycle;
\draw (156.000000, 15.000000) node {$H$};
\end{scope}
% Line 14: l2 +l1
\draw (180.000000,30.000000) -- (180.000000,15.000000);
\filldraw (180.000000, 15.000000) circle(1.500000pt);
\begin{scope}
\draw[fill=white] (180.000000, 30.000000) circle(3.000000pt);
\clip (180.000000, 30.000000) circle(3.000000pt);
\draw (177.000000, 30.000000) -- (183.000000, 30.000000);
\draw (180.000000, 27.000000) -- (180.000000, 33.000000);
\end{scope}
% Line 15: l3 H
\begin{scope}
\draw[fill=white] (180.000000, -0.000000) +(-45.000000:8.485281pt and 8.485281pt) -- +(45.000000:8.485281pt and 8.485281pt) -- +(135.000000:8.485281pt and 8.485281pt) -- +(225.000000:8.485281pt and 8.485281pt) -- cycle;
\clip (180.000000, -0.000000) +(-45.000000:8.485281pt and 8.485281pt) -- +(45.000000:8.485281pt and 8.485281pt) -- +(135.000000:8.485281pt and 8.485281pt) -- +(225.000000:8.485281pt and 8.485281pt) -- cycle;
\draw (180.000000, -0.000000) node {$H$};
\end{scope}
% Line 17: l3 +l2
\draw (201.000000,15.000000) -- (201.000000,0.000000);
\filldraw (201.000000, 0.000000) circle(1.500000pt);
\begin{scope}
\draw[fill=white] (201.000000, 15.000000) circle(3.000000pt);
\clip (201.000000, 15.000000) circle(3.000000pt);
\draw (198.000000, 15.000000) -- (204.000000, 15.000000);
\draw (201.000000, 12.000000) -- (201.000000, 18.000000);
\end{scope}
% Line 18: l2 +l1
\draw (219.000000,30.000000) -- (219.000000,15.000000);
\filldraw (219.000000, 15.000000) circle(1.500000pt);
\begin{scope}
\draw[fill=white] (219.000000, 30.000000) circle(3.000000pt);
\clip (219.000000, 30.000000) circle(3.000000pt);
\draw (216.000000, 30.000000) -- (222.000000, 30.000000);
\draw (219.000000, 27.000000) -- (219.000000, 33.000000);
\end{scope}
% Line 19: l3 +l2
\draw (240.000000,15.000000) -- (240.000000,0.000000);
\filldraw (240.000000, 0.000000) circle(1.500000pt);
\begin{scope}
\draw[fill=white] (240.000000, 15.000000) circle(3.000000pt);
\clip (240.000000, 15.000000) circle(3.000000pt);
\draw (237.000000, 15.000000) -- (243.000000, 15.000000);
\draw (240.000000, 12.000000) -- (240.000000, 18.000000);
\end{scope}
% Line 21: l1 H
\begin{scope}
\draw[fill=white] (240.000000, 30.000000) +(-45.000000:8.485281pt and 8.485281pt) -- +(45.000000:8.485281pt and 8.485281pt) -- +(135.000000:8.485281pt and 8.485281pt) -- +(225.000000:8.485281pt and 8.485281pt) -- cycle;
\clip (240.000000, 30.000000) +(-45.000000:8.485281pt and 8.485281pt) -- +(45.000000:8.485281pt and 8.485281pt) -- +(135.000000:8.485281pt and 8.485281pt) -- +(225.000000:8.485281pt and 8.485281pt) -- cycle;
\draw (240.000000, 30.000000) node {$H$};
\end{scope}
% Line 20: l3 H
\begin{scope}
\draw[fill=white] (264.000000, -0.000000) +(-45.000000:8.485281pt and 8.485281pt) -- +(45.000000:8.485281pt and 8.485281pt) -- +(135.000000:8.485281pt and 8.485281pt) -- +(225.000000:8.485281pt and 8.485281pt) -- cycle;
\clip (264.000000, -0.000000) +(-45.000000:8.485281pt and 8.485281pt) -- +(45.000000:8.485281pt and 8.485281pt) -- +(135.000000:8.485281pt and 8.485281pt) -- +(225.000000:8.485281pt and 8.485281pt) -- cycle;
\draw (264.000000, -0.000000) node {$H$};
\end{scope}
% Line 22: l2 H
\begin{scope}
\draw[fill=white] (264.000000, 15.000000) +(-45.000000:8.485281pt and 8.485281pt) -- +(45.000000:8.485281pt and 8.485281pt) -- +(135.000000:8.485281pt and 8.485281pt) -- +(225.000000:8.485281pt and 8.485281pt) -- cycle;
\clip (264.000000, 15.000000) +(-45.000000:8.485281pt and 8.485281pt) -- +(45.000000:8.485281pt and 8.485281pt) -- +(135.000000:8.485281pt and 8.485281pt) -- +(225.000000:8.485281pt and 8.485281pt) -- cycle;
\draw (264.000000, 15.000000) node {$H$};
\end{scope}
% Done with gates; drawing ending labels
% Done with ending labels; drawing cut lines and comments
% Done with comments
\end{tikzpicture}

%% file: QX4template2.tikz
\begin{tikzpicture}[scale=1.000000,x=1pt,y=1pt]
\filldraw[color=white] (0.000000, -7.500000) rectangle (258.000000, 37.500000);
% Drawing wires
% Line 1: l1 W Q1
\draw[color=black] (0.000000,30.000000) -- (258.000000,30.000000);
\draw[color=black] (0.000000,30.000000) node[left] {$Q1$};
% Line 2: l2 W Q2
\draw[color=black] (0.000000,15.000000) -- (258.000000,15.000000);
\draw[color=black] (0.000000,15.000000) node[left] {$Q2$};
% Line 3: l3 W Q3
\draw[color=black] (0.000000,0.000000) -- (258.000000,0.000000);
\draw[color=black] (0.000000,0.000000) node[left] {$Q3$};
% Done with wires; drawing gates
% Line 4: l1 +l3
\draw (9.000000,30.000000) -- (9.000000,0.000000);
\filldraw (9.000000, 30.000000) circle(1.500000pt);
\begin{scope}
\draw[fill=white] (9.000000, 0.000000) circle(3.000000pt);
\clip (9.000000, 0.000000) circle(3.000000pt);
\draw (6.000000, 0.000000) -- (12.000000, 0.000000);
\draw (9.000000, -3.000000) -- (9.000000, 3.000000);
\end{scope}
% Line 5: =
\draw[fill=white,color=white] (24.000000, -6.000000) rectangle (39.000000, 36.000000);
\draw (31.500000, 15.000000) node {$=$};
% Line 6: l1 H
\begin{scope}
\draw[fill=white] (57.000000, 30.000000) +(-45.000000:8.485281pt and 8.485281pt) -- +(45.000000:8.485281pt and 8.485281pt) -- +(135.000000:8.485281pt and 8.485281pt) -- +(225.000000:8.485281pt and 8.485281pt) -- cycle;
\clip (57.000000, 30.000000) +(-45.000000:8.485281pt and 8.485281pt) -- +(45.000000:8.485281pt and 8.485281pt) -- +(135.000000:8.485281pt and 8.485281pt) -- +(225.000000:8.485281pt and 8.485281pt) -- cycle;
\draw (57.000000, 30.000000) node {$H$};
\end{scope}
% Line 7: l3 H
\begin{scope}
\draw[fill=white] (57.000000, -0.000000) +(-45.000000:8.485281pt and 8.485281pt) -- +(45.000000:8.485281pt and 8.485281pt) -- +(135.000000:8.485281pt and 8.485281pt) -- +(225.000000:8.485281pt and 8.485281pt) -- cycle;
\clip (57.000000, -0.000000) +(-45.000000:8.485281pt and 8.485281pt) -- +(45.000000:8.485281pt and 8.485281pt) -- +(135.000000:8.485281pt and 8.485281pt) -- +(225.000000:8.485281pt and 8.485281pt) -- cycle;
\draw (57.000000, -0.000000) node {$H$};
\end{scope}
% Line 8: l3 +l1
\draw (78.000000,30.000000) -- (78.000000,0.000000);
\filldraw (78.000000, 0.000000) circle(1.500000pt);
\begin{scope}
\draw[fill=white] (78.000000, 30.000000) circle(3.000000pt);
\clip (78.000000, 30.000000) circle(3.000000pt);
\draw (75.000000, 30.000000) -- (81.000000, 30.000000);
\draw (78.000000, 27.000000) -- (78.000000, 33.000000);
\end{scope}
% Line 9: l1 H
\begin{scope}
\draw[fill=white] (99.000000, 30.000000) +(-45.000000:8.485281pt and 8.485281pt) -- +(45.000000:8.485281pt and 8.485281pt) -- +(135.000000:8.485281pt and 8.485281pt) -- +(225.000000:8.485281pt and 8.485281pt) -- cycle;
\clip (99.000000, 30.000000) +(-45.000000:8.485281pt and 8.485281pt) -- +(45.000000:8.485281pt and 8.485281pt) -- +(135.000000:8.485281pt and 8.485281pt) -- +(225.000000:8.485281pt and 8.485281pt) -- cycle;
\draw (99.000000, 30.000000) node {$H$};
\end{scope}
% Line 10: l3 H
\begin{scope}
\draw[fill=white] (99.000000, -0.000000) +(-45.000000:8.485281pt and 8.485281pt) -- +(45.000000:8.485281pt and 8.485281pt) -- +(135.000000:8.485281pt and 8.485281pt) -- +(225.000000:8.485281pt and 8.485281pt) -- cycle;
\clip (99.000000, -0.000000) +(-45.000000:8.485281pt and 8.485281pt) -- +(45.000000:8.485281pt and 8.485281pt) -- +(135.000000:8.485281pt and 8.485281pt) -- +(225.000000:8.485281pt and 8.485281pt) -- cycle;
\draw (99.000000, -0.000000) node {$H$};
\end{scope}
% Line 11: =
\draw[fill=white,color=white] (117.000000, -6.000000) rectangle (132.000000, 36.000000);
\draw (124.500000, 15.000000) node {$=$};
% Line 12: l1 H
\begin{scope}
\draw[fill=white] (150.000000, 30.000000) +(-45.000000:8.485281pt and 8.485281pt) -- +(45.000000:8.485281pt and 8.485281pt) -- +(135.000000:8.485281pt and 8.485281pt) -- +(225.000000:8.485281pt and 8.485281pt) -- cycle;
\clip (150.000000, 30.000000) +(-45.000000:8.485281pt and 8.485281pt) -- +(45.000000:8.485281pt and 8.485281pt) -- +(135.000000:8.485281pt and 8.485281pt) -- +(225.000000:8.485281pt and 8.485281pt) -- cycle;
\draw (150.000000, 30.000000) node {$H$};
\end{scope}
% Line 13: l3 H
\begin{scope}
\draw[fill=white] (150.000000, -0.000000) +(-45.000000:8.485281pt and 8.485281pt) -- +(45.000000:8.485281pt and 8.485281pt) -- +(135.000000:8.485281pt and 8.485281pt) -- +(225.000000:8.485281pt and 8.485281pt) -- cycle;
\clip (150.000000, -0.000000) +(-45.000000:8.485281pt and 8.485281pt) -- +(45.000000:8.485281pt and 8.485281pt) -- +(135.000000:8.485281pt and 8.485281pt) -- +(225.000000:8.485281pt and 8.485281pt) -- cycle;
\draw (150.000000, -0.000000) node {$H$};
\end{scope}
% Line 14: l2 +l1
\draw (171.000000,30.000000) -- (171.000000,15.000000);
\filldraw (171.000000, 15.000000) circle(1.500000pt);
\begin{scope}
\draw[fill=white] (171.000000, 30.000000) circle(3.000000pt);
\clip (171.000000, 30.000000) circle(3.000000pt);
\draw (168.000000, 30.000000) -- (174.000000, 30.000000);
\draw (171.000000, 27.000000) -- (171.000000, 33.000000);
\end{scope}
% Line 15: l3 +l2
\draw (189.000000,15.000000) -- (189.000000,0.000000);
\filldraw (189.000000, 0.000000) circle(1.500000pt);
\begin{scope}
\draw[fill=white] (189.000000, 15.000000) circle(3.000000pt);
\clip (189.000000, 15.000000) circle(3.000000pt);
\draw (186.000000, 15.000000) -- (192.000000, 15.000000);
\draw (189.000000, 12.000000) -- (189.000000, 18.000000);
\end{scope}
% Line 16: l2 +l1
\draw (207.000000,30.000000) -- (207.000000,15.000000);
\filldraw (207.000000, 15.000000) circle(1.500000pt);
\begin{scope}
\draw[fill=white] (207.000000, 30.000000) circle(3.000000pt);
\clip (207.000000, 30.000000) circle(3.000000pt);
\draw (204.000000, 30.000000) -- (210.000000, 30.000000);
\draw (207.000000, 27.000000) -- (207.000000, 33.000000);
\end{scope}
% Line 17: l3 +l2
\draw (225.000000,15.000000) -- (225.000000,0.000000);
\filldraw (225.000000, 0.000000) circle(1.500000pt);
\begin{scope}
\draw[fill=white] (225.000000, 15.000000) circle(3.000000pt);
\clip (225.000000, 15.000000) circle(3.000000pt);
\draw (222.000000, 15.000000) -- (228.000000, 15.000000);
\draw (225.000000, 12.000000) -- (225.000000, 18.000000);
\end{scope}
% Line 19: l1 H
\begin{scope}
\draw[fill=white] (246.000000, 30.000000) +(-45.000000:8.485281pt and 8.485281pt) -- +(45.000000:8.485281pt and 8.485281pt) -- +(135.000000:8.485281pt and 8.485281pt) -- +(225.000000:8.485281pt and 8.485281pt) -- cycle;
\clip (246.000000, 30.000000) +(-45.000000:8.485281pt and 8.485281pt) -- +(45.000000:8.485281pt and 8.485281pt) -- +(135.000000:8.485281pt and 8.485281pt) -- +(225.000000:8.485281pt and 8.485281pt) -- cycle;
\draw (246.000000, 30.000000) node {$H$};
\end{scope}
% Line 20: l3 H
\begin{scope}
\draw[fill=white] (246.000000, -0.000000) +(-45.000000:8.485281pt and 8.485281pt) -- +(45.000000:8.485281pt and 8.485281pt) -- +(135.000000:8.485281pt and 8.485281pt) -- +(225.000000:8.485281pt and 8.485281pt) -- cycle;
\clip (246.000000, -0.000000) +(-45.000000:8.485281pt and 8.485281pt) -- +(45.000000:8.485281pt and 8.485281pt) -- +(135.000000:8.485281pt and 8.485281pt) -- +(225.000000:8.485281pt and 8.485281pt) -- cycle;
\draw (246.000000, -0.000000) node {$H$};
\end{scope}
% Done with gates; drawing ending labels
% Done with ending labels; drawing cut lines and comments
% Done with comments
\end{tikzpicture}

%% file: fig4_orig.tikz
\begin{tikzpicture}[scale=1.000000,x=1pt,y=1pt]
\filldraw[color=white] (0.000000, -7.500000) rectangle (426.000000, 37.500000);
% Drawing wires
% Line 1: l0 W Q0 Q0
\draw[color=black] (0.000000,30.000000) -- (426.000000,30.000000);
\draw[color=black] (0.000000,30.000000) node[left] {$Q0$};
% Line 2: l1 W Q1 Q1
\draw[color=black] (0.000000,15.000000) -- (426.000000,15.000000);
\draw[color=black] (0.000000,15.000000) node[left] {$Q1$};
% Line 3: l2 W Q2 Q2
\draw[color=black] (0.000000,0.000000) -- (426.000000,0.000000);
\draw[color=black] (0.000000,0.000000) node[left] {$Q2$};
% Done with wires; drawing gates
% Line 4: l0 G {$S^\dagger$}
\begin{scope}
\draw[fill=white] (12.000000, 30.000000) +(-45.000000:8.485281pt and 8.485281pt) -- +(45.000000:8.485281pt and 8.485281pt) -- +(135.000000:8.485281pt and 8.485281pt) -- +(225.000000:8.485281pt and 8.485281pt) -- cycle;
\clip (12.000000, 30.000000) +(-45.000000:8.485281pt and 8.485281pt) -- +(45.000000:8.485281pt and 8.485281pt) -- +(135.000000:8.485281pt and 8.485281pt) -- +(225.000000:8.485281pt and 8.485281pt) -- cycle;
\draw (12.000000, 30.000000) node {{$S^\dagger$}};
\end{scope}
% Line 5: l0 G {$S^\dagger$}
\begin{scope}
\draw[fill=white] (36.000000, 30.000000) +(-45.000000:8.485281pt and 8.485281pt) -- +(45.000000:8.485281pt and 8.485281pt) -- +(135.000000:8.485281pt and 8.485281pt) -- +(225.000000:8.485281pt and 8.485281pt) -- cycle;
\clip (36.000000, 30.000000) +(-45.000000:8.485281pt and 8.485281pt) -- +(45.000000:8.485281pt and 8.485281pt) -- +(135.000000:8.485281pt and 8.485281pt) -- +(225.000000:8.485281pt and 8.485281pt) -- cycle;
\draw (36.000000, 30.000000) node {{$S^\dagger$}};
\end{scope}
% Line 6: l0 H
\begin{scope}
\draw[fill=white] (60.000000, 30.000000) +(-45.000000:8.485281pt and 8.485281pt) -- +(45.000000:8.485281pt and 8.485281pt) -- +(135.000000:8.485281pt and 8.485281pt) -- +(225.000000:8.485281pt and 8.485281pt) -- cycle;
\clip (60.000000, 30.000000) +(-45.000000:8.485281pt and 8.485281pt) -- +(45.000000:8.485281pt and 8.485281pt) -- +(135.000000:8.485281pt and 8.485281pt) -- +(225.000000:8.485281pt and 8.485281pt) -- cycle;
\draw (60.000000, 30.000000) node {$H$};
\end{scope}
% Line 7: l0 G {$S$}
\begin{scope}
\draw[fill=white] (84.000000, 30.000000) +(-45.000000:8.485281pt and 8.485281pt) -- +(45.000000:8.485281pt and 8.485281pt) -- +(135.000000:8.485281pt and 8.485281pt) -- +(225.000000:8.485281pt and 8.485281pt) -- cycle;
\clip (84.000000, 30.000000) +(-45.000000:8.485281pt and 8.485281pt) -- +(45.000000:8.485281pt and 8.485281pt) -- +(135.000000:8.485281pt and 8.485281pt) -- +(225.000000:8.485281pt and 8.485281pt) -- cycle;
\draw (84.000000, 30.000000) node {{$S$}};
\end{scope}
% Line 8: l0 H
\begin{scope}
\draw[fill=white] (108.000000, 30.000000) +(-45.000000:8.485281pt and 8.485281pt) -- +(45.000000:8.485281pt and 8.485281pt) -- +(135.000000:8.485281pt and 8.485281pt) -- +(225.000000:8.485281pt and 8.485281pt) -- cycle;
\clip (108.000000, 30.000000) +(-45.000000:8.485281pt and 8.485281pt) -- +(45.000000:8.485281pt and 8.485281pt) -- +(135.000000:8.485281pt and 8.485281pt) -- +(225.000000:8.485281pt and 8.485281pt) -- cycle;
\draw (108.000000, 30.000000) node {$H$};
\end{scope}
% Line 10: l0 G {$S$}
\begin{scope}
\draw[fill=white] (132.000000, 30.000000) +(-45.000000:8.485281pt and 8.485281pt) -- +(45.000000:8.485281pt and 8.485281pt) -- +(135.000000:8.485281pt and 8.485281pt) -- +(225.000000:8.485281pt and 8.485281pt) -- cycle;
\clip (132.000000, 30.000000) +(-45.000000:8.485281pt and 8.485281pt) -- +(45.000000:8.485281pt and 8.485281pt) -- +(135.000000:8.485281pt and 8.485281pt) -- +(225.000000:8.485281pt and 8.485281pt) -- cycle;
\draw (132.000000, 30.000000) node {{$S$}};
\end{scope}
% Line 11: l2 G {$S^\dagger$}
\begin{scope}
\draw[fill=white] (132.000000, -0.000000) +(-45.000000:8.485281pt and 8.485281pt) -- +(45.000000:8.485281pt and 8.485281pt) -- +(135.000000:8.485281pt and 8.485281pt) -- +(225.000000:8.485281pt and 8.485281pt) -- cycle;
\clip (132.000000, -0.000000) +(-45.000000:8.485281pt and 8.485281pt) -- +(45.000000:8.485281pt and 8.485281pt) -- +(135.000000:8.485281pt and 8.485281pt) -- +(225.000000:8.485281pt and 8.485281pt) -- cycle;
\draw (132.000000, -0.000000) node {{$S^\dagger$}};
\end{scope}
% Line 13: l2 H
\begin{scope}
\draw[fill=white] (156.000000, -0.000000) +(-45.000000:8.485281pt and 8.485281pt) -- +(45.000000:8.485281pt and 8.485281pt) -- +(135.000000:8.485281pt and 8.485281pt) -- +(225.000000:8.485281pt and 8.485281pt) -- cycle;
\clip (156.000000, -0.000000) +(-45.000000:8.485281pt and 8.485281pt) -- +(45.000000:8.485281pt and 8.485281pt) -- +(135.000000:8.485281pt and 8.485281pt) -- +(225.000000:8.485281pt and 8.485281pt) -- cycle;
\draw (156.000000, -0.000000) node {$H$};
\end{scope}
% Line 14: l1 +l0
\draw (156.000000,30.000000) -- (156.000000,15.000000);
\filldraw (156.000000, 15.000000) circle(1.500000pt);
\begin{scope}
\draw[fill=white] (156.000000, 30.000000) circle(3.000000pt);
\clip (156.000000, 30.000000) circle(3.000000pt);
\draw (153.000000, 30.000000) -- (159.000000, 30.000000);
\draw (156.000000, 27.000000) -- (156.000000, 33.000000);
\end{scope}
% Line 15: l2 G {$T$}
\begin{scope}
\draw[fill=white] (180.000000, -0.000000) +(-45.000000:8.485281pt and 8.485281pt) -- +(45.000000:8.485281pt and 8.485281pt) -- +(135.000000:8.485281pt and 8.485281pt) -- +(225.000000:8.485281pt and 8.485281pt) -- cycle;
\clip (180.000000, -0.000000) +(-45.000000:8.485281pt and 8.485281pt) -- +(45.000000:8.485281pt and 8.485281pt) -- +(135.000000:8.485281pt and 8.485281pt) -- +(225.000000:8.485281pt and 8.485281pt) -- cycle;
\draw (180.000000, -0.000000) node {{$T$}};
\end{scope}
% Line 16: l1 H
\begin{scope}
\draw[fill=white] (180.000000, 15.000000) +(-45.000000:8.485281pt and 8.485281pt) -- +(45.000000:8.485281pt and 8.485281pt) -- +(135.000000:8.485281pt and 8.485281pt) -- +(225.000000:8.485281pt and 8.485281pt) -- cycle;
\clip (180.000000, 15.000000) +(-45.000000:8.485281pt and 8.485281pt) -- +(45.000000:8.485281pt and 8.485281pt) -- +(135.000000:8.485281pt and 8.485281pt) -- +(225.000000:8.485281pt and 8.485281pt) -- cycle;
\draw (180.000000, 15.000000) node {$H$};
\end{scope}
% Line 19: l0 G {$S^\dagger$}
\begin{scope}
\draw[fill=white] (180.000000, 30.000000) +(-45.000000:8.485281pt and 8.485281pt) -- +(45.000000:8.485281pt and 8.485281pt) -- +(135.000000:8.485281pt and 8.485281pt) -- +(225.000000:8.485281pt and 8.485281pt) -- cycle;
\clip (180.000000, 30.000000) +(-45.000000:8.485281pt and 8.485281pt) -- +(45.000000:8.485281pt and 8.485281pt) -- +(135.000000:8.485281pt and 8.485281pt) -- +(225.000000:8.485281pt and 8.485281pt) -- cycle;
\draw (180.000000, 30.000000) node {{$S^\dagger$}};
\end{scope}
% Line 17: l1 G {$S^\dagger$}
\begin{scope}
\draw[fill=white] (204.000000, 15.000000) +(-45.000000:8.485281pt and 8.485281pt) -- +(45.000000:8.485281pt and 8.485281pt) -- +(135.000000:8.485281pt and 8.485281pt) -- +(225.000000:8.485281pt and 8.485281pt) -- cycle;
\clip (204.000000, 15.000000) +(-45.000000:8.485281pt and 8.485281pt) -- +(45.000000:8.485281pt and 8.485281pt) -- +(135.000000:8.485281pt and 8.485281pt) -- +(225.000000:8.485281pt and 8.485281pt) -- cycle;
\draw (204.000000, 15.000000) node {{$S^\dagger$}};
\end{scope}
% Line 18: l2 G {$S$}
\begin{scope}
\draw[fill=white] (204.000000, -0.000000) +(-45.000000:8.485281pt and 8.485281pt) -- +(45.000000:8.485281pt and 8.485281pt) -- +(135.000000:8.485281pt and 8.485281pt) -- +(225.000000:8.485281pt and 8.485281pt) -- cycle;
\clip (204.000000, -0.000000) +(-45.000000:8.485281pt and 8.485281pt) -- +(45.000000:8.485281pt and 8.485281pt) -- +(135.000000:8.485281pt and 8.485281pt) -- +(225.000000:8.485281pt and 8.485281pt) -- cycle;
\draw (204.000000, -0.000000) node {{$S$}};
\end{scope}
% Line 22: l0 H
\begin{scope}
\draw[fill=white] (204.000000, 30.000000) +(-45.000000:8.485281pt and 8.485281pt) -- +(45.000000:8.485281pt and 8.485281pt) -- +(135.000000:8.485281pt and 8.485281pt) -- +(225.000000:8.485281pt and 8.485281pt) -- cycle;
\clip (204.000000, 30.000000) +(-45.000000:8.485281pt and 8.485281pt) -- +(45.000000:8.485281pt and 8.485281pt) -- +(135.000000:8.485281pt and 8.485281pt) -- +(225.000000:8.485281pt and 8.485281pt) -- cycle;
\draw (204.000000, 30.000000) node {$H$};
\end{scope}
% Line 20: l1 H
\begin{scope}
\draw[fill=white] (228.000000, 15.000000) +(-45.000000:8.485281pt and 8.485281pt) -- +(45.000000:8.485281pt and 8.485281pt) -- +(135.000000:8.485281pt and 8.485281pt) -- +(225.000000:8.485281pt and 8.485281pt) -- cycle;
\clip (228.000000, 15.000000) +(-45.000000:8.485281pt and 8.485281pt) -- +(45.000000:8.485281pt and 8.485281pt) -- +(135.000000:8.485281pt and 8.485281pt) -- +(225.000000:8.485281pt and 8.485281pt) -- cycle;
\draw (228.000000, 15.000000) node {$H$};
\end{scope}
% Line 21: l2 H
\begin{scope}
\draw[fill=white] (228.000000, -0.000000) +(-45.000000:8.485281pt and 8.485281pt) -- +(45.000000:8.485281pt and 8.485281pt) -- +(135.000000:8.485281pt and 8.485281pt) -- +(225.000000:8.485281pt and 8.485281pt) -- cycle;
\clip (228.000000, -0.000000) +(-45.000000:8.485281pt and 8.485281pt) -- +(45.000000:8.485281pt and 8.485281pt) -- +(135.000000:8.485281pt and 8.485281pt) -- +(225.000000:8.485281pt and 8.485281pt) -- cycle;
\draw (228.000000, -0.000000) node {$H$};
\end{scope}
% Line 25: l0 G {$T$}
\begin{scope}
\draw[fill=white] (228.000000, 30.000000) +(-45.000000:8.485281pt and 8.485281pt) -- +(45.000000:8.485281pt and 8.485281pt) -- +(135.000000:8.485281pt and 8.485281pt) -- +(225.000000:8.485281pt and 8.485281pt) -- cycle;
\clip (228.000000, 30.000000) +(-45.000000:8.485281pt and 8.485281pt) -- +(45.000000:8.485281pt and 8.485281pt) -- +(135.000000:8.485281pt and 8.485281pt) -- +(225.000000:8.485281pt and 8.485281pt) -- cycle;
\draw (228.000000, 30.000000) node {{$T$}};
\end{scope}
% Line 23: l1 G {$Y$}
\begin{scope}
\draw[fill=white] (252.000000, 15.000000) +(-45.000000:8.485281pt and 8.485281pt) -- +(45.000000:8.485281pt and 8.485281pt) -- +(135.000000:8.485281pt and 8.485281pt) -- +(225.000000:8.485281pt and 8.485281pt) -- cycle;
\clip (252.000000, 15.000000) +(-45.000000:8.485281pt and 8.485281pt) -- +(45.000000:8.485281pt and 8.485281pt) -- +(135.000000:8.485281pt and 8.485281pt) -- +(225.000000:8.485281pt and 8.485281pt) -- cycle;
\draw (252.000000, 15.000000) node {{$Y$}};
\end{scope}
% Line 24: l2 G {$S$}
\begin{scope}
\draw[fill=white] (252.000000, -0.000000) +(-45.000000:8.485281pt and 8.485281pt) -- +(45.000000:8.485281pt and 8.485281pt) -- +(135.000000:8.485281pt and 8.485281pt) -- +(225.000000:8.485281pt and 8.485281pt) -- cycle;
\clip (252.000000, -0.000000) +(-45.000000:8.485281pt and 8.485281pt) -- +(45.000000:8.485281pt and 8.485281pt) -- +(135.000000:8.485281pt and 8.485281pt) -- +(225.000000:8.485281pt and 8.485281pt) -- cycle;
\draw (252.000000, -0.000000) node {{$S$}};
\end{scope}
% Line 26: l0 H
\begin{scope}
\draw[fill=white] (252.000000, 30.000000) +(-45.000000:8.485281pt and 8.485281pt) -- +(45.000000:8.485281pt and 8.485281pt) -- +(135.000000:8.485281pt and 8.485281pt) -- +(225.000000:8.485281pt and 8.485281pt) -- cycle;
\clip (252.000000, 30.000000) +(-45.000000:8.485281pt and 8.485281pt) -- +(45.000000:8.485281pt and 8.485281pt) -- +(135.000000:8.485281pt and 8.485281pt) -- +(225.000000:8.485281pt and 8.485281pt) -- cycle;
\draw (252.000000, 30.000000) node {$H$};
\end{scope}
% Line 27: l0 G {$S$}
\begin{scope}
\draw[fill=white] (276.000000, 30.000000) +(-45.000000:8.485281pt and 8.485281pt) -- +(45.000000:8.485281pt and 8.485281pt) -- +(135.000000:8.485281pt and 8.485281pt) -- +(225.000000:8.485281pt and 8.485281pt) -- cycle;
\clip (276.000000, 30.000000) +(-45.000000:8.485281pt and 8.485281pt) -- +(45.000000:8.485281pt and 8.485281pt) -- +(135.000000:8.485281pt and 8.485281pt) -- +(225.000000:8.485281pt and 8.485281pt) -- cycle;
\draw (276.000000, 30.000000) node {{$S$}};
\end{scope}
% Line 28: l0 G {$T$}
\begin{scope}
\draw[fill=white] (300.000000, 30.000000) +(-45.000000:8.485281pt and 8.485281pt) -- +(45.000000:8.485281pt and 8.485281pt) -- +(135.000000:8.485281pt and 8.485281pt) -- +(225.000000:8.485281pt and 8.485281pt) -- cycle;
\clip (300.000000, 30.000000) +(-45.000000:8.485281pt and 8.485281pt) -- +(45.000000:8.485281pt and 8.485281pt) -- +(135.000000:8.485281pt and 8.485281pt) -- +(225.000000:8.485281pt and 8.485281pt) -- cycle;
\draw (300.000000, 30.000000) node {{$T$}};
\end{scope}
% Line 29: l0 G {$S$}
\begin{scope}
\draw[fill=white] (324.000000, 30.000000) +(-45.000000:8.485281pt and 8.485281pt) -- +(45.000000:8.485281pt and 8.485281pt) -- +(135.000000:8.485281pt and 8.485281pt) -- +(225.000000:8.485281pt and 8.485281pt) -- cycle;
\clip (324.000000, 30.000000) +(-45.000000:8.485281pt and 8.485281pt) -- +(45.000000:8.485281pt and 8.485281pt) -- +(135.000000:8.485281pt and 8.485281pt) -- +(225.000000:8.485281pt and 8.485281pt) -- cycle;
\draw (324.000000, 30.000000) node {{$S$}};
\end{scope}
% Line 30: l0 H
\begin{scope}
\draw[fill=white] (348.000000, 30.000000) +(-45.000000:8.485281pt and 8.485281pt) -- +(45.000000:8.485281pt and 8.485281pt) -- +(135.000000:8.485281pt and 8.485281pt) -- +(225.000000:8.485281pt and 8.485281pt) -- cycle;
\clip (348.000000, 30.000000) +(-45.000000:8.485281pt and 8.485281pt) -- +(45.000000:8.485281pt and 8.485281pt) -- +(135.000000:8.485281pt and 8.485281pt) -- +(225.000000:8.485281pt and 8.485281pt) -- cycle;
\draw (348.000000, 30.000000) node {$H$};
\end{scope}
% Line 31: l0 G {$T^\dagger$}
\begin{scope}
\draw[fill=white] (372.000000, 30.000000) +(-45.000000:8.485281pt and 8.485281pt) -- +(45.000000:8.485281pt and 8.485281pt) -- +(135.000000:8.485281pt and 8.485281pt) -- +(225.000000:8.485281pt and 8.485281pt) -- cycle;
\clip (372.000000, 30.000000) +(-45.000000:8.485281pt and 8.485281pt) -- +(45.000000:8.485281pt and 8.485281pt) -- +(135.000000:8.485281pt and 8.485281pt) -- +(225.000000:8.485281pt and 8.485281pt) -- cycle;
\draw (372.000000, 30.000000) node {{$T^\dagger$}};
\end{scope}
% Line 32: l0 H
\begin{scope}
\draw[fill=white] (396.000000, 30.000000) +(-45.000000:8.485281pt and 8.485281pt) -- +(45.000000:8.485281pt and 8.485281pt) -- +(135.000000:8.485281pt and 8.485281pt) -- +(225.000000:8.485281pt and 8.485281pt) -- cycle;
\clip (396.000000, 30.000000) +(-45.000000:8.485281pt and 8.485281pt) -- +(45.000000:8.485281pt and 8.485281pt) -- +(135.000000:8.485281pt and 8.485281pt) -- +(225.000000:8.485281pt and 8.485281pt) -- cycle;
\draw (396.000000, 30.000000) node {$H$};
\end{scope}
% Line 33: l0 +l2
\draw (417.000000,30.000000) -- (417.000000,0.000000);
\filldraw (417.000000, 30.000000) circle(1.500000pt);
\begin{scope}
\draw[fill=white] (417.000000, 0.000000) circle(3.000000pt);
\clip (417.000000, 0.000000) circle(3.000000pt);
\draw (414.000000, 0.000000) -- (420.000000, 0.000000);
\draw (417.000000, -3.000000) -- (417.000000, 3.000000);
\end{scope}
% Done with gates; drawing ending labels
\draw[color=black] (426.000000,30.000000) node[right] {$Q0$};
\draw[color=black] (426.000000,15.000000) node[right] {$Q1$};
\draw[color=black] (426.000000,0.000000) node[right] {$Q2$};
% Done with ending labels; drawing cut lines and comments
% Done with comments
\end{tikzpicture}

%% file: fig4_orig_opt.tikz
\begin{tikzpicture}[scale=1.000000,x=1pt,y=1pt]
\filldraw[color=white] (0.000000, -7.500000) rectangle (420.000000, 37.500000);
% Drawing wires
% Line 1: l0 W Q0 Q0
\draw[color=black] (0.000000,30.000000) -- (420.000000,30.000000);
\draw[color=black] (0.000000,30.000000) node[left] {$Q0$};
% Line 2: l1 W Q2 Q2
\draw[color=black] (0.000000,15.000000) -- (420.000000,15.000000);
\draw[color=black] (0.000000,15.000000) node[left] {$Q2$};
% Line 3: l2 W Q1 Q1
\draw[color=black] (0.000000,0.000000) -- (420.000000,0.000000);
\draw[color=black] (0.000000,0.000000) node[left] {$Q1$};
% Done with wires; drawing gates
% Line 4: l0 G {$Z$}
\begin{scope}
\draw[fill=white] (12.000000, 30.000000) +(-45.000000:8.485281pt and 8.485281pt) -- +(45.000000:8.485281pt and 8.485281pt) -- +(135.000000:8.485281pt and 8.485281pt) -- +(225.000000:8.485281pt and 8.485281pt) -- cycle;
\clip (12.000000, 30.000000) +(-45.000000:8.485281pt and 8.485281pt) -- +(45.000000:8.485281pt and 8.485281pt) -- +(135.000000:8.485281pt and 8.485281pt) -- +(225.000000:8.485281pt and 8.485281pt) -- cycle;
\draw (12.000000, 30.000000) node {{$Z$}};
\end{scope}
% Line 5: l0 H
\begin{scope}
\draw[fill=white] (36.000000, 30.000000) +(-45.000000:8.485281pt and 8.485281pt) -- +(45.000000:8.485281pt and 8.485281pt) -- +(135.000000:8.485281pt and 8.485281pt) -- +(225.000000:8.485281pt and 8.485281pt) -- cycle;
\clip (36.000000, 30.000000) +(-45.000000:8.485281pt and 8.485281pt) -- +(45.000000:8.485281pt and 8.485281pt) -- +(135.000000:8.485281pt and 8.485281pt) -- +(225.000000:8.485281pt and 8.485281pt) -- cycle;
\draw (36.000000, 30.000000) node {$H$};
\end{scope}
% Line 6: l0 G {$S$}
\begin{scope}
\draw[fill=white] (60.000000, 30.000000) +(-45.000000:8.485281pt and 8.485281pt) -- +(45.000000:8.485281pt and 8.485281pt) -- +(135.000000:8.485281pt and 8.485281pt) -- +(225.000000:8.485281pt and 8.485281pt) -- cycle;
\clip (60.000000, 30.000000) +(-45.000000:8.485281pt and 8.485281pt) -- +(45.000000:8.485281pt and 8.485281pt) -- +(135.000000:8.485281pt and 8.485281pt) -- +(225.000000:8.485281pt and 8.485281pt) -- cycle;
\draw (60.000000, 30.000000) node {{$S$}};
\end{scope}
% Line 7: l0 H
\begin{scope}
\draw[fill=white] (84.000000, 30.000000) +(-45.000000:8.485281pt and 8.485281pt) -- +(45.000000:8.485281pt and 8.485281pt) -- +(135.000000:8.485281pt and 8.485281pt) -- +(225.000000:8.485281pt and 8.485281pt) -- cycle;
\clip (84.000000, 30.000000) +(-45.000000:8.485281pt and 8.485281pt) -- +(45.000000:8.485281pt and 8.485281pt) -- +(135.000000:8.485281pt and 8.485281pt) -- +(225.000000:8.485281pt and 8.485281pt) -- cycle;
\draw (84.000000, 30.000000) node {$H$};
\end{scope}
% Line 9: l0 G {$S$}
\begin{scope}
\draw[fill=white] (108.000000, 30.000000) +(-45.000000:8.485281pt and 8.485281pt) -- +(45.000000:8.485281pt and 8.485281pt) -- +(135.000000:8.485281pt and 8.485281pt) -- +(225.000000:8.485281pt and 8.485281pt) -- cycle;
\clip (108.000000, 30.000000) +(-45.000000:8.485281pt and 8.485281pt) -- +(45.000000:8.485281pt and 8.485281pt) -- +(135.000000:8.485281pt and 8.485281pt) -- +(225.000000:8.485281pt and 8.485281pt) -- cycle;
\draw (108.000000, 30.000000) node {{$S$}};
\end{scope}
% Line 10: l1 G {$S^\dagger$}
\begin{scope}
\draw[fill=white] (108.000000, 15.000000) +(-45.000000:8.485281pt and 8.485281pt) -- +(45.000000:8.485281pt and 8.485281pt) -- +(135.000000:8.485281pt and 8.485281pt) -- +(225.000000:8.485281pt and 8.485281pt) -- cycle;
\clip (108.000000, 15.000000) +(-45.000000:8.485281pt and 8.485281pt) -- +(45.000000:8.485281pt and 8.485281pt) -- +(135.000000:8.485281pt and 8.485281pt) -- +(225.000000:8.485281pt and 8.485281pt) -- cycle;
\draw (108.000000, 15.000000) node {{$S^\dagger$}};
\end{scope}
% Line 13: l1 H
\begin{scope}
\draw[fill=white] (132.000000, 15.000000) +(-45.000000:8.485281pt and 8.485281pt) -- +(45.000000:8.485281pt and 8.485281pt) -- +(135.000000:8.485281pt and 8.485281pt) -- +(225.000000:8.485281pt and 8.485281pt) -- cycle;
\clip (132.000000, 15.000000) +(-45.000000:8.485281pt and 8.485281pt) -- +(45.000000:8.485281pt and 8.485281pt) -- +(135.000000:8.485281pt and 8.485281pt) -- +(225.000000:8.485281pt and 8.485281pt) -- cycle;
\draw (132.000000, 15.000000) node {$H$};
\end{scope}
% Line 14: l2 H
\begin{scope}
\draw[fill=white] (132.000000, -0.000000) +(-45.000000:8.485281pt and 8.485281pt) -- +(45.000000:8.485281pt and 8.485281pt) -- +(135.000000:8.485281pt and 8.485281pt) -- +(225.000000:8.485281pt and 8.485281pt) -- cycle;
\clip (132.000000, -0.000000) +(-45.000000:8.485281pt and 8.485281pt) -- +(45.000000:8.485281pt and 8.485281pt) -- +(135.000000:8.485281pt and 8.485281pt) -- +(225.000000:8.485281pt and 8.485281pt) -- cycle;
\draw (132.000000, -0.000000) node {$H$};
\end{scope}
% Line 15: l0 H
\begin{scope}
\draw[fill=white] (132.000000, 30.000000) +(-45.000000:8.485281pt and 8.485281pt) -- +(45.000000:8.485281pt and 8.485281pt) -- +(135.000000:8.485281pt and 8.485281pt) -- +(225.000000:8.485281pt and 8.485281pt) -- cycle;
\clip (132.000000, 30.000000) +(-45.000000:8.485281pt and 8.485281pt) -- +(45.000000:8.485281pt and 8.485281pt) -- +(135.000000:8.485281pt and 8.485281pt) -- +(225.000000:8.485281pt and 8.485281pt) -- cycle;
\draw (132.000000, 30.000000) node {$H$};
\end{scope}
% Line 17: l0 +l2
\draw (153.000000,30.000000) -- (153.000000,0.000000);
\filldraw (153.000000, 30.000000) circle(1.500000pt);
\begin{scope}
\draw[fill=white] (153.000000, 0.000000) circle(3.000000pt);
\clip (153.000000, 0.000000) circle(3.000000pt);
\draw (150.000000, 0.000000) -- (156.000000, 0.000000);
\draw (153.000000, -3.000000) -- (153.000000, 3.000000);
\end{scope}
% Line 19: l0 H
\begin{scope}
\draw[fill=white] (174.000000, 30.000000) +(-45.000000:8.485281pt and 8.485281pt) -- +(45.000000:8.485281pt and 8.485281pt) -- +(135.000000:8.485281pt and 8.485281pt) -- +(225.000000:8.485281pt and 8.485281pt) -- cycle;
\clip (174.000000, 30.000000) +(-45.000000:8.485281pt and 8.485281pt) -- +(45.000000:8.485281pt and 8.485281pt) -- +(135.000000:8.485281pt and 8.485281pt) -- +(225.000000:8.485281pt and 8.485281pt) -- cycle;
\draw (174.000000, 30.000000) node {$H$};
\end{scope}
% Line 20: l1 G {$T$}
\begin{scope}
\draw[fill=white] (174.000000, 15.000000) +(-45.000000:8.485281pt and 8.485281pt) -- +(45.000000:8.485281pt and 8.485281pt) -- +(135.000000:8.485281pt and 8.485281pt) -- +(225.000000:8.485281pt and 8.485281pt) -- cycle;
\clip (174.000000, 15.000000) +(-45.000000:8.485281pt and 8.485281pt) -- +(45.000000:8.485281pt and 8.485281pt) -- +(135.000000:8.485281pt and 8.485281pt) -- +(225.000000:8.485281pt and 8.485281pt) -- cycle;
\draw (174.000000, 15.000000) node {{$T$}};
\end{scope}
% Line 21: l2 G {$S^\dagger$}
\begin{scope}
\draw[fill=white] (174.000000, -0.000000) +(-45.000000:8.485281pt and 8.485281pt) -- +(45.000000:8.485281pt and 8.485281pt) -- +(135.000000:8.485281pt and 8.485281pt) -- +(225.000000:8.485281pt and 8.485281pt) -- cycle;
\clip (174.000000, -0.000000) +(-45.000000:8.485281pt and 8.485281pt) -- +(45.000000:8.485281pt and 8.485281pt) -- +(135.000000:8.485281pt and 8.485281pt) -- +(225.000000:8.485281pt and 8.485281pt) -- cycle;
\draw (174.000000, -0.000000) node {{$S^\dagger$}};
\end{scope}
% Line 23: l1 G {$S$}
\begin{scope}
\draw[fill=white] (198.000000, 15.000000) +(-45.000000:8.485281pt and 8.485281pt) -- +(45.000000:8.485281pt and 8.485281pt) -- +(135.000000:8.485281pt and 8.485281pt) -- +(225.000000:8.485281pt and 8.485281pt) -- cycle;
\clip (198.000000, 15.000000) +(-45.000000:8.485281pt and 8.485281pt) -- +(45.000000:8.485281pt and 8.485281pt) -- +(135.000000:8.485281pt and 8.485281pt) -- +(225.000000:8.485281pt and 8.485281pt) -- cycle;
\draw (198.000000, 15.000000) node {{$S$}};
\end{scope}
% Line 24: l0 G {$S^\dagger$}
\begin{scope}
\draw[fill=white] (198.000000, 30.000000) +(-45.000000:8.485281pt and 8.485281pt) -- +(45.000000:8.485281pt and 8.485281pt) -- +(135.000000:8.485281pt and 8.485281pt) -- +(225.000000:8.485281pt and 8.485281pt) -- cycle;
\clip (198.000000, 30.000000) +(-45.000000:8.485281pt and 8.485281pt) -- +(45.000000:8.485281pt and 8.485281pt) -- +(135.000000:8.485281pt and 8.485281pt) -- +(225.000000:8.485281pt and 8.485281pt) -- cycle;
\draw (198.000000, 30.000000) node {{$S^\dagger$}};
\end{scope}
% Line 25: l2 H
\begin{scope}
\draw[fill=white] (198.000000, -0.000000) +(-45.000000:8.485281pt and 8.485281pt) -- +(45.000000:8.485281pt and 8.485281pt) -- +(135.000000:8.485281pt and 8.485281pt) -- +(225.000000:8.485281pt and 8.485281pt) -- cycle;
\clip (198.000000, -0.000000) +(-45.000000:8.485281pt and 8.485281pt) -- +(45.000000:8.485281pt and 8.485281pt) -- +(135.000000:8.485281pt and 8.485281pt) -- +(225.000000:8.485281pt and 8.485281pt) -- cycle;
\draw (198.000000, -0.000000) node {$H$};
\end{scope}
% Line 26: l1 H
\begin{scope}
\draw[fill=white] (222.000000, 15.000000) +(-45.000000:8.485281pt and 8.485281pt) -- +(45.000000:8.485281pt and 8.485281pt) -- +(135.000000:8.485281pt and 8.485281pt) -- +(225.000000:8.485281pt and 8.485281pt) -- cycle;
\clip (222.000000, 15.000000) +(-45.000000:8.485281pt and 8.485281pt) -- +(45.000000:8.485281pt and 8.485281pt) -- +(135.000000:8.485281pt and 8.485281pt) -- +(225.000000:8.485281pt and 8.485281pt) -- cycle;
\draw (222.000000, 15.000000) node {$H$};
\end{scope}
% Line 27: l0 H
\begin{scope}
\draw[fill=white] (222.000000, 30.000000) +(-45.000000:8.485281pt and 8.485281pt) -- +(45.000000:8.485281pt and 8.485281pt) -- +(135.000000:8.485281pt and 8.485281pt) -- +(225.000000:8.485281pt and 8.485281pt) -- cycle;
\clip (222.000000, 30.000000) +(-45.000000:8.485281pt and 8.485281pt) -- +(45.000000:8.485281pt and 8.485281pt) -- +(135.000000:8.485281pt and 8.485281pt) -- +(225.000000:8.485281pt and 8.485281pt) -- cycle;
\draw (222.000000, 30.000000) node {$H$};
\end{scope}
% Line 28: l2 G {$Y$}
\begin{scope}
\draw[fill=white] (222.000000, -0.000000) +(-45.000000:8.485281pt and 8.485281pt) -- +(45.000000:8.485281pt and 8.485281pt) -- +(135.000000:8.485281pt and 8.485281pt) -- +(225.000000:8.485281pt and 8.485281pt) -- cycle;
\clip (222.000000, -0.000000) +(-45.000000:8.485281pt and 8.485281pt) -- +(45.000000:8.485281pt and 8.485281pt) -- +(135.000000:8.485281pt and 8.485281pt) -- +(225.000000:8.485281pt and 8.485281pt) -- cycle;
\draw (222.000000, -0.000000) node {{$Y$}};
\end{scope}
% Line 29: l1 G {$S$}
\begin{scope}
\draw[fill=white] (246.000000, 15.000000) +(-45.000000:8.485281pt and 8.485281pt) -- +(45.000000:8.485281pt and 8.485281pt) -- +(135.000000:8.485281pt and 8.485281pt) -- +(225.000000:8.485281pt and 8.485281pt) -- cycle;
\clip (246.000000, 15.000000) +(-45.000000:8.485281pt and 8.485281pt) -- +(45.000000:8.485281pt and 8.485281pt) -- +(135.000000:8.485281pt and 8.485281pt) -- +(225.000000:8.485281pt and 8.485281pt) -- cycle;
\draw (246.000000, 15.000000) node {{$S$}};
\end{scope}
% Line 30: l0 G {$T$}
\begin{scope}
\draw[fill=white] (246.000000, 30.000000) +(-45.000000:8.485281pt and 8.485281pt) -- +(45.000000:8.485281pt and 8.485281pt) -- +(135.000000:8.485281pt and 8.485281pt) -- +(225.000000:8.485281pt and 8.485281pt) -- cycle;
\clip (246.000000, 30.000000) +(-45.000000:8.485281pt and 8.485281pt) -- +(45.000000:8.485281pt and 8.485281pt) -- +(135.000000:8.485281pt and 8.485281pt) -- +(225.000000:8.485281pt and 8.485281pt) -- cycle;
\draw (246.000000, 30.000000) node {{$T$}};
\end{scope}
% Line 31: l0 H
\begin{scope}
\draw[fill=white] (270.000000, 30.000000) +(-45.000000:8.485281pt and 8.485281pt) -- +(45.000000:8.485281pt and 8.485281pt) -- +(135.000000:8.485281pt and 8.485281pt) -- +(225.000000:8.485281pt and 8.485281pt) -- cycle;
\clip (270.000000, 30.000000) +(-45.000000:8.485281pt and 8.485281pt) -- +(45.000000:8.485281pt and 8.485281pt) -- +(135.000000:8.485281pt and 8.485281pt) -- +(225.000000:8.485281pt and 8.485281pt) -- cycle;
\draw (270.000000, 30.000000) node {$H$};
\end{scope}
% Line 32: l0 G {$Z$}
\begin{scope}
\draw[fill=white] (294.000000, 30.000000) +(-45.000000:8.485281pt and 8.485281pt) -- +(45.000000:8.485281pt and 8.485281pt) -- +(135.000000:8.485281pt and 8.485281pt) -- +(225.000000:8.485281pt and 8.485281pt) -- cycle;
\clip (294.000000, 30.000000) +(-45.000000:8.485281pt and 8.485281pt) -- +(45.000000:8.485281pt and 8.485281pt) -- +(135.000000:8.485281pt and 8.485281pt) -- +(225.000000:8.485281pt and 8.485281pt) -- cycle;
\draw (294.000000, 30.000000) node {{$Z$}};
\end{scope}
% Line 33: l0 G {$T$}
\begin{scope}
\draw[fill=white] (318.000000, 30.000000) +(-45.000000:8.485281pt and 8.485281pt) -- +(45.000000:8.485281pt and 8.485281pt) -- +(135.000000:8.485281pt and 8.485281pt) -- +(225.000000:8.485281pt and 8.485281pt) -- cycle;
\clip (318.000000, 30.000000) +(-45.000000:8.485281pt and 8.485281pt) -- +(45.000000:8.485281pt and 8.485281pt) -- +(135.000000:8.485281pt and 8.485281pt) -- +(225.000000:8.485281pt and 8.485281pt) -- cycle;
\draw (318.000000, 30.000000) node {{$T$}};
\end{scope}
% Line 34: l0 H
\begin{scope}
\draw[fill=white] (342.000000, 30.000000) +(-45.000000:8.485281pt and 8.485281pt) -- +(45.000000:8.485281pt and 8.485281pt) -- +(135.000000:8.485281pt and 8.485281pt) -- +(225.000000:8.485281pt and 8.485281pt) -- cycle;
\clip (342.000000, 30.000000) +(-45.000000:8.485281pt and 8.485281pt) -- +(45.000000:8.485281pt and 8.485281pt) -- +(135.000000:8.485281pt and 8.485281pt) -- +(225.000000:8.485281pt and 8.485281pt) -- cycle;
\draw (342.000000, 30.000000) node {$H$};
\end{scope}
% Line 35: l0 G {$T^\dagger$}
\begin{scope}
\draw[fill=white] (366.000000, 30.000000) +(-45.000000:8.485281pt and 8.485281pt) -- +(45.000000:8.485281pt and 8.485281pt) -- +(135.000000:8.485281pt and 8.485281pt) -- +(225.000000:8.485281pt and 8.485281pt) -- cycle;
\clip (366.000000, 30.000000) +(-45.000000:8.485281pt and 8.485281pt) -- +(45.000000:8.485281pt and 8.485281pt) -- +(135.000000:8.485281pt and 8.485281pt) -- +(225.000000:8.485281pt and 8.485281pt) -- cycle;
\draw (366.000000, 30.000000) node {{$T^\dagger$}};
\end{scope}
% Line 36: l0 H
\begin{scope}
\draw[fill=white] (390.000000, 30.000000) +(-45.000000:8.485281pt and 8.485281pt) -- +(45.000000:8.485281pt and 8.485281pt) -- +(135.000000:8.485281pt and 8.485281pt) -- +(225.000000:8.485281pt and 8.485281pt) -- cycle;
\clip (390.000000, 30.000000) +(-45.000000:8.485281pt and 8.485281pt) -- +(45.000000:8.485281pt and 8.485281pt) -- +(135.000000:8.485281pt and 8.485281pt) -- +(225.000000:8.485281pt and 8.485281pt) -- cycle;
\draw (390.000000, 30.000000) node {$H$};
\end{scope}
% Line 37: l0 +l1
\draw (411.000000,30.000000) -- (411.000000,15.000000);
\filldraw (411.000000, 30.000000) circle(1.500000pt);
\begin{scope}
\draw[fill=white] (411.000000, 15.000000) circle(3.000000pt);
\clip (411.000000, 15.000000) circle(3.000000pt);
\draw (408.000000, 15.000000) -- (414.000000, 15.000000);
\draw (411.000000, 12.000000) -- (411.000000, 18.000000);
\end{scope}
% Done with gates; drawing ending labels
\draw[color=black] (420.000000,30.000000) node[right] {$Q0$};
\draw[color=black] (420.000000,15.000000) node[right] {$Q2$};
\draw[color=black] (420.000000,0.000000) node[right] {$Q1$};
% Done with ending labels; drawing cut lines and comments
% Done with comments
\end{tikzpicture}

%% file: cnotTcnot.tikz
\begin{tikzpicture}[scale=1.000000,x=1pt,y=1pt]
\filldraw[color=white] (0.000000, -7.500000) rectangle (147.000000, 22.500000);
% Drawing wires
% Line 1: 1 +0
\draw[color=black] (0.000000,15.000000) -- (147.000000,15.000000);
% Line 1: 1 +0
\draw[color=black] (0.000000,0.000000) -- (147.000000,0.000000);
% Done with wires; drawing gates
% Line 1: 1 +0
\draw (9.000000,15.000000) -- (9.000000,0.000000);
\filldraw (9.000000, 0.000000) circle(1.500000pt);
\begin{scope}
\draw[fill=white] (9.000000, 15.000000) circle(3.000000pt);
\clip (9.000000, 15.000000) circle(3.000000pt);
\draw (6.000000, 15.000000) -- (12.000000, 15.000000);
\draw (9.000000, 12.000000) -- (9.000000, 18.000000);
\end{scope}
% Line 2: 0 G {$T$}
\begin{scope}
\draw[fill=white] (30.000000, 15.000000) +(-45.000000:8.485281pt and 8.485281pt) -- +(45.000000:8.485281pt and 8.485281pt) -- +(135.000000:8.485281pt and 8.485281pt) -- +(225.000000:8.485281pt and 8.485281pt) -- cycle;
\clip (30.000000, 15.000000) +(-45.000000:8.485281pt and 8.485281pt) -- +(45.000000:8.485281pt and 8.485281pt) -- +(135.000000:8.485281pt and 8.485281pt) -- +(225.000000:8.485281pt and 8.485281pt) -- cycle;
\draw (30.000000, 15.000000) node {{$T$}};
\end{scope}
% Line 3: 1 +0
\draw (51.000000,15.000000) -- (51.000000,0.000000);
\filldraw (51.000000, 0.000000) circle(1.500000pt);
\begin{scope}
\draw[fill=white] (51.000000, 15.000000) circle(3.000000pt);
\clip (51.000000, 15.000000) circle(3.000000pt);
\draw (48.000000, 15.000000) -- (54.000000, 15.000000);
\draw (51.000000, 12.000000) -- (51.000000, 18.000000);
\end{scope}
% Line 4: =
\draw[fill=white,color=white] (66.000000, -6.000000) rectangle (81.000000, 21.000000);
\draw (73.500000, 7.500000) node {$=$};
% Line 5: 0 +1
\draw (96.000000,15.000000) -- (96.000000,0.000000);
\filldraw (96.000000, 15.000000) circle(1.500000pt);
\begin{scope}
\draw[fill=white] (96.000000, 0.000000) circle(3.000000pt);
\clip (96.000000, 0.000000) circle(3.000000pt);
\draw (93.000000, 0.000000) -- (99.000000, 0.000000);
\draw (96.000000, -3.000000) -- (96.000000, 3.000000);
\end{scope}
% Line 6: 1 G {$T$}
\begin{scope}
\draw[fill=white] (117.000000, -0.000000) +(-45.000000:8.485281pt and 8.485281pt) -- +(45.000000:8.485281pt and 8.485281pt) -- +(135.000000:8.485281pt and 8.485281pt) -- +(225.000000:8.485281pt and 8.485281pt) -- cycle;
\clip (117.000000, -0.000000) +(-45.000000:8.485281pt and 8.485281pt) -- +(45.000000:8.485281pt and 8.485281pt) -- +(135.000000:8.485281pt and 8.485281pt) -- +(225.000000:8.485281pt and 8.485281pt) -- cycle;
\draw (117.000000, -0.000000) node {{$T$}};
\end{scope}
% Line 7: 0 +1
\draw (138.000000,15.000000) -- (138.000000,0.000000);
\filldraw (138.000000, 15.000000) circle(1.500000pt);
\begin{scope}
\draw[fill=white] (138.000000, 0.000000) circle(3.000000pt);
\clip (138.000000, 0.000000) circle(3.000000pt);
\draw (135.000000, 0.000000) -- (141.000000, 0.000000);
\draw (138.000000, -3.000000) -- (138.000000, 3.000000);
\end{scope}
% Done with gates; drawing ending labels
% Done with ending labels; drawing cut lines and comments
% Done with comments
\end{tikzpicture}

%% file: Full_Adder_c.tikz
\begin{tikzpicture}[scale=1.000000,x=1pt,y=1pt]
\filldraw[color=white] (0.000000, -7.500000) rectangle (348.000000, 52.500000);
% Drawing wires
% Line 1: l0 W i0 o0
\draw[color=black] (0.000000,45.000000) -- (348.000000,45.000000);
\draw[color=black] (0.000000,45.000000) node[left] {$i0$};
% Line 2: l1 W i1 o1
\draw[color=black] (0.000000,30.000000) -- (348.000000,30.000000);
\draw[color=black] (0.000000,30.000000) node[left] {$i1$};
% Line 3: l2 W i3 o3
\draw[color=black] (0.000000,15.000000) -- (348.000000,15.000000);
\draw[color=black] (0.000000,15.000000) node[left] {$i3$};
% Line 4: l3 W i2 o2
\draw[color=black] (0.000000,0.000000) -- (348.000000,0.000000);
\draw[color=black] (0.000000,0.000000) node[left] {$i2$};
% Done with wires; drawing gates
% Line 5: l2 H
\begin{scope}
\draw[fill=white] (12.000000, 15.000000) +(-45.000000:8.485281pt and 8.485281pt) -- +(45.000000:8.485281pt and 8.485281pt) -- +(135.000000:8.485281pt and 8.485281pt) -- +(225.000000:8.485281pt and 8.485281pt) -- cycle;
\clip (12.000000, 15.000000) +(-45.000000:8.485281pt and 8.485281pt) -- +(45.000000:8.485281pt and 8.485281pt) -- +(135.000000:8.485281pt and 8.485281pt) -- +(225.000000:8.485281pt and 8.485281pt) -- cycle;
\draw (12.000000, 15.000000) node {$H$};
\end{scope}
% Line 6: l3 +l2
\draw (33.000000,15.000000) -- (33.000000,0.000000);
\filldraw (33.000000, 0.000000) circle(1.500000pt);
\begin{scope}
\draw[fill=white] (33.000000, 15.000000) circle(3.000000pt);
\clip (33.000000, 15.000000) circle(3.000000pt);
\draw (30.000000, 15.000000) -- (36.000000, 15.000000);
\draw (33.000000, 12.000000) -- (33.000000, 18.000000);
\end{scope}
% Line 7: l2 G {$T^\dagger$}
\begin{scope}
\draw[fill=white] (54.000000, 15.000000) +(-45.000000:8.485281pt and 8.485281pt) -- +(45.000000:8.485281pt and 8.485281pt) -- +(135.000000:8.485281pt and 8.485281pt) -- +(225.000000:8.485281pt and 8.485281pt) -- cycle;
\clip (54.000000, 15.000000) +(-45.000000:8.485281pt and 8.485281pt) -- +(45.000000:8.485281pt and 8.485281pt) -- +(135.000000:8.485281pt and 8.485281pt) -- +(225.000000:8.485281pt and 8.485281pt) -- cycle;
\draw (54.000000, 15.000000) node {{$T^\dagger$}};
\end{scope}
% Line 8: l3 +l2
\draw (75.000000,15.000000) -- (75.000000,0.000000);
\filldraw (75.000000, 0.000000) circle(1.500000pt);
\begin{scope}
\draw[fill=white] (75.000000, 15.000000) circle(3.000000pt);
\clip (75.000000, 15.000000) circle(3.000000pt);
\draw (72.000000, 15.000000) -- (78.000000, 15.000000);
\draw (75.000000, 12.000000) -- (75.000000, 18.000000);
\end{scope}
% Line 9: l1 +l2
\draw (93.000000,30.000000) -- (93.000000,15.000000);
\filldraw (93.000000, 30.000000) circle(1.500000pt);
\begin{scope}
\draw[fill=white] (93.000000, 15.000000) circle(3.000000pt);
\clip (93.000000, 15.000000) circle(3.000000pt);
\draw (90.000000, 15.000000) -- (96.000000, 15.000000);
\draw (93.000000, 12.000000) -- (93.000000, 18.000000);
\end{scope}
% Line 10: l2 G {$T^\dagger$}
\begin{scope}
\draw[fill=white] (114.000000, 15.000000) +(-45.000000:8.485281pt and 8.485281pt) -- +(45.000000:8.485281pt and 8.485281pt) -- +(135.000000:8.485281pt and 8.485281pt) -- +(225.000000:8.485281pt and 8.485281pt) -- cycle;
\clip (114.000000, 15.000000) +(-45.000000:8.485281pt and 8.485281pt) -- +(45.000000:8.485281pt and 8.485281pt) -- +(135.000000:8.485281pt and 8.485281pt) -- +(225.000000:8.485281pt and 8.485281pt) -- cycle;
\draw (114.000000, 15.000000) node {{$T^\dagger$}};
\end{scope}
% Line 11: l1 +l2
\draw (135.000000,30.000000) -- (135.000000,15.000000);
\filldraw (135.000000, 30.000000) circle(1.500000pt);
\begin{scope}
\draw[fill=white] (135.000000, 15.000000) circle(3.000000pt);
\clip (135.000000, 15.000000) circle(3.000000pt);
\draw (132.000000, 15.000000) -- (138.000000, 15.000000);
\draw (135.000000, 12.000000) -- (135.000000, 18.000000);
\end{scope}
% Line 12: l0 +l2
\draw (153.000000,45.000000) -- (153.000000,15.000000);
\filldraw (153.000000, 45.000000) circle(1.500000pt);
\begin{scope}
\draw[fill=white] (153.000000, 15.000000) circle(3.000000pt);
\clip (153.000000, 15.000000) circle(3.000000pt);
\draw (150.000000, 15.000000) -- (156.000000, 15.000000);
\draw (153.000000, 12.000000) -- (153.000000, 18.000000);
\end{scope}
% Line 14: l2 G {$T^\dagger$}
\begin{scope}
\draw[fill=white] (174.000000, 15.000000) +(-45.000000:8.485281pt and 8.485281pt) -- +(45.000000:8.485281pt and 8.485281pt) -- +(135.000000:8.485281pt and 8.485281pt) -- +(225.000000:8.485281pt and 8.485281pt) -- cycle;
\clip (174.000000, 15.000000) +(-45.000000:8.485281pt and 8.485281pt) -- +(45.000000:8.485281pt and 8.485281pt) -- +(135.000000:8.485281pt and 8.485281pt) -- +(225.000000:8.485281pt and 8.485281pt) -- cycle;
\draw (174.000000, 15.000000) node {{$T^\dagger$}};
\end{scope}
% Line 15: l3 G {$T$}
\begin{scope}
\draw[fill=white] (174.000000, -0.000000) +(-45.000000:8.485281pt and 8.485281pt) -- +(45.000000:8.485281pt and 8.485281pt) -- +(135.000000:8.485281pt and 8.485281pt) -- +(225.000000:8.485281pt and 8.485281pt) -- cycle;
\clip (174.000000, -0.000000) +(-45.000000:8.485281pt and 8.485281pt) -- +(45.000000:8.485281pt and 8.485281pt) -- +(135.000000:8.485281pt and 8.485281pt) -- +(225.000000:8.485281pt and 8.485281pt) -- cycle;
\draw (174.000000, -0.000000) node {{$T$}};
\end{scope}
% Line 16: l1 G {$T$}
\begin{scope}
\draw[fill=white] (174.000000, 30.000000) +(-45.000000:8.485281pt and 8.485281pt) -- +(45.000000:8.485281pt and 8.485281pt) -- +(135.000000:8.485281pt and 8.485281pt) -- +(225.000000:8.485281pt and 8.485281pt) -- cycle;
\clip (174.000000, 30.000000) +(-45.000000:8.485281pt and 8.485281pt) -- +(45.000000:8.485281pt and 8.485281pt) -- +(135.000000:8.485281pt and 8.485281pt) -- +(225.000000:8.485281pt and 8.485281pt) -- cycle;
\draw (174.000000, 30.000000) node {{$T$}};
\end{scope}
% Line 18: l0 +l2
\draw (195.000000,45.000000) -- (195.000000,15.000000);
\filldraw (195.000000, 45.000000) circle(1.500000pt);
\begin{scope}
\draw[fill=white] (195.000000, 15.000000) circle(3.000000pt);
\clip (195.000000, 15.000000) circle(3.000000pt);
\draw (192.000000, 15.000000) -- (198.000000, 15.000000);
\draw (195.000000, 12.000000) -- (195.000000, 18.000000);
\end{scope}
% Line 19: l0 +l1
\draw (213.000000,45.000000) -- (213.000000,30.000000);
\filldraw (213.000000, 45.000000) circle(1.500000pt);
\begin{scope}
\draw[fill=white] (213.000000, 30.000000) circle(3.000000pt);
\clip (213.000000, 30.000000) circle(3.000000pt);
\draw (210.000000, 30.000000) -- (216.000000, 30.000000);
\draw (213.000000, 27.000000) -- (213.000000, 33.000000);
\end{scope}
% Line 20: l1 +l3 color=red
\begin{scope}[color=red]
\draw (231.000000,30.000000) -- (231.000000,0.000000);
\filldraw (231.000000, 30.000000) circle(1.500000pt);
\begin{scope}
\draw[fill=white] (231.000000, 0.000000) circle(3.000000pt);
\clip (231.000000, 0.000000) circle(3.000000pt);
\draw (228.000000, 0.000000) -- (234.000000, 0.000000);
\draw (231.000000, -3.000000) -- (231.000000, 3.000000);
\end{scope}
\end{scope}
% Line 21: l3 +l2
\draw (249.000000,15.000000) -- (249.000000,0.000000);
\filldraw (249.000000, 0.000000) circle(1.500000pt);
\begin{scope}
\draw[fill=white] (249.000000, 15.000000) circle(3.000000pt);
\clip (249.000000, 15.000000) circle(3.000000pt);
\draw (246.000000, 15.000000) -- (252.000000, 15.000000);
\draw (249.000000, 12.000000) -- (249.000000, 18.000000);
\end{scope}
% Line 22: l2 G {$T$}
\begin{scope}
\draw[fill=white] (270.000000, 15.000000) +(-45.000000:8.485281pt and 8.485281pt) -- +(45.000000:8.485281pt and 8.485281pt) -- +(135.000000:8.485281pt and 8.485281pt) -- +(225.000000:8.485281pt and 8.485281pt) -- cycle;
\clip (270.000000, 15.000000) +(-45.000000:8.485281pt and 8.485281pt) -- +(45.000000:8.485281pt and 8.485281pt) -- +(135.000000:8.485281pt and 8.485281pt) -- +(225.000000:8.485281pt and 8.485281pt) -- cycle;
\draw (270.000000, 15.000000) node {{$T$}};
\end{scope}
% Line 23: l3 +l2
\draw (291.000000,15.000000) -- (291.000000,0.000000);
\filldraw (291.000000, 0.000000) circle(1.500000pt);
\begin{scope}
\draw[fill=white] (291.000000, 15.000000) circle(3.000000pt);
\clip (291.000000, 15.000000) circle(3.000000pt);
\draw (288.000000, 15.000000) -- (294.000000, 15.000000);
\draw (291.000000, 12.000000) -- (291.000000, 18.000000);
\end{scope}
% Line 24: l3 G {$T^\dagger$}
\begin{scope}
\draw[fill=white] (312.000000, -0.000000) +(-45.000000:8.485281pt and 8.485281pt) -- +(45.000000:8.485281pt and 8.485281pt) -- +(135.000000:8.485281pt and 8.485281pt) -- +(225.000000:8.485281pt and 8.485281pt) -- cycle;
\clip (312.000000, -0.000000) +(-45.000000:8.485281pt and 8.485281pt) -- +(45.000000:8.485281pt and 8.485281pt) -- +(135.000000:8.485281pt and 8.485281pt) -- +(225.000000:8.485281pt and 8.485281pt) -- cycle;
\draw (312.000000, -0.000000) node {{$T^\dagger$}};
\end{scope}
% Line 25: l2 G {$S$}
\begin{scope}
\draw[fill=white] (312.000000, 15.000000) +(-45.000000:8.485281pt and 8.485281pt) -- +(45.000000:8.485281pt and 8.485281pt) -- +(135.000000:8.485281pt and 8.485281pt) -- +(225.000000:8.485281pt and 8.485281pt) -- cycle;
\clip (312.000000, 15.000000) +(-45.000000:8.485281pt and 8.485281pt) -- +(45.000000:8.485281pt and 8.485281pt) -- +(135.000000:8.485281pt and 8.485281pt) -- +(225.000000:8.485281pt and 8.485281pt) -- cycle;
\draw (312.000000, 15.000000) node {{$S$}};
\end{scope}
% Line 26: l2 H
\begin{scope}
\draw[fill=white] (336.000000, 15.000000) +(-45.000000:8.485281pt and 8.485281pt) -- +(45.000000:8.485281pt and 8.485281pt) -- +(135.000000:8.485281pt and 8.485281pt) -- +(225.000000:8.485281pt and 8.485281pt) -- cycle;
\clip (336.000000, 15.000000) +(-45.000000:8.485281pt and 8.485281pt) -- +(45.000000:8.485281pt and 8.485281pt) -- +(135.000000:8.485281pt and 8.485281pt) -- +(225.000000:8.485281pt and 8.485281pt) -- cycle;
\draw (336.000000, 15.000000) node {$H$};
\end{scope}
% Done with gates; drawing ending labels
\draw[color=black] (348.000000,45.000000) node[right] {$o0$};
\draw[color=black] (348.000000,30.000000) node[right] {$o1$};
\draw[color=black] (348.000000,15.000000) node[right] {$o3$};
\draw[color=black] (348.000000,0.000000) node[right] {$o2$};
% Done with ending labels; drawing cut lines and comments
% Done with comments
\end{tikzpicture}

%% file: Full_Adder_c_a.tikz
\begin{tikzpicture}[scale=1.000000,x=1pt,y=1pt]
\filldraw[color=white] (0.000000, -7.500000) rectangle (432.000000, 52.500000);
% Drawing wires
% Line 1: l0 W i0 o0
\draw[color=black] (0.000000,45.000000) -- (432.000000,45.000000);
\draw[color=black] (0.000000,45.000000) node[left] {$i0$};
% Line 2: l1 W i1 o1
\draw[color=black] (0.000000,30.000000) -- (432.000000,30.000000);
\draw[color=black] (0.000000,30.000000) node[left] {$i1$};
% Line 3: l2 W i3 o2
\draw[color=black] (0.000000,15.000000) -- (432.000000,15.000000);
\draw[color=black] (0.000000,15.000000) node[left] {$i3$};
% Line 4: l3 W i2 o3
\draw[color=black] (0.000000,0.000000) -- (432.000000,0.000000);
\draw[color=black] (0.000000,0.000000) node[left] {$i2$};
% Done with wires; drawing gates
% Line 5: l2 H
\begin{scope}
\draw[fill=white] (12.000000, 15.000000) +(-45.000000:8.485281pt and 8.485281pt) -- +(45.000000:8.485281pt and 8.485281pt) -- +(135.000000:8.485281pt and 8.485281pt) -- +(225.000000:8.485281pt and 8.485281pt) -- cycle;
\clip (12.000000, 15.000000) +(-45.000000:8.485281pt and 8.485281pt) -- +(45.000000:8.485281pt and 8.485281pt) -- +(135.000000:8.485281pt and 8.485281pt) -- +(225.000000:8.485281pt and 8.485281pt) -- cycle;
\draw (12.000000, 15.000000) node {$H$};
\end{scope}
% Line 6: l3 +l2
\draw (33.000000,15.000000) -- (33.000000,0.000000);
\filldraw (33.000000, 0.000000) circle(1.500000pt);
\begin{scope}
\draw[fill=white] (33.000000, 15.000000) circle(3.000000pt);
\clip (33.000000, 15.000000) circle(3.000000pt);
\draw (30.000000, 15.000000) -- (36.000000, 15.000000);
\draw (33.000000, 12.000000) -- (33.000000, 18.000000);
\end{scope}
% Line 7: l2 G {$T^\dagger$}
\begin{scope}
\draw[fill=white] (54.000000, 15.000000) +(-45.000000:8.485281pt and 8.485281pt) -- +(45.000000:8.485281pt and 8.485281pt) -- +(135.000000:8.485281pt and 8.485281pt) -- +(225.000000:8.485281pt and 8.485281pt) -- cycle;
\clip (54.000000, 15.000000) +(-45.000000:8.485281pt and 8.485281pt) -- +(45.000000:8.485281pt and 8.485281pt) -- +(135.000000:8.485281pt and 8.485281pt) -- +(225.000000:8.485281pt and 8.485281pt) -- cycle;
\draw (54.000000, 15.000000) node {{$T^\dagger$}};
\end{scope}
% Line 8: l3 +l2
\draw (75.000000,15.000000) -- (75.000000,0.000000);
\filldraw (75.000000, 0.000000) circle(1.500000pt);
\begin{scope}
\draw[fill=white] (75.000000, 15.000000) circle(3.000000pt);
\clip (75.000000, 15.000000) circle(3.000000pt);
\draw (72.000000, 15.000000) -- (78.000000, 15.000000);
\draw (75.000000, 12.000000) -- (75.000000, 18.000000);
\end{scope}
% Line 9: l1 +l2
\draw (93.000000,30.000000) -- (93.000000,15.000000);
\filldraw (93.000000, 30.000000) circle(1.500000pt);
\begin{scope}
\draw[fill=white] (93.000000, 15.000000) circle(3.000000pt);
\clip (93.000000, 15.000000) circle(3.000000pt);
\draw (90.000000, 15.000000) -- (96.000000, 15.000000);
\draw (93.000000, 12.000000) -- (93.000000, 18.000000);
\end{scope}
% Line 10: l2 G {$T^\dagger$}
\begin{scope}
\draw[fill=white] (114.000000, 15.000000) +(-45.000000:8.485281pt and 8.485281pt) -- +(45.000000:8.485281pt and 8.485281pt) -- +(135.000000:8.485281pt and 8.485281pt) -- +(225.000000:8.485281pt and 8.485281pt) -- cycle;
\clip (114.000000, 15.000000) +(-45.000000:8.485281pt and 8.485281pt) -- +(45.000000:8.485281pt and 8.485281pt) -- +(135.000000:8.485281pt and 8.485281pt) -- +(225.000000:8.485281pt and 8.485281pt) -- cycle;
\draw (114.000000, 15.000000) node {{$T^\dagger$}};
\end{scope}
% Line 11: l1 +l2
\draw (135.000000,30.000000) -- (135.000000,15.000000);
\filldraw (135.000000, 30.000000) circle(1.500000pt);
\begin{scope}
\draw[fill=white] (135.000000, 15.000000) circle(3.000000pt);
\clip (135.000000, 15.000000) circle(3.000000pt);
\draw (132.000000, 15.000000) -- (138.000000, 15.000000);
\draw (135.000000, 12.000000) -- (135.000000, 18.000000);
\end{scope}
% Line 12: l0 +l2
\draw (153.000000,45.000000) -- (153.000000,15.000000);
\filldraw (153.000000, 45.000000) circle(1.500000pt);
\begin{scope}
\draw[fill=white] (153.000000, 15.000000) circle(3.000000pt);
\clip (153.000000, 15.000000) circle(3.000000pt);
\draw (150.000000, 15.000000) -- (156.000000, 15.000000);
\draw (153.000000, 12.000000) -- (153.000000, 18.000000);
\end{scope}
% Line 14: l2 G {$T^\dagger$}
\begin{scope}
\draw[fill=white] (174.000000, 15.000000) +(-45.000000:8.485281pt and 8.485281pt) -- +(45.000000:8.485281pt and 8.485281pt) -- +(135.000000:8.485281pt and 8.485281pt) -- +(225.000000:8.485281pt and 8.485281pt) -- cycle;
\clip (174.000000, 15.000000) +(-45.000000:8.485281pt and 8.485281pt) -- +(45.000000:8.485281pt and 8.485281pt) -- +(135.000000:8.485281pt and 8.485281pt) -- +(225.000000:8.485281pt and 8.485281pt) -- cycle;
\draw (174.000000, 15.000000) node {{$T^\dagger$}};
\end{scope}
% Line 15: l3 G {$T$}
\begin{scope}
\draw[fill=white] (174.000000, -0.000000) +(-45.000000:8.485281pt and 8.485281pt) -- +(45.000000:8.485281pt and 8.485281pt) -- +(135.000000:8.485281pt and 8.485281pt) -- +(225.000000:8.485281pt and 8.485281pt) -- cycle;
\clip (174.000000, -0.000000) +(-45.000000:8.485281pt and 8.485281pt) -- +(45.000000:8.485281pt and 8.485281pt) -- +(135.000000:8.485281pt and 8.485281pt) -- +(225.000000:8.485281pt and 8.485281pt) -- cycle;
\draw (174.000000, -0.000000) node {{$T$}};
\end{scope}
% Line 16: l1 G {$T$}
\begin{scope}
\draw[fill=white] (174.000000, 30.000000) +(-45.000000:8.485281pt and 8.485281pt) -- +(45.000000:8.485281pt and 8.485281pt) -- +(135.000000:8.485281pt and 8.485281pt) -- +(225.000000:8.485281pt and 8.485281pt) -- cycle;
\clip (174.000000, 30.000000) +(-45.000000:8.485281pt and 8.485281pt) -- +(45.000000:8.485281pt and 8.485281pt) -- +(135.000000:8.485281pt and 8.485281pt) -- +(225.000000:8.485281pt and 8.485281pt) -- cycle;
\draw (174.000000, 30.000000) node {{$T$}};
\end{scope}
% Line 18: l0 +l2
\draw (195.000000,45.000000) -- (195.000000,15.000000);
\filldraw (195.000000, 45.000000) circle(1.500000pt);
\begin{scope}
\draw[fill=white] (195.000000, 15.000000) circle(3.000000pt);
\clip (195.000000, 15.000000) circle(3.000000pt);
\draw (192.000000, 15.000000) -- (198.000000, 15.000000);
\draw (195.000000, 12.000000) -- (195.000000, 18.000000);
\end{scope}
% Line 19: l0 +l1
\draw (213.000000,45.000000) -- (213.000000,30.000000);
\filldraw (213.000000, 45.000000) circle(1.500000pt);
\begin{scope}
\draw[fill=white] (213.000000, 30.000000) circle(3.000000pt);
\clip (213.000000, 30.000000) circle(3.000000pt);
\draw (210.000000, 30.000000) -- (216.000000, 30.000000);
\draw (213.000000, 27.000000) -- (213.000000, 33.000000);
\end{scope}
% Line 21: l3 +l2 color=purple
\begin{scope}[color=purple]
\draw (213.000000,15.000000) -- (213.000000,0.000000);
\filldraw (213.000000, 0.000000) circle(1.500000pt);
\begin{scope}
\draw[fill=white] (213.000000, 15.000000) circle(3.000000pt);
\clip (213.000000, 15.000000) circle(3.000000pt);
\draw (210.000000, 15.000000) -- (216.000000, 15.000000);
\draw (213.000000, 12.000000) -- (213.000000, 18.000000);
\end{scope}
\end{scope}
% Line 22: l2 H color=purple
\begin{scope}[color=purple]
\begin{scope}[color=purple]
\begin{scope}
\draw[fill=white] (234.000000, 15.000000) +(-45.000000:8.485281pt and 8.485281pt) -- +(45.000000:8.485281pt and 8.485281pt) -- +(135.000000:8.485281pt and 8.485281pt) -- +(225.000000:8.485281pt and 8.485281pt) -- cycle;
\clip (234.000000, 15.000000) +(-45.000000:8.485281pt and 8.485281pt) -- +(45.000000:8.485281pt and 8.485281pt) -- +(135.000000:8.485281pt and 8.485281pt) -- +(225.000000:8.485281pt and 8.485281pt) -- cycle;
\draw (234.000000, 15.000000) node {$H$};
\end{scope}
\end{scope}
\end{scope}
% Line 23: l3 H color=purple
\begin{scope}[color=purple]
\begin{scope}[color=purple]
\begin{scope}
\draw[fill=white] (234.000000, -0.000000) +(-45.000000:8.485281pt and 8.485281pt) -- +(45.000000:8.485281pt and 8.485281pt) -- +(135.000000:8.485281pt and 8.485281pt) -- +(225.000000:8.485281pt and 8.485281pt) -- cycle;
\clip (234.000000, -0.000000) +(-45.000000:8.485281pt and 8.485281pt) -- +(45.000000:8.485281pt and 8.485281pt) -- +(135.000000:8.485281pt and 8.485281pt) -- +(225.000000:8.485281pt and 8.485281pt) -- cycle;
\draw (234.000000, -0.000000) node {$H$};
\end{scope}
\end{scope}
\end{scope}
% Line 24: l3 +l2 color=purple
\begin{scope}[color=purple]
\draw (255.000000,15.000000) -- (255.000000,0.000000);
\filldraw (255.000000, 0.000000) circle(1.500000pt);
\begin{scope}
\draw[fill=white] (255.000000, 15.000000) circle(3.000000pt);
\clip (255.000000, 15.000000) circle(3.000000pt);
\draw (252.000000, 15.000000) -- (258.000000, 15.000000);
\draw (255.000000, 12.000000) -- (255.000000, 18.000000);
\end{scope}
\end{scope}
% Line 25: l2 H color=purple
\begin{scope}[color=purple]
\begin{scope}[color=purple]
\begin{scope}
\draw[fill=white] (276.000000, 15.000000) +(-45.000000:8.485281pt and 8.485281pt) -- +(45.000000:8.485281pt and 8.485281pt) -- +(135.000000:8.485281pt and 8.485281pt) -- +(225.000000:8.485281pt and 8.485281pt) -- cycle;
\clip (276.000000, 15.000000) +(-45.000000:8.485281pt and 8.485281pt) -- +(45.000000:8.485281pt and 8.485281pt) -- +(135.000000:8.485281pt and 8.485281pt) -- +(225.000000:8.485281pt and 8.485281pt) -- cycle;
\draw (276.000000, 15.000000) node {$H$};
\end{scope}
\end{scope}
\end{scope}
% Line 26: l3 H color=purple
\begin{scope}[color=purple]
\begin{scope}[color=purple]
\begin{scope}
\draw[fill=white] (276.000000, -0.000000) +(-45.000000:8.485281pt and 8.485281pt) -- +(45.000000:8.485281pt and 8.485281pt) -- +(135.000000:8.485281pt and 8.485281pt) -- +(225.000000:8.485281pt and 8.485281pt) -- cycle;
\clip (276.000000, -0.000000) +(-45.000000:8.485281pt and 8.485281pt) -- +(45.000000:8.485281pt and 8.485281pt) -- +(135.000000:8.485281pt and 8.485281pt) -- +(225.000000:8.485281pt and 8.485281pt) -- cycle;
\draw (276.000000, -0.000000) node {$H$};
\end{scope}
\end{scope}
\end{scope}
% Line 27: l3 +l2 color=purple
\begin{scope}[color=purple]
\draw (297.000000,15.000000) -- (297.000000,0.000000);
\filldraw (297.000000, 0.000000) circle(1.500000pt);
\begin{scope}
\draw[fill=white] (297.000000, 15.000000) circle(3.000000pt);
\clip (297.000000, 15.000000) circle(3.000000pt);
\draw (294.000000, 15.000000) -- (300.000000, 15.000000);
\draw (297.000000, 12.000000) -- (297.000000, 18.000000);
\end{scope}
\end{scope}
% Line 29: l1 +l2 color=red
\begin{scope}[color=red]
\draw (315.000000,30.000000) -- (315.000000,15.000000);
\filldraw (315.000000, 30.000000) circle(1.500000pt);
\begin{scope}
\draw[fill=white] (315.000000, 15.000000) circle(3.000000pt);
\clip (315.000000, 15.000000) circle(3.000000pt);
\draw (312.000000, 15.000000) -- (318.000000, 15.000000);
\draw (315.000000, 12.000000) -- (315.000000, 18.000000);
\end{scope}
\end{scope}
% Line 30: l2 +l3
\draw (333.000000,15.000000) -- (333.000000,0.000000);
\filldraw (333.000000, 15.000000) circle(1.500000pt);
\begin{scope}
\draw[fill=white] (333.000000, 0.000000) circle(3.000000pt);
\clip (333.000000, 0.000000) circle(3.000000pt);
\draw (330.000000, 0.000000) -- (336.000000, 0.000000);
\draw (333.000000, -3.000000) -- (333.000000, 3.000000);
\end{scope}
% Line 31: l3 G {$T$}
\begin{scope}
\draw[fill=white] (354.000000, -0.000000) +(-45.000000:8.485281pt and 8.485281pt) -- +(45.000000:8.485281pt and 8.485281pt) -- +(135.000000:8.485281pt and 8.485281pt) -- +(225.000000:8.485281pt and 8.485281pt) -- cycle;
\clip (354.000000, -0.000000) +(-45.000000:8.485281pt and 8.485281pt) -- +(45.000000:8.485281pt and 8.485281pt) -- +(135.000000:8.485281pt and 8.485281pt) -- +(225.000000:8.485281pt and 8.485281pt) -- cycle;
\draw (354.000000, -0.000000) node {{$T$}};
\end{scope}
% Line 32: l2 +l3
\draw (375.000000,15.000000) -- (375.000000,0.000000);
\filldraw (375.000000, 15.000000) circle(1.500000pt);
\begin{scope}
\draw[fill=white] (375.000000, 0.000000) circle(3.000000pt);
\clip (375.000000, 0.000000) circle(3.000000pt);
\draw (372.000000, 0.000000) -- (378.000000, 0.000000);
\draw (375.000000, -3.000000) -- (375.000000, 3.000000);
\end{scope}
% Line 33: l2 G {$T^\dagger$}
\begin{scope}
\draw[fill=white] (396.000000, 15.000000) +(-45.000000:8.485281pt and 8.485281pt) -- +(45.000000:8.485281pt and 8.485281pt) -- +(135.000000:8.485281pt and 8.485281pt) -- +(225.000000:8.485281pt and 8.485281pt) -- cycle;
\clip (396.000000, 15.000000) +(-45.000000:8.485281pt and 8.485281pt) -- +(45.000000:8.485281pt and 8.485281pt) -- +(135.000000:8.485281pt and 8.485281pt) -- +(225.000000:8.485281pt and 8.485281pt) -- cycle;
\draw (396.000000, 15.000000) node {{$T^\dagger$}};
\end{scope}
% Line 34: l3 G {$S$}
\begin{scope}
\draw[fill=white] (396.000000, -0.000000) +(-45.000000:8.485281pt and 8.485281pt) -- +(45.000000:8.485281pt and 8.485281pt) -- +(135.000000:8.485281pt and 8.485281pt) -- +(225.000000:8.485281pt and 8.485281pt) -- cycle;
\clip (396.000000, -0.000000) +(-45.000000:8.485281pt and 8.485281pt) -- +(45.000000:8.485281pt and 8.485281pt) -- +(135.000000:8.485281pt and 8.485281pt) -- +(225.000000:8.485281pt and 8.485281pt) -- cycle;
\draw (396.000000, -0.000000) node {{$S$}};
\end{scope}
% Line 35: l3 H
\begin{scope}
\draw[fill=white] (420.000000, -0.000000) +(-45.000000:8.485281pt and 8.485281pt) -- +(45.000000:8.485281pt and 8.485281pt) -- +(135.000000:8.485281pt and 8.485281pt) -- +(225.000000:8.485281pt and 8.485281pt) -- cycle;
\clip (420.000000, -0.000000) +(-45.000000:8.485281pt and 8.485281pt) -- +(45.000000:8.485281pt and 8.485281pt) -- +(135.000000:8.485281pt and 8.485281pt) -- +(225.000000:8.485281pt and 8.485281pt) -- cycle;
\draw (420.000000, -0.000000) node {$H$};
\end{scope}
% Done with gates; drawing ending labels
\draw[color=black] (432.000000,45.000000) node[right] {$o0$};
\draw[color=black] (432.000000,30.000000) node[right] {$o1$};
\draw[color=black] (432.000000,15.000000) node[right] {$o2$};
\draw[color=black] (432.000000,0.000000) node[right] {$o3$};
% Done with ending labels; drawing cut lines and comments
% Done with comments
\end{tikzpicture}

%% file: Full_Adder_c_b.tikz
\begin{tikzpicture}[scale=1.000000,x=1pt,y=1pt]
\filldraw[color=white] (0.000000, -7.500000) rectangle (432.000000, 52.500000);
% Drawing wires
% Line 1: l0 W i0 o0
\draw[color=black] (0.000000,45.000000) -- (432.000000,45.000000);
\draw[color=black] (0.000000,45.000000) node[left] {$i0$};
% Line 2: l1 W i1 o1
\draw[color=black] (0.000000,30.000000) -- (432.000000,30.000000);
\draw[color=black] (0.000000,30.000000) node[left] {$i1$};
% Line 3: l2 W i3 o2
\draw[color=black] (0.000000,15.000000) -- (432.000000,15.000000);
\draw[color=black] (0.000000,15.000000) node[left] {$i3$};
% Line 4: l3 W i2 o3
\draw[color=black] (0.000000,0.000000) -- (432.000000,0.000000);
\draw[color=black] (0.000000,0.000000) node[left] {$i2$};
% Done with wires; drawing gates
% Line 5: l2 H
\begin{scope}
\draw[fill=white] (12.000000, 15.000000) +(-45.000000:8.485281pt and 8.485281pt) -- +(45.000000:8.485281pt and 8.485281pt) -- +(135.000000:8.485281pt and 8.485281pt) -- +(225.000000:8.485281pt and 8.485281pt) -- cycle;
\clip (12.000000, 15.000000) +(-45.000000:8.485281pt and 8.485281pt) -- +(45.000000:8.485281pt and 8.485281pt) -- +(135.000000:8.485281pt and 8.485281pt) -- +(225.000000:8.485281pt and 8.485281pt) -- cycle;
\draw (12.000000, 15.000000) node {$H$};
\end{scope}
% Line 6: l3 +l2
\draw (33.000000,15.000000) -- (33.000000,0.000000);
\filldraw (33.000000, 0.000000) circle(1.500000pt);
\begin{scope}
\draw[fill=white] (33.000000, 15.000000) circle(3.000000pt);
\clip (33.000000, 15.000000) circle(3.000000pt);
\draw (30.000000, 15.000000) -- (36.000000, 15.000000);
\draw (33.000000, 12.000000) -- (33.000000, 18.000000);
\end{scope}
% Line 7: l2 G {$T^\dagger$}
\begin{scope}
\draw[fill=white] (54.000000, 15.000000) +(-45.000000:8.485281pt and 8.485281pt) -- +(45.000000:8.485281pt and 8.485281pt) -- +(135.000000:8.485281pt and 8.485281pt) -- +(225.000000:8.485281pt and 8.485281pt) -- cycle;
\clip (54.000000, 15.000000) +(-45.000000:8.485281pt and 8.485281pt) -- +(45.000000:8.485281pt and 8.485281pt) -- +(135.000000:8.485281pt and 8.485281pt) -- +(225.000000:8.485281pt and 8.485281pt) -- cycle;
\draw (54.000000, 15.000000) node {{$T^\dagger$}};
\end{scope}
% Line 8: l3 +l2
\draw (75.000000,15.000000) -- (75.000000,0.000000);
\filldraw (75.000000, 0.000000) circle(1.500000pt);
\begin{scope}
\draw[fill=white] (75.000000, 15.000000) circle(3.000000pt);
\clip (75.000000, 15.000000) circle(3.000000pt);
\draw (72.000000, 15.000000) -- (78.000000, 15.000000);
\draw (75.000000, 12.000000) -- (75.000000, 18.000000);
\end{scope}
% Line 9: l1 +l2
\draw (93.000000,30.000000) -- (93.000000,15.000000);
\filldraw (93.000000, 30.000000) circle(1.500000pt);
\begin{scope}
\draw[fill=white] (93.000000, 15.000000) circle(3.000000pt);
\clip (93.000000, 15.000000) circle(3.000000pt);
\draw (90.000000, 15.000000) -- (96.000000, 15.000000);
\draw (93.000000, 12.000000) -- (93.000000, 18.000000);
\end{scope}
% Line 10: l2 G {$T^\dagger$}
\begin{scope}
\draw[fill=white] (114.000000, 15.000000) +(-45.000000:8.485281pt and 8.485281pt) -- +(45.000000:8.485281pt and 8.485281pt) -- +(135.000000:8.485281pt and 8.485281pt) -- +(225.000000:8.485281pt and 8.485281pt) -- cycle;
\clip (114.000000, 15.000000) +(-45.000000:8.485281pt and 8.485281pt) -- +(45.000000:8.485281pt and 8.485281pt) -- +(135.000000:8.485281pt and 8.485281pt) -- +(225.000000:8.485281pt and 8.485281pt) -- cycle;
\draw (114.000000, 15.000000) node {{$T^\dagger$}};
\end{scope}
% Line 11: l1 +l2
\draw (135.000000,30.000000) -- (135.000000,15.000000);
\filldraw (135.000000, 30.000000) circle(1.500000pt);
\begin{scope}
\draw[fill=white] (135.000000, 15.000000) circle(3.000000pt);
\clip (135.000000, 15.000000) circle(3.000000pt);
\draw (132.000000, 15.000000) -- (138.000000, 15.000000);
\draw (135.000000, 12.000000) -- (135.000000, 18.000000);
\end{scope}
% Line 12: l0 +l2
\draw (153.000000,45.000000) -- (153.000000,15.000000);
\filldraw (153.000000, 45.000000) circle(1.500000pt);
\begin{scope}
\draw[fill=white] (153.000000, 15.000000) circle(3.000000pt);
\clip (153.000000, 15.000000) circle(3.000000pt);
\draw (150.000000, 15.000000) -- (156.000000, 15.000000);
\draw (153.000000, 12.000000) -- (153.000000, 18.000000);
\end{scope}
% Line 14: l2 G {$T^\dagger$}
\begin{scope}
\draw[fill=white] (174.000000, 15.000000) +(-45.000000:8.485281pt and 8.485281pt) -- +(45.000000:8.485281pt and 8.485281pt) -- +(135.000000:8.485281pt and 8.485281pt) -- +(225.000000:8.485281pt and 8.485281pt) -- cycle;
\clip (174.000000, 15.000000) +(-45.000000:8.485281pt and 8.485281pt) -- +(45.000000:8.485281pt and 8.485281pt) -- +(135.000000:8.485281pt and 8.485281pt) -- +(225.000000:8.485281pt and 8.485281pt) -- cycle;
\draw (174.000000, 15.000000) node {{$T^\dagger$}};
\end{scope}
% Line 15: l3 G {$T$}
\begin{scope}
\draw[fill=white] (174.000000, -0.000000) +(-45.000000:8.485281pt and 8.485281pt) -- +(45.000000:8.485281pt and 8.485281pt) -- +(135.000000:8.485281pt and 8.485281pt) -- +(225.000000:8.485281pt and 8.485281pt) -- cycle;
\clip (174.000000, -0.000000) +(-45.000000:8.485281pt and 8.485281pt) -- +(45.000000:8.485281pt and 8.485281pt) -- +(135.000000:8.485281pt and 8.485281pt) -- +(225.000000:8.485281pt and 8.485281pt) -- cycle;
\draw (174.000000, -0.000000) node {{$T$}};
\end{scope}
% Line 16: l1 G {$T$}
\begin{scope}
\draw[fill=white] (174.000000, 30.000000) +(-45.000000:8.485281pt and 8.485281pt) -- +(45.000000:8.485281pt and 8.485281pt) -- +(135.000000:8.485281pt and 8.485281pt) -- +(225.000000:8.485281pt and 8.485281pt) -- cycle;
\clip (174.000000, 30.000000) +(-45.000000:8.485281pt and 8.485281pt) -- +(45.000000:8.485281pt and 8.485281pt) -- +(135.000000:8.485281pt and 8.485281pt) -- +(225.000000:8.485281pt and 8.485281pt) -- cycle;
\draw (174.000000, 30.000000) node {{$T$}};
\end{scope}
% Line 18: l0 +l2
\draw (195.000000,45.000000) -- (195.000000,15.000000);
\filldraw (195.000000, 45.000000) circle(1.500000pt);
\begin{scope}
\draw[fill=white] (195.000000, 15.000000) circle(3.000000pt);
\clip (195.000000, 15.000000) circle(3.000000pt);
\draw (192.000000, 15.000000) -- (198.000000, 15.000000);
\draw (195.000000, 12.000000) -- (195.000000, 18.000000);
\end{scope}
% Line 19: l0 +l1
\draw (213.000000,45.000000) -- (213.000000,30.000000);
\filldraw (213.000000, 45.000000) circle(1.500000pt);
\begin{scope}
\draw[fill=white] (213.000000, 30.000000) circle(3.000000pt);
\clip (213.000000, 30.000000) circle(3.000000pt);
\draw (210.000000, 30.000000) -- (216.000000, 30.000000);
\draw (213.000000, 27.000000) -- (213.000000, 33.000000);
\end{scope}
% Line 21: l3 +l2
\draw (213.000000,15.000000) -- (213.000000,0.000000);
\filldraw (213.000000, 0.000000) circle(1.500000pt);
\begin{scope}
\draw[fill=white] (213.000000, 15.000000) circle(3.000000pt);
\clip (213.000000, 15.000000) circle(3.000000pt);
\draw (210.000000, 15.000000) -- (216.000000, 15.000000);
\draw (213.000000, 12.000000) -- (213.000000, 18.000000);
\end{scope}
% Line 22: l2 H
\begin{scope}
\draw[fill=white] (234.000000, 15.000000) +(-45.000000:8.485281pt and 8.485281pt) -- +(45.000000:8.485281pt and 8.485281pt) -- +(135.000000:8.485281pt and 8.485281pt) -- +(225.000000:8.485281pt and 8.485281pt) -- cycle;
\clip (234.000000, 15.000000) +(-45.000000:8.485281pt and 8.485281pt) -- +(45.000000:8.485281pt and 8.485281pt) -- +(135.000000:8.485281pt and 8.485281pt) -- +(225.000000:8.485281pt and 8.485281pt) -- cycle;
\draw (234.000000, 15.000000) node {$H$};
\end{scope}
% Line 23: l3 H
\begin{scope}
\draw[fill=white] (234.000000, -0.000000) +(-45.000000:8.485281pt and 8.485281pt) -- +(45.000000:8.485281pt and 8.485281pt) -- +(135.000000:8.485281pt and 8.485281pt) -- +(225.000000:8.485281pt and 8.485281pt) -- cycle;
\clip (234.000000, -0.000000) +(-45.000000:8.485281pt and 8.485281pt) -- +(45.000000:8.485281pt and 8.485281pt) -- +(135.000000:8.485281pt and 8.485281pt) -- +(225.000000:8.485281pt and 8.485281pt) -- cycle;
\draw (234.000000, -0.000000) node {$H$};
\end{scope}
% Line 24: l3 +l2
\draw (255.000000,15.000000) -- (255.000000,0.000000);
\filldraw (255.000000, 0.000000) circle(1.500000pt);
\begin{scope}
\draw[fill=white] (255.000000, 15.000000) circle(3.000000pt);
\clip (255.000000, 15.000000) circle(3.000000pt);
\draw (252.000000, 15.000000) -- (258.000000, 15.000000);
\draw (255.000000, 12.000000) -- (255.000000, 18.000000);
\end{scope}
% Line 25: l2 H
\begin{scope}
\draw[fill=white] (276.000000, 15.000000) +(-45.000000:8.485281pt and 8.485281pt) -- +(45.000000:8.485281pt and 8.485281pt) -- +(135.000000:8.485281pt and 8.485281pt) -- +(225.000000:8.485281pt and 8.485281pt) -- cycle;
\clip (276.000000, 15.000000) +(-45.000000:8.485281pt and 8.485281pt) -- +(45.000000:8.485281pt and 8.485281pt) -- +(135.000000:8.485281pt and 8.485281pt) -- +(225.000000:8.485281pt and 8.485281pt) -- cycle;
\draw (276.000000, 15.000000) node {$H$};
\end{scope}
% Line 26: l3 H
\begin{scope}
\draw[fill=white] (276.000000, -0.000000) +(-45.000000:8.485281pt and 8.485281pt) -- +(45.000000:8.485281pt and 8.485281pt) -- +(135.000000:8.485281pt and 8.485281pt) -- +(225.000000:8.485281pt and 8.485281pt) -- cycle;
\clip (276.000000, -0.000000) +(-45.000000:8.485281pt and 8.485281pt) -- +(45.000000:8.485281pt and 8.485281pt) -- +(135.000000:8.485281pt and 8.485281pt) -- +(225.000000:8.485281pt and 8.485281pt) -- cycle;
\draw (276.000000, -0.000000) node {$H$};
\end{scope}
% Line 27: l3 +l2 color=blue
\begin{scope}[color=blue]
\draw (297.000000,15.000000) -- (297.000000,0.000000);
\filldraw (297.000000, 0.000000) circle(1.500000pt);
\begin{scope}
\draw[fill=white] (297.000000, 15.000000) circle(3.000000pt);
\clip (297.000000, 15.000000) circle(3.000000pt);
\draw (294.000000, 15.000000) -- (300.000000, 15.000000);
\draw (297.000000, 12.000000) -- (297.000000, 18.000000);
\end{scope}
\end{scope}
% Line 29: l1 +l2 color=red
\begin{scope}[color=red]
\draw (315.000000,30.000000) -- (315.000000,15.000000);
\filldraw (315.000000, 30.000000) circle(1.500000pt);
\begin{scope}
\draw[fill=white] (315.000000, 15.000000) circle(3.000000pt);
\clip (315.000000, 15.000000) circle(3.000000pt);
\draw (312.000000, 15.000000) -- (318.000000, 15.000000);
\draw (315.000000, 12.000000) -- (315.000000, 18.000000);
\end{scope}
\end{scope}
% Line 30: l3 +l2 color=blue
\begin{scope}[color=blue]
\draw (333.000000,15.000000) -- (333.000000,0.000000);
\filldraw (333.000000, 0.000000) circle(1.500000pt);
\begin{scope}
\draw[fill=white] (333.000000, 15.000000) circle(3.000000pt);
\clip (333.000000, 15.000000) circle(3.000000pt);
\draw (330.000000, 15.000000) -- (336.000000, 15.000000);
\draw (333.000000, 12.000000) -- (333.000000, 18.000000);
\end{scope}
\end{scope}
% Line 31: l2 G {$T$}
\begin{scope}
\draw[fill=white] (354.000000, 15.000000) +(-45.000000:8.485281pt and 8.485281pt) -- +(45.000000:8.485281pt and 8.485281pt) -- +(135.000000:8.485281pt and 8.485281pt) -- +(225.000000:8.485281pt and 8.485281pt) -- cycle;
\clip (354.000000, 15.000000) +(-45.000000:8.485281pt and 8.485281pt) -- +(45.000000:8.485281pt and 8.485281pt) -- +(135.000000:8.485281pt and 8.485281pt) -- +(225.000000:8.485281pt and 8.485281pt) -- cycle;
\draw (354.000000, 15.000000) node {{$T$}};
\end{scope}
% Line 32: l3 +l2
\draw (375.000000,15.000000) -- (375.000000,0.000000);
\filldraw (375.000000, 0.000000) circle(1.500000pt);
\begin{scope}
\draw[fill=white] (375.000000, 15.000000) circle(3.000000pt);
\clip (375.000000, 15.000000) circle(3.000000pt);
\draw (372.000000, 15.000000) -- (378.000000, 15.000000);
\draw (375.000000, 12.000000) -- (375.000000, 18.000000);
\end{scope}
% Line 33: l2 G {$T^\dagger$}
\begin{scope}
\draw[fill=white] (396.000000, 15.000000) +(-45.000000:8.485281pt and 8.485281pt) -- +(45.000000:8.485281pt and 8.485281pt) -- +(135.000000:8.485281pt and 8.485281pt) -- +(225.000000:8.485281pt and 8.485281pt) -- cycle;
\clip (396.000000, 15.000000) +(-45.000000:8.485281pt and 8.485281pt) -- +(45.000000:8.485281pt and 8.485281pt) -- +(135.000000:8.485281pt and 8.485281pt) -- +(225.000000:8.485281pt and 8.485281pt) -- cycle;
\draw (396.000000, 15.000000) node {{$T^\dagger$}};
\end{scope}
% Line 34: l3 G {$S$}
\begin{scope}
\draw[fill=white] (396.000000, -0.000000) +(-45.000000:8.485281pt and 8.485281pt) -- +(45.000000:8.485281pt and 8.485281pt) -- +(135.000000:8.485281pt and 8.485281pt) -- +(225.000000:8.485281pt and 8.485281pt) -- cycle;
\clip (396.000000, -0.000000) +(-45.000000:8.485281pt and 8.485281pt) -- +(45.000000:8.485281pt and 8.485281pt) -- +(135.000000:8.485281pt and 8.485281pt) -- +(225.000000:8.485281pt and 8.485281pt) -- cycle;
\draw (396.000000, -0.000000) node {{$S$}};
\end{scope}
% Line 35: l3 H
\begin{scope}
\draw[fill=white] (420.000000, -0.000000) +(-45.000000:8.485281pt and 8.485281pt) -- +(45.000000:8.485281pt and 8.485281pt) -- +(135.000000:8.485281pt and 8.485281pt) -- +(225.000000:8.485281pt and 8.485281pt) -- cycle;
\clip (420.000000, -0.000000) +(-45.000000:8.485281pt and 8.485281pt) -- +(45.000000:8.485281pt and 8.485281pt) -- +(135.000000:8.485281pt and 8.485281pt) -- +(225.000000:8.485281pt and 8.485281pt) -- cycle;
\draw (420.000000, -0.000000) node {$H$};
\end{scope}
% Done with gates; drawing ending labels
\draw[color=black] (432.000000,45.000000) node[right] {$o0$};
\draw[color=black] (432.000000,30.000000) node[right] {$o1$};
\draw[color=black] (432.000000,15.000000) node[right] {$o2$};
\draw[color=black] (432.000000,0.000000) node[right] {$o3$};
% Done with ending labels; drawing cut lines and comments
% Done with comments
\end{tikzpicture}

%% file: Full_Adder_c_c.tikz
\begin{tikzpicture}[scale=1.000000,x=1pt,y=1pt]
\filldraw[color=white] (0.000000, -7.500000) rectangle (396.000000, 52.500000);
% Drawing wires
% Line 1: l0 W i0 o0
\draw[color=black] (0.000000,45.000000) -- (396.000000,45.000000);
\draw[color=black] (0.000000,45.000000) node[left] {$i0$};
% Line 2: l1 W i1 o1
\draw[color=black] (0.000000,30.000000) -- (396.000000,30.000000);
\draw[color=black] (0.000000,30.000000) node[left] {$i1$};
% Line 3: l2 W i3 o2
\draw[color=black] (0.000000,15.000000) -- (396.000000,15.000000);
\draw[color=black] (0.000000,15.000000) node[left] {$i3$};
% Line 4: l3 W i2 o3
\draw[color=black] (0.000000,0.000000) -- (396.000000,0.000000);
\draw[color=black] (0.000000,0.000000) node[left] {$i2$};
% Done with wires; drawing gates
% Line 5: l2 H
\begin{scope}
\draw[fill=white] (12.000000, 15.000000) +(-45.000000:8.485281pt and 8.485281pt) -- +(45.000000:8.485281pt and 8.485281pt) -- +(135.000000:8.485281pt and 8.485281pt) -- +(225.000000:8.485281pt and 8.485281pt) -- cycle;
\clip (12.000000, 15.000000) +(-45.000000:8.485281pt and 8.485281pt) -- +(45.000000:8.485281pt and 8.485281pt) -- +(135.000000:8.485281pt and 8.485281pt) -- +(225.000000:8.485281pt and 8.485281pt) -- cycle;
\draw (12.000000, 15.000000) node {$H$};
\end{scope}
% Line 6: l3 +l2
\draw (33.000000,15.000000) -- (33.000000,0.000000);
\filldraw (33.000000, 0.000000) circle(1.500000pt);
\begin{scope}
\draw[fill=white] (33.000000, 15.000000) circle(3.000000pt);
\clip (33.000000, 15.000000) circle(3.000000pt);
\draw (30.000000, 15.000000) -- (36.000000, 15.000000);
\draw (33.000000, 12.000000) -- (33.000000, 18.000000);
\end{scope}
% Line 7: l2 G {$T^\dagger$}
\begin{scope}
\draw[fill=white] (54.000000, 15.000000) +(-45.000000:8.485281pt and 8.485281pt) -- +(45.000000:8.485281pt and 8.485281pt) -- +(135.000000:8.485281pt and 8.485281pt) -- +(225.000000:8.485281pt and 8.485281pt) -- cycle;
\clip (54.000000, 15.000000) +(-45.000000:8.485281pt and 8.485281pt) -- +(45.000000:8.485281pt and 8.485281pt) -- +(135.000000:8.485281pt and 8.485281pt) -- +(225.000000:8.485281pt and 8.485281pt) -- cycle;
\draw (54.000000, 15.000000) node {{$T^\dagger$}};
\end{scope}
% Line 8: l3 +l2
\draw (75.000000,15.000000) -- (75.000000,0.000000);
\filldraw (75.000000, 0.000000) circle(1.500000pt);
\begin{scope}
\draw[fill=white] (75.000000, 15.000000) circle(3.000000pt);
\clip (75.000000, 15.000000) circle(3.000000pt);
\draw (72.000000, 15.000000) -- (78.000000, 15.000000);
\draw (75.000000, 12.000000) -- (75.000000, 18.000000);
\end{scope}
% Line 9: l1 +l2
\draw (93.000000,30.000000) -- (93.000000,15.000000);
\filldraw (93.000000, 30.000000) circle(1.500000pt);
\begin{scope}
\draw[fill=white] (93.000000, 15.000000) circle(3.000000pt);
\clip (93.000000, 15.000000) circle(3.000000pt);
\draw (90.000000, 15.000000) -- (96.000000, 15.000000);
\draw (93.000000, 12.000000) -- (93.000000, 18.000000);
\end{scope}
% Line 10: l2 G {$T^\dagger$}
\begin{scope}
\draw[fill=white] (114.000000, 15.000000) +(-45.000000:8.485281pt and 8.485281pt) -- +(45.000000:8.485281pt and 8.485281pt) -- +(135.000000:8.485281pt and 8.485281pt) -- +(225.000000:8.485281pt and 8.485281pt) -- cycle;
\clip (114.000000, 15.000000) +(-45.000000:8.485281pt and 8.485281pt) -- +(45.000000:8.485281pt and 8.485281pt) -- +(135.000000:8.485281pt and 8.485281pt) -- +(225.000000:8.485281pt and 8.485281pt) -- cycle;
\draw (114.000000, 15.000000) node {{$T^\dagger$}};
\end{scope}
% Line 11: l1 +l2
\draw (135.000000,30.000000) -- (135.000000,15.000000);
\filldraw (135.000000, 30.000000) circle(1.500000pt);
\begin{scope}
\draw[fill=white] (135.000000, 15.000000) circle(3.000000pt);
\clip (135.000000, 15.000000) circle(3.000000pt);
\draw (132.000000, 15.000000) -- (138.000000, 15.000000);
\draw (135.000000, 12.000000) -- (135.000000, 18.000000);
\end{scope}
% Line 12: l0 +l2
\draw (153.000000,45.000000) -- (153.000000,15.000000);
\filldraw (153.000000, 45.000000) circle(1.500000pt);
\begin{scope}
\draw[fill=white] (153.000000, 15.000000) circle(3.000000pt);
\clip (153.000000, 15.000000) circle(3.000000pt);
\draw (150.000000, 15.000000) -- (156.000000, 15.000000);
\draw (153.000000, 12.000000) -- (153.000000, 18.000000);
\end{scope}
% Line 14: l2 G {$T^\dagger$}
\begin{scope}
\draw[fill=white] (174.000000, 15.000000) +(-45.000000:8.485281pt and 8.485281pt) -- +(45.000000:8.485281pt and 8.485281pt) -- +(135.000000:8.485281pt and 8.485281pt) -- +(225.000000:8.485281pt and 8.485281pt) -- cycle;
\clip (174.000000, 15.000000) +(-45.000000:8.485281pt and 8.485281pt) -- +(45.000000:8.485281pt and 8.485281pt) -- +(135.000000:8.485281pt and 8.485281pt) -- +(225.000000:8.485281pt and 8.485281pt) -- cycle;
\draw (174.000000, 15.000000) node {{$T^\dagger$}};
\end{scope}
% Line 15: l3 G {$T$}
\begin{scope}
\draw[fill=white] (174.000000, -0.000000) +(-45.000000:8.485281pt and 8.485281pt) -- +(45.000000:8.485281pt and 8.485281pt) -- +(135.000000:8.485281pt and 8.485281pt) -- +(225.000000:8.485281pt and 8.485281pt) -- cycle;
\clip (174.000000, -0.000000) +(-45.000000:8.485281pt and 8.485281pt) -- +(45.000000:8.485281pt and 8.485281pt) -- +(135.000000:8.485281pt and 8.485281pt) -- +(225.000000:8.485281pt and 8.485281pt) -- cycle;
\draw (174.000000, -0.000000) node {{$T$}};
\end{scope}
% Line 16: l1 G {$T$}
\begin{scope}
\draw[fill=white] (174.000000, 30.000000) +(-45.000000:8.485281pt and 8.485281pt) -- +(45.000000:8.485281pt and 8.485281pt) -- +(135.000000:8.485281pt and 8.485281pt) -- +(225.000000:8.485281pt and 8.485281pt) -- cycle;
\clip (174.000000, 30.000000) +(-45.000000:8.485281pt and 8.485281pt) -- +(45.000000:8.485281pt and 8.485281pt) -- +(135.000000:8.485281pt and 8.485281pt) -- +(225.000000:8.485281pt and 8.485281pt) -- cycle;
\draw (174.000000, 30.000000) node {{$T$}};
\end{scope}
% Line 18: l0 +l2
\draw (195.000000,45.000000) -- (195.000000,15.000000);
\filldraw (195.000000, 45.000000) circle(1.500000pt);
\begin{scope}
\draw[fill=white] (195.000000, 15.000000) circle(3.000000pt);
\clip (195.000000, 15.000000) circle(3.000000pt);
\draw (192.000000, 15.000000) -- (198.000000, 15.000000);
\draw (195.000000, 12.000000) -- (195.000000, 18.000000);
\end{scope}
% Line 19: l0 +l1
\draw (213.000000,45.000000) -- (213.000000,30.000000);
\filldraw (213.000000, 45.000000) circle(1.500000pt);
\begin{scope}
\draw[fill=white] (213.000000, 30.000000) circle(3.000000pt);
\clip (213.000000, 30.000000) circle(3.000000pt);
\draw (210.000000, 30.000000) -- (216.000000, 30.000000);
\draw (213.000000, 27.000000) -- (213.000000, 33.000000);
\end{scope}
% Line 21: l3 +l2
\draw (213.000000,15.000000) -- (213.000000,0.000000);
\filldraw (213.000000, 0.000000) circle(1.500000pt);
\begin{scope}
\draw[fill=white] (213.000000, 15.000000) circle(3.000000pt);
\clip (213.000000, 15.000000) circle(3.000000pt);
\draw (210.000000, 15.000000) -- (216.000000, 15.000000);
\draw (213.000000, 12.000000) -- (213.000000, 18.000000);
\end{scope}
% Line 22: l2 H
\begin{scope}
\draw[fill=white] (234.000000, 15.000000) +(-45.000000:8.485281pt and 8.485281pt) -- +(45.000000:8.485281pt and 8.485281pt) -- +(135.000000:8.485281pt and 8.485281pt) -- +(225.000000:8.485281pt and 8.485281pt) -- cycle;
\clip (234.000000, 15.000000) +(-45.000000:8.485281pt and 8.485281pt) -- +(45.000000:8.485281pt and 8.485281pt) -- +(135.000000:8.485281pt and 8.485281pt) -- +(225.000000:8.485281pt and 8.485281pt) -- cycle;
\draw (234.000000, 15.000000) node {$H$};
\end{scope}
% Line 23: l3 H
\begin{scope}
\draw[fill=white] (234.000000, -0.000000) +(-45.000000:8.485281pt and 8.485281pt) -- +(45.000000:8.485281pt and 8.485281pt) -- +(135.000000:8.485281pt and 8.485281pt) -- +(225.000000:8.485281pt and 8.485281pt) -- cycle;
\clip (234.000000, -0.000000) +(-45.000000:8.485281pt and 8.485281pt) -- +(45.000000:8.485281pt and 8.485281pt) -- +(135.000000:8.485281pt and 8.485281pt) -- +(225.000000:8.485281pt and 8.485281pt) -- cycle;
\draw (234.000000, -0.000000) node {$H$};
\end{scope}
% Line 24: l3 +l2
\draw (255.000000,15.000000) -- (255.000000,0.000000);
\filldraw (255.000000, 0.000000) circle(1.500000pt);
\begin{scope}
\draw[fill=white] (255.000000, 15.000000) circle(3.000000pt);
\clip (255.000000, 15.000000) circle(3.000000pt);
\draw (252.000000, 15.000000) -- (258.000000, 15.000000);
\draw (255.000000, 12.000000) -- (255.000000, 18.000000);
\end{scope}
% Line 25: l2 H
\begin{scope}
\draw[fill=white] (276.000000, 15.000000) +(-45.000000:8.485281pt and 8.485281pt) -- +(45.000000:8.485281pt and 8.485281pt) -- +(135.000000:8.485281pt and 8.485281pt) -- +(225.000000:8.485281pt and 8.485281pt) -- cycle;
\clip (276.000000, 15.000000) +(-45.000000:8.485281pt and 8.485281pt) -- +(45.000000:8.485281pt and 8.485281pt) -- +(135.000000:8.485281pt and 8.485281pt) -- +(225.000000:8.485281pt and 8.485281pt) -- cycle;
\draw (276.000000, 15.000000) node {$H$};
\end{scope}
% Line 26: l3 H
\begin{scope}
\draw[fill=white] (276.000000, -0.000000) +(-45.000000:8.485281pt and 8.485281pt) -- +(45.000000:8.485281pt and 8.485281pt) -- +(135.000000:8.485281pt and 8.485281pt) -- +(225.000000:8.485281pt and 8.485281pt) -- cycle;
\clip (276.000000, -0.000000) +(-45.000000:8.485281pt and 8.485281pt) -- +(45.000000:8.485281pt and 8.485281pt) -- +(135.000000:8.485281pt and 8.485281pt) -- +(225.000000:8.485281pt and 8.485281pt) -- cycle;
\draw (276.000000, -0.000000) node {$H$};
\end{scope}
% Line 29: l1 +l2
\draw (297.000000,30.000000) -- (297.000000,15.000000);
\filldraw (297.000000, 30.000000) circle(1.500000pt);
\begin{scope}
\draw[fill=white] (297.000000, 15.000000) circle(3.000000pt);
\clip (297.000000, 15.000000) circle(3.000000pt);
\draw (294.000000, 15.000000) -- (300.000000, 15.000000);
\draw (297.000000, 12.000000) -- (297.000000, 18.000000);
\end{scope}
% Line 31: l2 G {$T$}
\begin{scope}
\draw[fill=white] (318.000000, 15.000000) +(-45.000000:8.485281pt and 8.485281pt) -- +(45.000000:8.485281pt and 8.485281pt) -- +(135.000000:8.485281pt and 8.485281pt) -- +(225.000000:8.485281pt and 8.485281pt) -- cycle;
\clip (318.000000, 15.000000) +(-45.000000:8.485281pt and 8.485281pt) -- +(45.000000:8.485281pt and 8.485281pt) -- +(135.000000:8.485281pt and 8.485281pt) -- +(225.000000:8.485281pt and 8.485281pt) -- cycle;
\draw (318.000000, 15.000000) node {{$T$}};
\end{scope}
% Line 32: l3 +l2
\draw (339.000000,15.000000) -- (339.000000,0.000000);
\filldraw (339.000000, 0.000000) circle(1.500000pt);
\begin{scope}
\draw[fill=white] (339.000000, 15.000000) circle(3.000000pt);
\clip (339.000000, 15.000000) circle(3.000000pt);
\draw (336.000000, 15.000000) -- (342.000000, 15.000000);
\draw (339.000000, 12.000000) -- (339.000000, 18.000000);
\end{scope}
% Line 33: l2 G {$T^\dagger$}
\begin{scope}
\draw[fill=white] (360.000000, 15.000000) +(-45.000000:8.485281pt and 8.485281pt) -- +(45.000000:8.485281pt and 8.485281pt) -- +(135.000000:8.485281pt and 8.485281pt) -- +(225.000000:8.485281pt and 8.485281pt) -- cycle;
\clip (360.000000, 15.000000) +(-45.000000:8.485281pt and 8.485281pt) -- +(45.000000:8.485281pt and 8.485281pt) -- +(135.000000:8.485281pt and 8.485281pt) -- +(225.000000:8.485281pt and 8.485281pt) -- cycle;
\draw (360.000000, 15.000000) node {{$T^\dagger$}};
\end{scope}
% Line 34: l3 G {$S$}
\begin{scope}
\draw[fill=white] (360.000000, -0.000000) +(-45.000000:8.485281pt and 8.485281pt) -- +(45.000000:8.485281pt and 8.485281pt) -- +(135.000000:8.485281pt and 8.485281pt) -- +(225.000000:8.485281pt and 8.485281pt) -- cycle;
\clip (360.000000, -0.000000) +(-45.000000:8.485281pt and 8.485281pt) -- +(45.000000:8.485281pt and 8.485281pt) -- +(135.000000:8.485281pt and 8.485281pt) -- +(225.000000:8.485281pt and 8.485281pt) -- cycle;
\draw (360.000000, -0.000000) node {{$S$}};
\end{scope}
% Line 35: l3 H
\begin{scope}
\draw[fill=white] (384.000000, -0.000000) +(-45.000000:8.485281pt and 8.485281pt) -- +(45.000000:8.485281pt and 8.485281pt) -- +(135.000000:8.485281pt and 8.485281pt) -- +(225.000000:8.485281pt and 8.485281pt) -- cycle;
\clip (384.000000, -0.000000) +(-45.000000:8.485281pt and 8.485281pt) -- +(45.000000:8.485281pt and 8.485281pt) -- +(135.000000:8.485281pt and 8.485281pt) -- +(225.000000:8.485281pt and 8.485281pt) -- cycle;
\draw (384.000000, -0.000000) node {$H$};
\end{scope}
% Done with gates; drawing ending labels
\draw[color=black] (396.000000,45.000000) node[right] {$o0$};
\draw[color=black] (396.000000,30.000000) node[right] {$o1$};
\draw[color=black] (396.000000,15.000000) node[right] {$o2$};
\draw[color=black] (396.000000,0.000000) node[right] {$o3$};
% Done with ending labels; drawing cut lines and comments
% Done with comments
\end{tikzpicture}